\documentclass{aa}  
\bibpunct{(}{)}{;}{a}{}{,}
\usepackage{graphicx}
\usepackage{ gensymb }
\usepackage{ tipa }
\usepackage{ textcomp }
\usepackage{ wasysym }
\usepackage{float}
\usepackage{txfonts}
\usepackage{lscape}
\usepackage{natbib}
\begin{document}
   \title{CO ro-vibrational lines in HD100546
\thanks{Based on observations made with ESO Telescopes at the La Silla Paranal Observatory under programme ID 084.C-0605.}}

   \subtitle{A search for disc asymmetries and the role of fluorescence}

   \author{R. P. Hein Bertelsen
	\inst{1}
	\and
  	I. Kamp\inst{1}
	\and
	M. Goto\inst{2}
	\and
	G. van der Plas\inst{3}
	\and
	W.-F. Thi\inst{4}
	\and
	L. B. F. M. Waters\inst{5}\inst{,6}
	\and
	M. E. van den Ancker\inst{7}
	\and
	P. Woitke\inst{8}
			}

                 \institute{Kapteyn Astronomical Institute Rijks-universiteit Groningen (RuG),
              Landleven 12, Groningen 9747 Netherlands\\
\email{bertelsen@astro.rug.nl}
     		\and
Universit\"ats-Sternwarte M\"unchen, Scheinerstr. 1, D-8167 9 Munich, Germany, Department of Physics and Astronomy, 3400 N
		\and
Departamento de Astronom\'{\i}a, Universidad de Chile, Casilla 36-D, Santiago, Chile
		\and
UJF-Grenoble 1 / CNRS-INSU, Institut de Plan\'{e}tologie et d'Astrophysique (IPAG) UMR 5274, Grenoble, F-38041, France
		\and
SRON Netherlands Institute for Space Research, Sorbonnelaan 2, 3584 CA Utrecht, The Netherlands 
		\and
Anton Pannekoek Astronomical Institute, University of Amsterdam, PO Box 94249, 1090 GE Amsterdam, The Netherlands 
		\and
European Southern Observatory, Karl-Schwarzschild-Str.2, D 85748 Garching bei M\"unchen, Germany
		\and
SUPA, School of Physics \& Astronomy, University of St. Andrews, North Haugh, St. Andrews KY16 9SS, UK             			}

% \abstract{}{}{}{}{} 
% 5 {} token are mandatory
 
  \abstract
  % context heading (optional)
  {}   
     % aims heading (mandatory)
   {We have studied the emission of CO ro-vibrational lines in the disc around the Herbig Be star HD100546 to determine physical properties, disc asymmetries, the CO excitation mechanism, and the spatial extent of the emission, with the final goal of using the CO ro-vibrational lines as a diagnostic to understand inner disc structure in the context of planet formation. }
  % methods heading (mandatory)
   { High-spectral-resolution infrared spectra of CO ro-vibrational emission at eight different position angles were taken with CRIRES (CRyogenic high-resolution InfraRed Echelle Spectrograph) at the VLT (Very Large Telescope). From these spectra flux tables, line profiles for individual CO ro-vibrational transitions, co-added line profiles, and population diagrams were produced. We have investigated variations in the line profile shapes and line strengths as a function of slit position angle. 
We used the thermo-chemical disc modelling code ProDiMo based on the chemistry, radiation field, and temperature structure of a previously published model for HD100546. We calculated line fluxes and profiles for the whole set of observed CO ro-vibrational transitions using a large CO model molecule that includes the lowest two electronic states, each with 7 vibrational levels and within them 60 rotational levels.
 Comparing observations and the model, we investigated the possibility of disc asymmetries, the excitation mechanism (UV fluorescence), the geometry, and physical conditions of the inner disc. }
  % results heading (mandatory)
   {The observed CO ro-vibrational lines are largely emitted from the inner rim of the outer disc at 10-13 AU. The line shapes are similar for all v levels and line fluxes from all vibrational levels vary only within one order of magnitude. All line profile asymmetries and variations can be explained with a symmetric disc model to which a slit correction and pointing offset is applied. Because the angular size of the CO emitting region (10-13 AU) and the slit width are comparable the line profiles are very sensitive to the placing of the slit. The model reproduces the line shapes and the fluxes of the v=1-0 lines as well as the spatial extent of the CO ro-vibrational emission. It does not reproduce the observed band ratios of 0.5-0.2 with higher vibrational bands. We find that lower gas volume densities at the surface of the inner rim of the outer disc can make the fluorescence pumping more efficient and reproduce the observed band ratios.}
% 
  % conclusions heading (optional), leave it empty if necessary 
   {}

   \keywords{Protoplanetary discs, Line: profiles, Stars: individual: HD100546, circumstellar matter, Techniques: imaging spectroscopy
               }
\authorrunning{Hein Bertelsen}
\titlerunning{CO ro-vibrational lines in HD100546}

   \maketitle
%
%________________________________________________________________

\section{Introduction}

The inner regions of protoplanetary discs are excellent laboratories for studying the formation of planets. 
At high spectral resolution, line profiles of various gas species contain a wealth of physical, chemical, and kinematic information. 
{The CO ro-vibrational lines around 4.7 $\mu$m have been shown to originate in the innermost regions of the discs \citep{najita2000,brittain2003,blake2004}.}

Holes or gaps due to planet formation can be directly traced in the line profiles of these CO ro-vibrational transitions \citep{Regaly2010}.
If CO is to be used as a probe, it is crucial that we understand the CO ro-vibrational lines in terms of the chemistry and line radiative transfer. We need to understand what governs line strengths and shapes and where the emission originates.

Comparing modelled and observed data, \citet{Brittain2007} found that UV fluorescence can be an important excitation mechanism for the CO in dust-depleted discs.
From the implementation of a large CO model molecule in the radiative thermo-chemical protoplanetary disc code ProDiMo, \citet{thi2012} confirmed that UV fluorescence has a significant impact on the population of the ro-vibrational levels of the CO molecule. {The main effect of the UV pumping is populating the $v>1$ levels.}

 A particularly well known Herbig Be star is HD100546, spectral type B9Vne, with a protoplanetary disc. It has been observed and analysed by several authors. In coronagraphic imaging, HD100546 has shown a large-scale envelope and a disc that extends out to 515 AU with an asymmetric brightness profile \citep{grady2001}. 
From its position in the HR diagram the age of the star has been estimated to about $10$Myr \citep{ancker1997}. The infrared spectrum of HD100546 shows exceptionally strong emission from crystalline silicates, suggesting a highly processed grain population, probably associated with the inner rim region of the outer disc \citep{malfait1998}.
 The disc has been classified as transitional with a gap from $4-13$AU \citep{grady2005} and a disc wall at 10-13 AU (from here on referred to as the disc wall). The gap might be caused by the presence of a giant planet or substellar companion \citep{acke2006}. Located at a distance of $103$ pc \citep{ancker1998}, the disc is spatially resolved on scales of 0$\farcs$1 corresponding to 10 AU. The disc inclination has been constrained to $i=42 \pm5$ and the position angle (P.A.) to $P.A.=145\pm5$ \citep{ardila2007}.
{The presence of molecular gas in the outer disc has been confirmed, with observations of CO pure rotational lines \citep{panic2010}. Based on these observations and new data from the Herschel Space Observatory, \citet{bruderer2012} modelled the disc and concluded that the highest CO rotational lines are emitted from $\sim$20-50 AU, the mid $J$ lines from $\sim$40-90 AU, and the low $J$ lines trace the outer disc. Furthermore, they favour a disc model with a gas/dust ratio of 100 with only a small fraction of volatile carbon}. \citet{benisty2010} presented observations from the VLT Interferometer using the AMBER instrument and obtained 26 visibilities in the [2.06-2.46] $\mu$m wavelength range and derived basic characteristics of the NIR emission. Their visibility curve is almost flat with wavelength, while the uniform brightness ring predicts a slightly steeper slope.
The interferometric observations are consistent with a disc model that includes a gap until $\sim$13 AU from the star and a total dust mass of $\sim$0.008 lunar mass inside it.
\citet{tatulli2011} aimed to refine the disc model presented in \citet{benisty2010}. Using interferometric data from the AMBER/VLTI instrument in the $H$- and $K$-band, they spatially resolved the warm inner disc and constrained the structure. Combining these with photometric observations they analysed the data using a passive disc model based on three dimensional Monte-Carlo radiative transfer. They found that the spectral energy distribution (SED) from the UV to mm range and the near Infrared (NIR) data was adequately reproduced by their model composed of a tenuous inner disc (0.24-4 AU) with a dust mass of $\sim$1.75$\cdot$ 10$^{-10}$M$_{\astrosun}$, a gap devoid of dust and a massive outer disc (13-500 AU) with a dust mass of $\sim4.3\cdot10^{-4}$M$_{\astrosun}$. Recent dust observations have revealed an inner disc that extends to no farther than 0.7 AU from the star with a following gap of about 10 AU \citep{panic2012}. Furthermore, the authors found this inner disc to be asymmetric, while the disc wall at 10 AU is fully symmetric.

The CO ro-vibrational transition lines at $4.7$ $\rm\mu m$ emitted from HD 100546, have been observed and studied and a lack of CO emission from small radii (\textless10 AU) has been documented \citep{plas2009,brittain2009}. It has been shown that the lack of CO cannot be caused by a completely gas-free inner disc, since [OI] 6300 $\AA$ emission has been observed from the disc \citep{acke2006}. {The main formation mechanism of [OI] 6300 $\AA$ emission is thought to be dissociation of OH. A detection of [OI] 6300 $\AA$ emission thereby suggests the presence of molecular gas in the inner disc. However, \citet{liskowsky2012} presented an OH spectrum from the disc and do not detect OH from small radii.}
Thermo-chemical modelling of the CH$^+$ emission observed with the Herschel Space Observatory in HD100546 suggests that CH$^+$ is mostly located at the second rim (10-13 AU) \citep{thi2011}. The same could be the case for CO, since no emission has been observed at smaller radii. \citet{goto2012} resolved the CO vibrational line emission from HD100546 with 0$\farcs$1 angular resolution using the CRIRES instrument and found unambiguous evidence of a warm disc atmosphere far away from the central star including a CO emitting region extending out to as far as 50 AU. {Recently \citet{liskowsky2012} and \citet{brittain2013} confirm the lack of CO emission from small radii. With a 0$\farcs$34 slit and a spatial resolution of 0$\farcs$4-0$\farcs$8, they find the main part of the CO emission to be consistent with an axisymmetric disc. Meanwhile gathering observations spanning several years, a periodically occurring asymmetry was detected only in the v=1-0 lines \citep{brittain2013}}.

In this paper we compare a detailed model with observational data using the CRIRES observations of the CO ro-vibrational emission from HD100546 from \citet{goto2012} and the corresponding modelled emission predicted by the thermo-chemical protoplanetary disc model ProDiMo \citep*{woitke2009}. This includes the comparison of line shapes, line strengths, and line ratios. {The possibility of disc asymmetries is investigated.} The goal is to test our understanding of the complex coupling between chemistry, IR and UV fluorescence, and radiative transfer, to assess the potential of the CO ro-vibrational emission lines as probes for the inner disc geometry.

First we present our CRIRES observations of HD 100546 (Sect. \ref{sec:obs}) and describe our data reduction method (Sect. \ref{sec:datared}). We then produce average line profiles and derive population diagrams (Sect. \ref{sec:res}). In Section \ref{sec:model}, we present the ProDiMo model of HD100546 with a large CO model molecule combined with a slit simulator. We explore the variety of line shapes that could come from changes in slit position and we produce modelled line profiles and population diagrams to be compared to the observational data. We finish the paper by comparing the observed results with the modelled results (Sect. \ref{sec:model} and \ref{discus}) and discuss possible solutions to some remaining inconsistencies between model and observations.
{We present our observational slit analysis and our modelled slit effects in the appendix.}

\section{Observations} \label{sec:obs}
\begin{figure}[!htbp]
\begin{center}
  \includegraphics[width=0.5\textwidth]{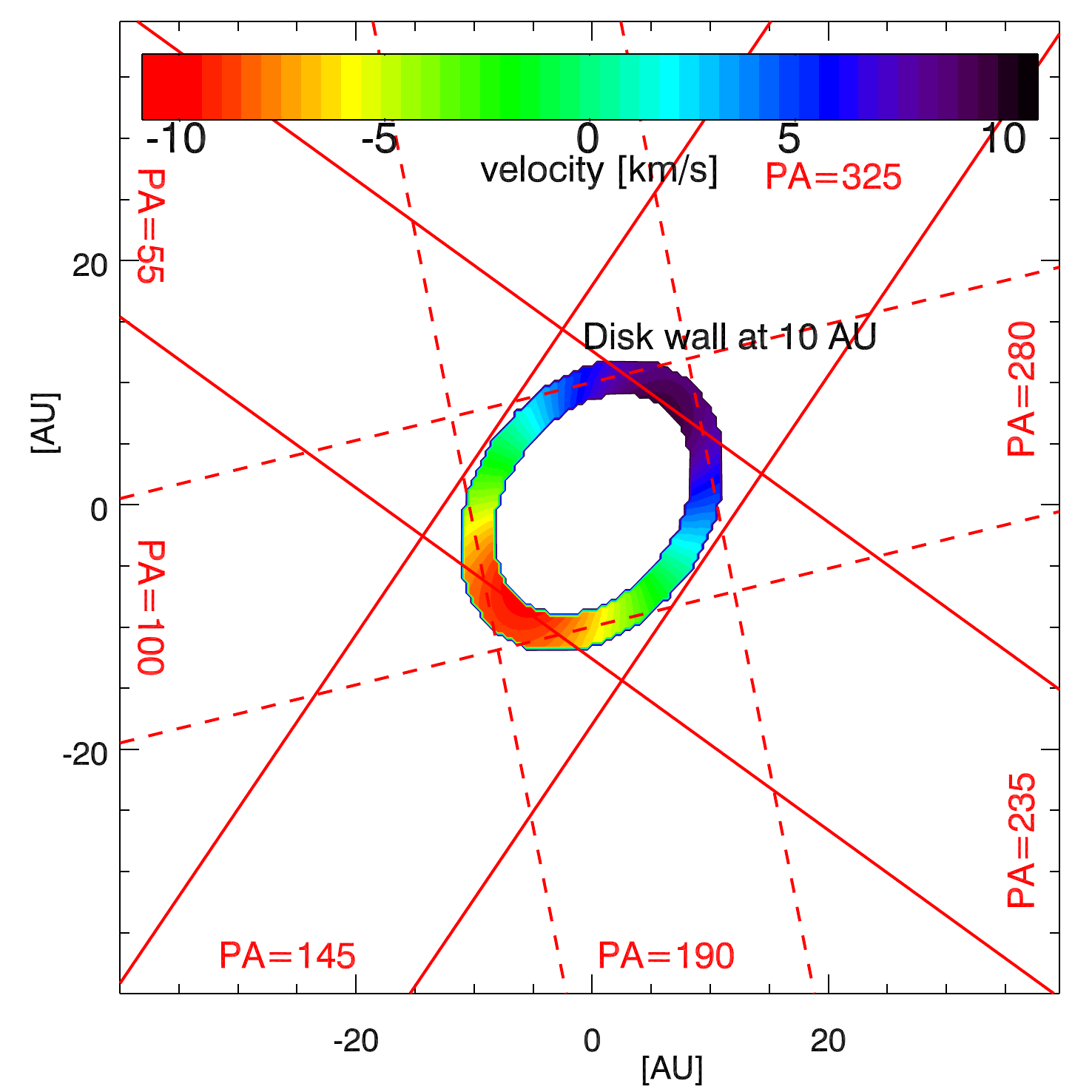}
\end{center}
\caption{A mock up velocity map and zoom in {on the surface of the disc wall (10-13 AU)} with the slit at P.A.=145$\degree$/325$\degree$ and P.A.=55$\degree$/235$\degree$ overplotted as solid red lines and the slit at P.A.=10$\degree$/190$\degree$ and P.A.=100$\degree$/280$\degree$ overplotted as dashed red lines. In the case of 145$\degree$/325$\degree$, we obtain clear double peaked profiles and at 55$\degree$/235$\degree$ flat topped profiles. The disc inclination is 42$\degree$ and the P.A. of the disc on the sky is 145$\degree$.}
\label{fig:velocity}
\end{figure}

High resolution spectra of HD100546 ($\Delta v$ = 3 km~s$^{-1}$, R=100,000) were obtained on 29 and 30 March 2010 UT with the VLT cryogenic high-resolution infrared echelle spectrograph (CRIRES)\citep{goto2012}. The wavelength interval between 4.6 to 5.0 $\mu$m was continuously covered with six different grating settings. The spectra were recorded by rotating the slit (slit width=0$\farcs$2) to P.A.=145\degr, 55\degr, 10\degr, 100\degr \, and their respective anti-parallel positions to increase the spatial coverage (see Fig. \ref{fig:velocity} for a view of the slit coverage). The telescope was nodded by 10" along the slit after each second exposure to subtract the sky emission and the dark current. The bright telluric standard stars, HR~6556 (A5III) and HR~6879 (B9.5II), were observed to remove the telluric absorption lines. The spectroscopic flat fields were collected in the morning after the observations with the same instrumental settings as used for the science observations. Table \ref{table:PA} provides a summary of the observations and {Fig. \ref{fig:velocity} shows a schematic drawing of how the slit is positioned with respect to the disc wall at the various P.A.}

\begin{table}[!htbp]
\caption{Observational settings and conditions}             % title of Table
\label{table:PA}      % is used to refer this table in the text
\centering                          % used for centering table
\begin{tabular}{c c c c c }        % centered columns (4 columns)
\hline            
 \hline
Date & $\lambda_{ref}$& 	P.A.&it		&STD	\\   
 & [$\mu$m] & 	&[min] 		&	\\   

\hline                           
$29/3$ 	 & 	$4662$	 & $145\degree/325\degree/55\degree/235\degree$	& 1	&HR~6556	\\   
2010 & $4.676$ & $145\degree/325\degree/55\degree/235\degree$	& 1	&HR~6556	\\   
 & $4.929$& $145\degree/325\degree/55\degree/235\degree$	& 1	&HR~6556	\\   
 & $4.957$ & $145\degree/325\degree/55\degree/235\degree$	& 1	&HR~6556	\\   

PSF$\sim$&0.169''&	&&				\\   

\hline\hline      
Date & $\lambda_{ref}$& 	P.A.&it		&STD	\\   
 & [$\mu$m] & 	&[min] 		&	\\   
\hline                           

$30/3$ 	 & $4.662$&$10\degree/190\degree/100\degree/280\degree$&1 &HR~6879\\   
2010  & $4.676$ &$10\degree/190\degree/100\degree/280\degree$ 	& 1	&HR~6879\\   
  & $4.782$& $145\degree/325\degree/55\degree/235\degree$ 	& 1	&HR~6879\\   
  & $4.796$& $145\degree/325\degree/55\degree/235\degree$ 	& 1	&HR~6879\\   

PSF$\sim$&0.186''& 		& &		\\   

\hline                           
\end{tabular}
\tablefoot{For each science spectra the associated telluric standard stars (STD) are listed. For each of the two observing nights the respective four wavelength settings are listed. For each wavelength setting four different position angles were taken. The PSF FWHM was measured at several locations in the STD spectra for each wavelength settings and the table lists an average, for each day.} 
\end{table}

{In our dataset, we found clear variations in the shape of line profiles observed at different position angles. These variations are caused by a varying slit loss. The details of this are discussed in Appendix \ref{sec:obs_com_prof}.}

\section{Data reduction and analysis} \label{sec:datared}

\subsection{Spectroscopy}
The science data was reduced using the CRIRES pipeline recipes ver. 1.11.0 on {\it esorex} platform ver. 3.8.1 and was corrected for the detector non-linearity and response. The observations where done in nodding mode: The spectra are nodded between two positions (A and B) 10" apart on the sky in the pattern ABBA so that the source is in different positions on the chip. The sky emission was removed by subtracting the nodded A and B spectra. The spectrograms were registered and combined, and one dimensional spectra were extracted for each grating setting and  slit position angle. The rectangular extraction method was used. {The width of the extracted rectangle was 41 pixels = 3$\farcs$53.} The data of the spectroscopic standard stars (HR 6556 and HR 6879) were reduced in the same way. The spectra of HD 100546 were divided by those of the standard stars after making small adjustments in the optical depth of the telluric lines, the wavelength, and the spectral resolution. These adjustments are done to remove the telluric lines as clean as possible. The wavelength calibration was performed by matching the telluric lines to the model atmospheric transmission spectra calculated using the LBLRTM code \citep{clough2005}. The calibration accuracy is better than 1 km s$^{-1}$. 

\subsection{Flux calibration}
There are several records of $M$-band photometry published. In 1988 the $M$-band magnitude was measured to $m_{\rm{M}}$=3.75 $\pm$ 0.06 \citep{winter2001}, where $m_{\rm{M}}$ refers to the apparent magnitude. In 1991, the ESO 1 m Schmidt telescope ($\lambda_{mean}$=4.6 $\mu$m)  measured  $m_{\rm{M}}$=3.76 \citep{fouque1992}; In 1992, \citet{garcia1997} obtained $m_{\rm{M}}$=3.79 $\pm$ 0.07. In 2009, HD 100546 was observed by the Wide-Field Infrared Survey Explorer (WISE), and the W2 brightness ($\lambda_{mean}$=4.6 $\mu$m ) was measured to  $m_{\rm{M}}$=3.156$\pm$0.049. However for sources brighter than magnitude 6, the WISE band 2 fluxes suffer from saturation effects and are therefore unreliable (Padgett, private communication).  

For our analysis, we calibrate the continuum level of HD 100546 to the spectroscopic standards. The $M$-band photometry of the standard stars, HR~6556 ( $m_{\rm{M}}$=1.62 mag) and HR~6879 ( $m_{\rm{M}}$=1.70~mag), was observed at the ESO 1-m telescope with the standard ESO near-infrared filter set ($\lambda_{c}$=4.46 $\mu$m, $\Delta \lambda$=0.8 $\mu$m). The continuum flux of HD 100546 (for the spectra taken with P.A.=145$\degree$ at reference wavelength=4.782 $\mu$m) after division by the STD was 9.73 Jy at the mean wavelength of the filter 4.750 $\mu$m. This is about a factor of 1.7 brighter than the above listed $M$-band photometry measurements. The error expected from the difference in the continuum slopes in HD 100546 and the standard star is 10-15\%. If our flux calibration is correct, HD 100546 is getting brighter by $\sim$70\% over the baseline of 18 years. Herbig stars are known to be intrinsically variable, and even the disc luminosity of some Herbig Ae/Be stars is known to change on short timescales \citep{bibo1991,sitko1994,eiroa2002}. 
{For HD100546, in particular, NIR variability has been documented: In the $K$-band a 0.5 mag decrease over a span of seven years and in the $L$-band a 0.84 mag decrease over the span of 20 years \citep{brittain2013}. Furthermore, \citet{brittain2013} find the CO ro-vibrational hot band emission ($\Delta$$v$=1, $v'$\textgreater1) to brighten over a span of eight years, and suggest this could be due to a 0.4 mag variation in the $M$-band flux. Thus, changes in the $M$-band continuum are not unlikely, but a follow up observation measuring the $M$-band magnitude should be done in order to confirm this increase in luminosity.} 

Our data suffers from slit loss both for the continuum and the lines. If we calibrate to previous measurements of the continuum, we lose the natural variation in fluxes between spectra taken at different position angles. However, this is exactly what we aim to explore. We will therefore use the spectra calibrated to the spectroscopic standard for the analysis hereafter (Table \ref{tab:line flux}).

\subsection{Line detection and selection}
\begin{figure*}[!htbp]
   \includegraphics[width=\textwidth]{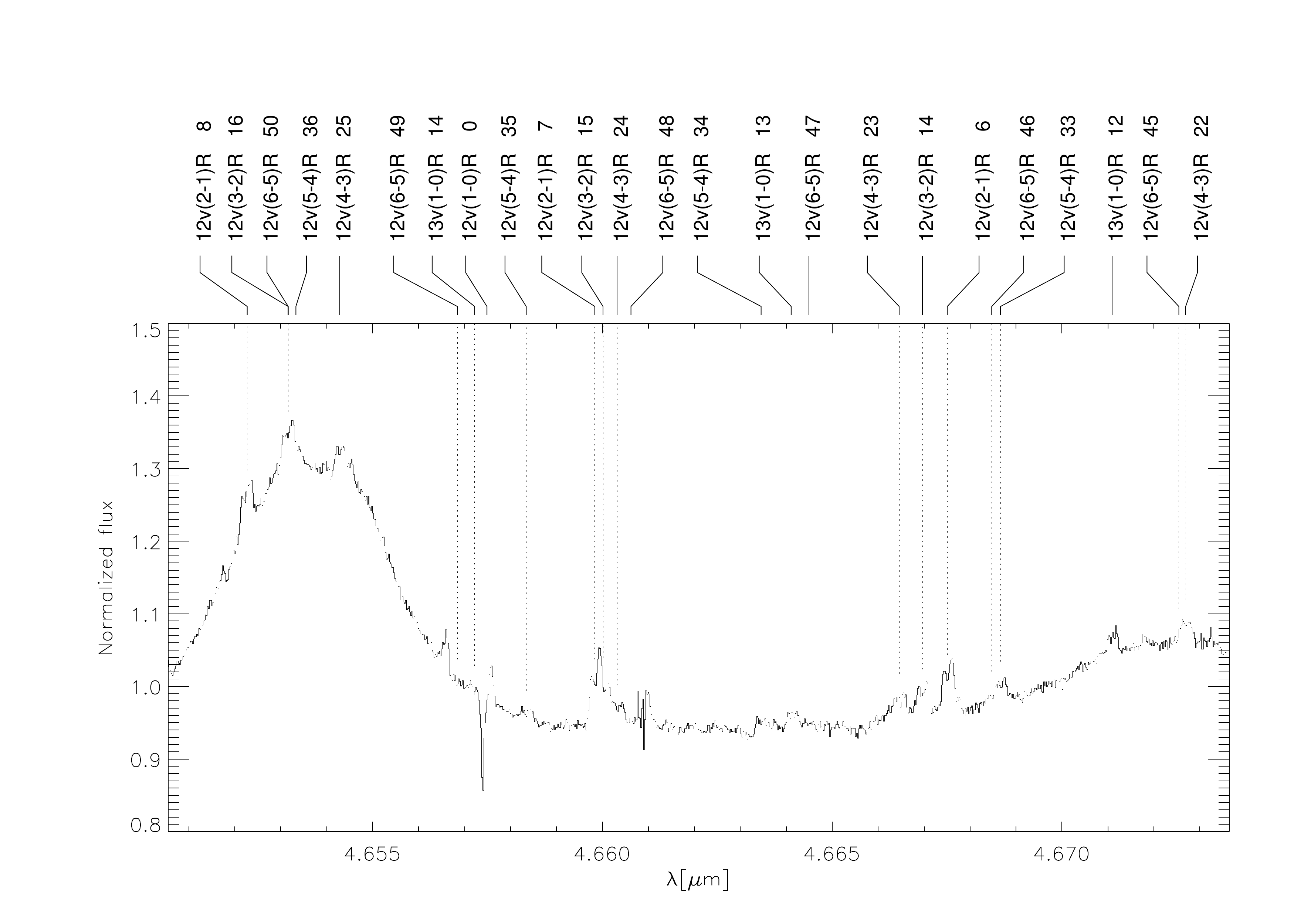}
      \caption{{The full spectrum collected at P.A.=145$\degree$ in the wavelength setting 4.662 $\mu$m (chip 3) with the CO ro-vibrational line identifications overplotted. The prefix 12 and 13 denote the different isotopologues $^{12}$CO and $^{13}$CO. Examples of imperfect telluric reductions are seen at: 4.6517, 4.6566, 4.6574 and 4.6610 $\mu$m.}}
         \label{fig:spectrum}
   \end{figure*}

For our full data set covering six different wavelength settings the quality of the data varies. For the 29th and the 30th respectively the Strehl ratio (the efficiency of the AO system) was $\sim$42$\%$ and $\sim$28$\%$ in the $K$-band. This corresponds to Strehl ratios at our wavelength settings of $\sim$80$\%$ on the 29th and a few percentages lower on the 30th. For CRIRES a good correction in the $K$-band typically corresponds to a Strehl ratio higher than 30$\%$ (CRIRES manual). The wavelength settings 4.662 $\mu$m and 4.676 $\mu$m observed on the 29th have the highest quality. 
The spectra from the wavelength settings 4782 and 4.796 $\mu$m were so noisy that it was necessary to convolve the signal with a Gaussian kernel in order to detect more than just a few lines.
{Within each spectrum} the detected lines also vary in quality. Chip 1 and 4 are more noisy than chip 2 and 3. Some lines fall closer to telluric lines and residuals can be left in the STD filtered spectra. Finally many lines are blended and hard to separate.
We thus manually select the best undistorted and unblended lines for further line profile analysis and comparison with models. For the two observing nights different wavelength regions are covered and the line samples from the two nights are thereby different. We detect lines all the way up to v=6-5, but the higher v-bands have low S/N profiles and not enough lines for good statistics.
{Fig. \ref{fig:spectrum} shows the full spectrum from the wavelength setting of 4.662 $\mu$m (chip 3) at a position angle of 145$\degree$.}

\begin{table}[!htbp]
\caption{CO ro-vibrational line sample}            
\label{tab:obsline}     
\centering                          % used for centering table
\begin{tabular}{lll }       
 \hline\hline                           
 29/3	 & 		 & 		\\   
v-band	 & wl.set[$\mu$m]		 & 	Transition	\\   
 \hline            
v=1-0	& 4.929		&P21,P26,P27,P30			\\   
		& 4.957		&P26,P22		\\   
 \hline            
v=2-1	& 4.662		&R4,R6,R8R10,R11,	R12,R14	\\   
		& 4.676		&R4,R5,R6,R8,R9,R10,R12,R13,R14		\\   
		& 4.929		&P21,P25				\\   
		& 4.957		&P23,P27		\\   
 \hline            
v=3-2	& 4.662		&P20,P23		\\   
		& 4.676		&R14,R17,R18,	R20,R23			\\   
		& 4.929		&P7,P14,P15			\\   
		& 4.957		&P11,P14,P15,P21		\\   
\hline            
 v=4-3	& 4.662		&R27,R29		\\   
		& 4.676		&R23,R27		\\   
		& 4.929		&P08			\\   
\hline\hline
30/3	 & 		 & 		\\   
v-band	 & wl.set[$\mu$m]		 & 	Transition	\\   
\hline            
v=1-0	& 4.782		&P12,P13,P14		\\   
		& 4.796		&P8,P11,P12,P14,P17		\\   
 \hline  
v=2-1	& 4.662		&R4,R6,R8,R10,R11,R12,R14,R16			\\   
		& 4.676		& R4,R5,R6,R8,R9,R10,R13,R14			\\   
		& 4.782		&P7		\\   
		& 4.796		&P8,P11		\\   
 \hline  
v=3-2	& 4.662		&R11,R14,R20,	R23,R26		\\   
		& 4.676		&R14,R17,R18,	R20,R23		\\   
		& 4.782		&P1,R5		\\   
		& 4.796		&P5,R4,R5		\\   
 \hline  
v=4-3	& 4.662		&R23,R27,R29			\\   
		& 4.676		&R23,R27		\\   
		& 4.782		&R10,R11,R13		\\   
		& 4.796		&R8,R12,R13		\\   
 \hline\hline  

\end{tabular}
\tablefoot{These are the lines used throughout the paper. Shown are only lines not contaminated by blends or telluric lines.}  
\end{table}

\section{Results} \label{sec:res}
Table \ref{tab:obsline} presents a full listing of our selected unblended and clean CO ro-vibrational lines. For all these CO ro-vibrational transitions, line profiles (flux as a function of velocity) have been extracted.
{The individual line profiles are normalized by their fitted continuum, shifted to zero continuum and the line profile is then normalized to the maximum flux value. 
Lines from different ro-vibrational transitions show very similar shape. Thus, we can make average profiles collecting all the individual transitions. 
The profiles are combined to median profiles (one for each P.A.). The details of this are described in Appendix \ref{sec:obs_com_prof} and the profiles are shown in Figure \ref{fig:avprofileb_obs}.} Integrating each line separately, individual line fluxes are derived. For P.A.=145$\degree$ these are listed in Table \ref{tab:line flux}.

\begin{table}[!htbp]
\caption{Line fluxes from data collected on the 29th at P.A.=145$\degree$}             % title of Table
\label{tab:line flux}     % is used to refer this table in the text
\centering                          % used for centering table
\begin{tabular}{c c c c }        % centered columns (4 columns)
\hline         
Line ID & $\lambda_{\rm{line}}$ &$F_{\rm{Total}}$&Error\\    [1ex]
 & [$\mu$m] &[10$\rm{^{-15}\frac{erg}{cm^2 s}}$]&[10$\rm{^{-15}\frac{erg}{cm^2 s}}$]\\   [1ex]
\hline         
      v(1-0)P21&4.8652  &    55.7 &    0.7  \\
      v(1-0)P22&4.8760  &    67.8 &    0.7\\
      v(1-0)P26&4.9204  &    42.7  &    1.1\\
      v(1-0)P26&4.9204  &    57.5  &   1.2\\
      v(1-0)P30&4.9668  &    61.4  &    1.4\\
      v(1-0)P27&4.9318  &    42.3  &    1.1\\
 \hline            
      v(2-1)P25&4.9716   &   22.3  &    0.9\\
      v(2-1)P21&4.9269  &    26.9  &    1.0\\
      v(2-1)P23&4.9490  &    24.6  &    0.5\\
      v(2-1)P27&4.9947  &    33.3  &    1.7\\
      v(2-1)R04&4.6831  &    30.5  &   1.6\\
      v(2-1)R06&4.6675  &    42.9  &   1.6\\
      v(2-1)R08&4.6523  &    37.4  &   1.3\\
      v(2-1)R09&4.6448  &    42.9  &    0.7\\
      v(2-1)R10&4.6374  &    37.1   &   0.5\\
      v(2-1)R10&4.6374  &    29.4   &   0.5\\
      v(2-1)R11&4.6301  &    28.4   &   1.1\\
      v(2-1)R12&4.6230  &    35.9   &  5.2\\
      v(2-1)R12&4.6230  &    46.1   &   1.7\\
      v(2-1)R13&4.6160  &    33.3   &   0.6\\
      v(2-1)R14&4.6090  &    24.3   &   1.0\\
      v(2-1)R14&4.6090  &    37.2   &   0.8\\
      v(2-1)R08&4.6523  &    39.9   &   1.8\\
      v(2-1)R04&4.6831  &    23.8   &   1.1\\
      v(2-1)R05&4.6752  &    31.7   &   1.2\\
      v(2-1)R06&4.6675  &    38.8   &   1.1\\
   \hline            
    v(3-2)P21&4.9901  &    21.1   &   1.3\\
      v(3-2)P15&4.9257  &    19.3   &  0.8\\
      v(3-2)P15&4.9257  &    30.2     & 1.0\\
      v(3-2)P14&4.9154  &    15.4     & 0.8\\
      v(3-2)P14&4.9154  &    16.3     & 0.5\\
      v(3-2)P11&4.8853  &    15.7   &   1.2\\
      v(3-2)P07&4.8468  &    15.3   &   1.1\\
      v(3-2)R14&4.6670  &    25.9   &   1.0\\
      v(3-2)R17&4.6464  &    21.1   &   0.5\\
      v(3-2)R18&4.6398  &    26.8   &   0.7\\
      v(3-2)R20&4.6268  &    21.0   &   0.3\\
      v(3-2)R20&4.6268  &    20.9   &   0.3\\
      v(3-2)R23&4.6080  &    21.8  &    0.8\\
      v(3-2)R23&4.6080  &    18.6  &    0.4\\
 \hline            
      v(4-3)P08&4.9185  &    12.9  &    1.0\\
      v(4-3)R29&4.6311  &    15.3  &    0.9\\
      v(4-3)R27&4.6425  &    20.9  &    0.5\\
      v(4-3)R27&4.6425  &    17.0  &    0.4\\
      v(4-3)R23&4.6665  &    14.4  &    0.7\\

 \hline            
    \hline            

\end{tabular}
\tablefoot{The flux error is found from the standard deviation (1$\sigma$) of the nearby continuum. The lines that occur twice are collected from different wavelength settings. The wavelengths of the individual lines are taken from Chandra et al. (1996).}
\end{table}

\subsection{Population diagrams}
\label{sec:boltz}
If (1) all CO comes from a region with a single temperature $T_{\rm{rot}}$=$T_{\rm{gas}}$, (2) the CO gas is in local thermodynamical equilibrium (LTE), and (3) the lines are optically thin, we can write the Boltzmann equation as:

\begin{equation}
 \frac{N_{vJ}}{g_J} = \frac{N_v}{Q_{\rm{rot}}(T_{\rm{rot}})} e^{\frac{-E_J}{kT_{\rm{rot}}}} \\
\end{equation}

We can calculate the ratio ${N_{vJ}/{g_J}}$  from the observed line flux:

\begin{equation}
\frac{N_{vJ}}{g_{vJ}}=\frac{4\pi F_{vJ}}{g_{vJ}h\nu_{vJ}A_{vJ} \cdot \Omega}
\end{equation}

In the above equations $N_{vJ}$ is the column density of the upper level, E$_{J}$ its energy, $g_{vJ}$ its statistical weight, $Q_{\rm{rot}}$ is the rotational partition function and $F_{vJ}$ is the observed line flux, $\nu_{vJ}$ is the frequency and $A_{vJ}$ is the Einstein A coefficient of the transition. $\Omega$ is the emitting area divided by the distance squared to the source i.e. the solid angle on the sky given in steradian. We combine Eq. (1) and (2):

\begin{equation}
\frac{N_v}{Q_{\rm{rot}}(T_{\rm{rot}})} e^{\frac{-E_J}{kT_{\rm{rot}}}}=\frac{4\pi F_{vJ}}{g_{vJ}h\nu_{vJ}A_{vJ} \cdot \Omega}
\end{equation}

If we then study $ln({N_{vJ}}/{g_{vJ}})$ versus the energy of the upper level for each transition, the slope becomes -1/$T_{\rm{rot}}$. Thus in the case of optically thin LTE emission of a single temperature gas, this rotational diagram can be fitted by a straight line, to give an estimate of $T_{\rm{rot}}$. 

Rotational diagrams have been compiled, for each vibrational level, using the integrated line fluxes derived from the CRIRES data separating data from different position angles. For each night, we show two orthogonal slit positions to check for P.A. variations (P.A.$=55\degree$ and P.A.$=145\degree$ on the 29th and P.A.$=10\degree$ and P.A.$=100\degree$ on the 30th) (see Figures \ref{fig:boltzmannobs29} and \ref{fig:boltzmannobs30} in Appendix A). The transition frequencies, Einstein A coefficients, statistical weights, and energies of the upper levels are taken from \citet{chandra1996}. We approximate the emitting region by a rectangle with the same width as the slit and length equal to the diameter of the emitting region of the CO (2x50AU, Goto et al. 2012): $\Omega$=0$\farcs$2$\cdot$1$\farcs$0$\cdot$($\pi$/180/3600). 
We note that the most optically thick lines, like the v=1-0 low J lines, are missing (these lines fall on strong telluric absorption features and are therefore excluded).

\begin{figure}[!htbp]
   \includegraphics[width=0.5\textwidth]{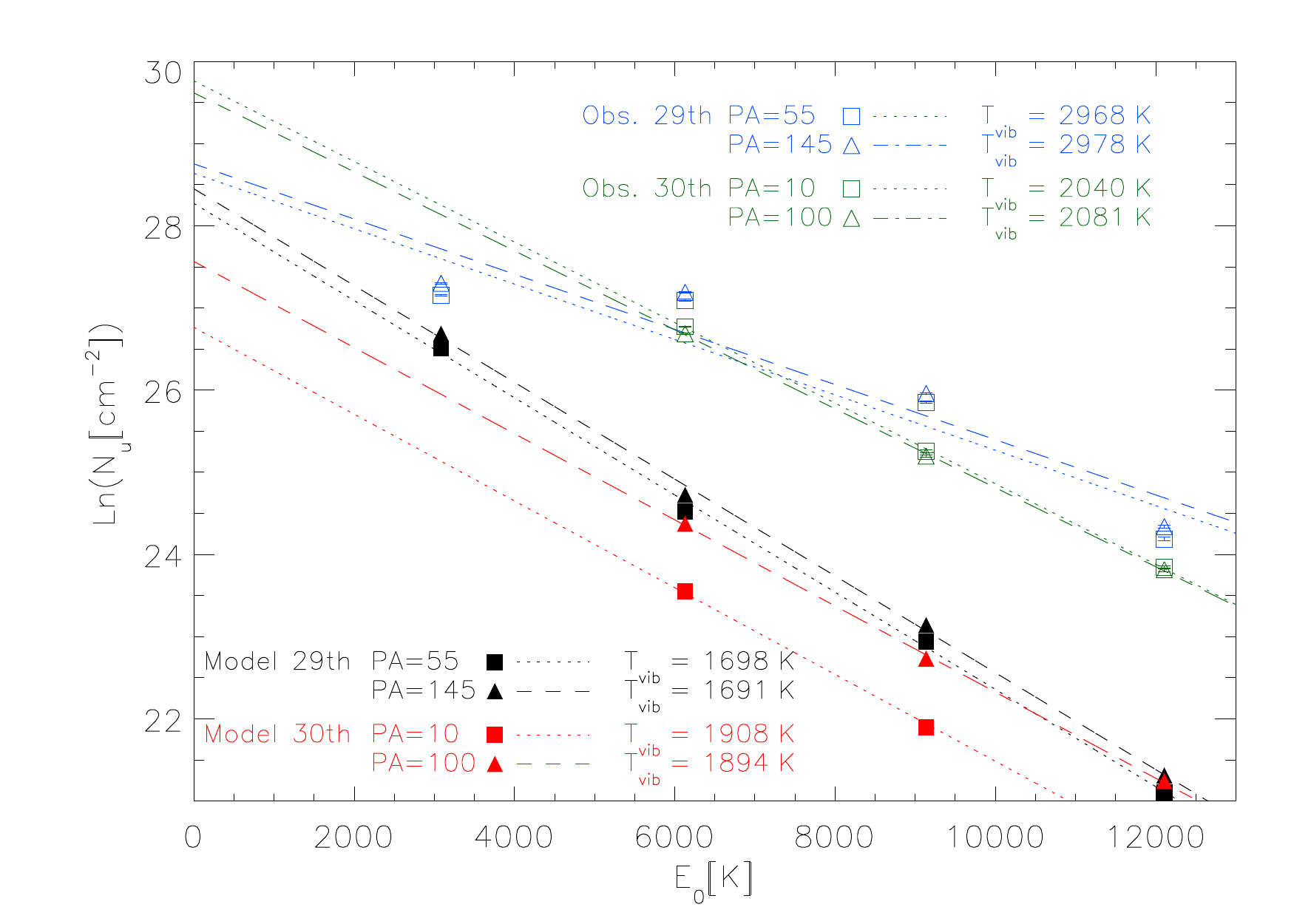}
      \caption{Vibrational diagram for all transitions observed on the 29th at P.A.=55$\degree/145\degree$ and on the 30th at P.A.=10$\degree/100\degree$ (the individual lines are listed in Table \ref{tab:obsline}). The error bars are smaller than the plotting symbols. For P.A.=10$\degree/100\degree$ the line sample does not include any v=1-0, transitions so the corresponding line has only been fitted from three points. The vibrational diagram from the two sets of simulated lines (different slit filter settings) from the model are are also included in the plot (derived and discussed in Section \ref{sec:mod_rot}).}
         \label{fig:vibobs29}
   \end{figure}

For the vibrational temperature of the CO gas, we obtain the following equation:
\begin{equation}
\frac{N_v}{g_v} = \sum_J \frac{N_{vJ}}{g_v} = \frac{N_{CO}}{Q_{\rm{vib}}(T_{\rm{vib}})}e^{\frac{-E_0(v)}{kT_{\rm{vib}}}} \\
\end{equation}
\noindent Thus, summing over $\frac{N_{vJ}}{g_v}$ from Eq. 2, for each v-band separately, we can construct a vibrational diagram of ln$(\sum_J \frac{N_{vJ}}{g_v})$ versus the ground level energy for each v level. We can then find the vibrational temperature $T_{vib}$ using the same approach as above for the rotational temperature.
It is important to state that we are not complete in J levels and therefore our $T_{\rm{vib}}$ determination is biased, but as long as we choose the same line selection for both model and observations, the plots and slopes we derive are useful for comparison purposes. {The effect of changing the line sample included in the fit was tested with the model and was found to be significant (see Sect. \ref{sec:mod_rot} and Fig. \ref{fig:modvib_full}). }
The vibrational diagram is shown in Fig. \ref{fig:vibobs29}.

We derive vibrational temperatures from our 29th line sample of $T_{\rm{vib}}$=2968 K and $T_{\rm{vib}}$=2978 K respectively for the P.A. settings 55$\degree$ and 145$\degree$ and for the 30th line sample  of $T_{\rm{vib}}$=2040 K and $T_{\rm{vib}}$=2081 K respectively for the P.A. settings 10$\degree$ and 100$\degree$. There is no significant difference in vibrational temperatures derived from the same date with orthogonal P.A. settings. The difference between the vibrational temperatures derived on the two separate nights comes from the lack of v=1-0 lines detected at the P.A. settings of 10$\degree$ and 100$\degree$. If we were to exclude the v=1-0 lines from the vibrational temperature fit done for the 29th we would find values similar to those derived from the 30th.

{Observations generally cover a small subset of all J levels within each v-band. This can have a strong impact on the slope of the fit in the vibrational diagram. Some authors use the partition function to estimate the full population of each v-band from the slope and y-intersection of the rotational diagram. In this way the population is not affected by the amount of lines detected at each v-band. Meanwhile, a sampling bias can still be present due to the range of J levels used in the rotational diagram (strong curvature present at low J levels).
In our sample, many detectable v=1-0 lines were discarded because of strong telluric absorption. Furthermore, we did not detect lines beyond J=30 and only a few lines of the v=5-4, v=6-5 bands (not included in the vibrational diagram). This has affected our derivation of vibrational temperatures. Since other papers use very different samples of lines and even numbers of lines, a direct comparison between vibrational temperatures become difficult. We do not provide error bars for our derived vibrational temperatures since we expect the error from sampling limitations to be significantly larger than the formal error bars. In Sect. \ref{sec:mod_rot} and Fig. \ref{fig:modvib_full} we show the effect that varying the line sample can have on a vibrational diagram.}

\section{Confronting a model with the observations} \label{sec:model}
To compare our observations with current theoretical disc models, we have derived CO ro-vibrational lines from an already existing model of HD100546 \citep{thi2011} and produced CO line data cubes. We then run the line cubes through a custom made IDL filtering script to simulate a CRIRES observation of our model. The final 'observed' model line fluxes and profiles are then compared to our actual CRIRES observations.

\subsection{ProDiMo/MCFOST}
A full model of the protoplanetary disc HD100546 already exists \citep{thi2011,tatulli2011,benisty2010}. 
The observed spectral energy distribution (SED) and near- and mid-infrared interferometric VLT-AMBER and MIDI data have been used to constrain the dust properties. The disc structure has been fitted with a parameterized disc. The fitting with MCFOST is described in \citet{benisty2010} and \citet{tatulli2011}. MCFOST is a three dimensional continuum and line radiative transfer code based on the Monte Carlo method \citep{pinte2006,pinte2009}. It calculates the dust temperature structure and radiation field, which can be used to produce images, polarization maps and SEDs with a ray-tracing method. 
\citet{thi2011} modelled the CH$^+$ chemistry in HD100546 using the thermo-chemical disc code ProDiMo on top of the above described dust model.
ProDiMo solves the gas heating and cooling, the gas chemistry and the disc vertical structure self-consistently \citep*{woitke2009}. However, we parametrize the vertical hydrostatic structure. The model uses detailed PAH heating/cooling in the chemistry \citep{woitke2011}.

The HD100546 model is composed of an inner disc, an outer disc, and an upper layer on the top of the outer disc. The inner and outer discs are separated by a gap from 4-13AU. The model is run on a 100x100 points numerical grid. The model parameters are shown in Table \ref{table:mod_param}. {We adopt a gas mass of 5.37$\cdot 10^{-4}$~M$_{\astrosun}$, while \citet{bruderer2012} use 4$\cdot 10^{-3}$~M$_{\astrosun}$ in their preferred model. Our model reproduces the CO ladder as observed by Herschel-PACS \citep{Sturm2010}, Herschel SPIRE \citep{wiel2013} and from the ground \citep{panic2010}. However, these CO lines are optically thick and cannot constrain the dust-to-gas mass ratio in discs. The lines we study in this paper probe only the very narrow region at the top of the inner rim an can thus not be used to infer global gas-to-dust mass ratios.} 

Since this is a previously published model, not fitted to our data or any other observations of CO ro-vibrational emission lines, the comparison with our data will provide an independent test of the reliability of the model predictions.  Recent observations of the inner disc have revealed a different geometry for the inner disc than previously assumed, but as we will show the CO emission is predominantly emitted from the inner rim of the outer disc and our analysis is therefore not affected by this. 

\begin{table}[!htbp]
\caption{Key parameters used in our modelling of HD100546}             % title of Table
\label{table:mod_param}      % is used to refer this table in the text
\centering                          % used for centering table
\begin{tabular}{l l l }        % centered columns (4 columns)
\hline

\hline            
&&\\
stellar mass&$M_{*}$&2.4 M$_{\astrosun}$\\
%&$T_{\rm{eff}}$&10500 K\\
star radius&$R_*$&1.54 R$_{\astrosun}$\\
stellar luminosity&$L_*$&26 L$_{\astrosun}$\\
gravity&log($g$)&4.36\\
%dust-to-gas mass ratio&dust/gas&0.71\\
dust mass&$M_{\rm{dust}}$&3.82 10$^{-4}$ M$_{\astrosun}$\\
gas mass&$M_{\rm{gas}}$&5.37 10$^{-4}$ M$_{\astrosun}$\\
inner radius&$R_{\rm{in}}$&0.19 AU\\
outer radius&$R_{\rm{out}}$&500 AU\\
&&\\
\hline\hline    
   
\end{tabular}
\tablefoot{The parameters not shown here are the same as in \citet{thi2011}.}
\end{table}

\subsection{CO ro-vibrational data}
For the modelling of CO ro-vibrational lines in this paper, we use the above described ProDiMo model for HD100546, with the complete CO ro-vibrational molecular model, described in \citet{thi2012}.
The collisional rate coefficients are gathered from the literature and scaled to extrapolate missing ones and the CO ro-vibrational transition probabilities are from \citet{chandra1996}.
We include fluorescence pumping to the $A^{1}\Pi$ electronic level and 60 rotational levels within 7 vibrational levels of both the ground electronic state $X^{1}\Sigma^{+}$ and the excited state $A^{1}\Pi$. We model the full set of observed lines listed in Table \ref{tab:obsline}.

\begin{figure*}[!htbp]
\begin{center}$
\begin{array}{cc}
   \includegraphics[width=0.425\textwidth]{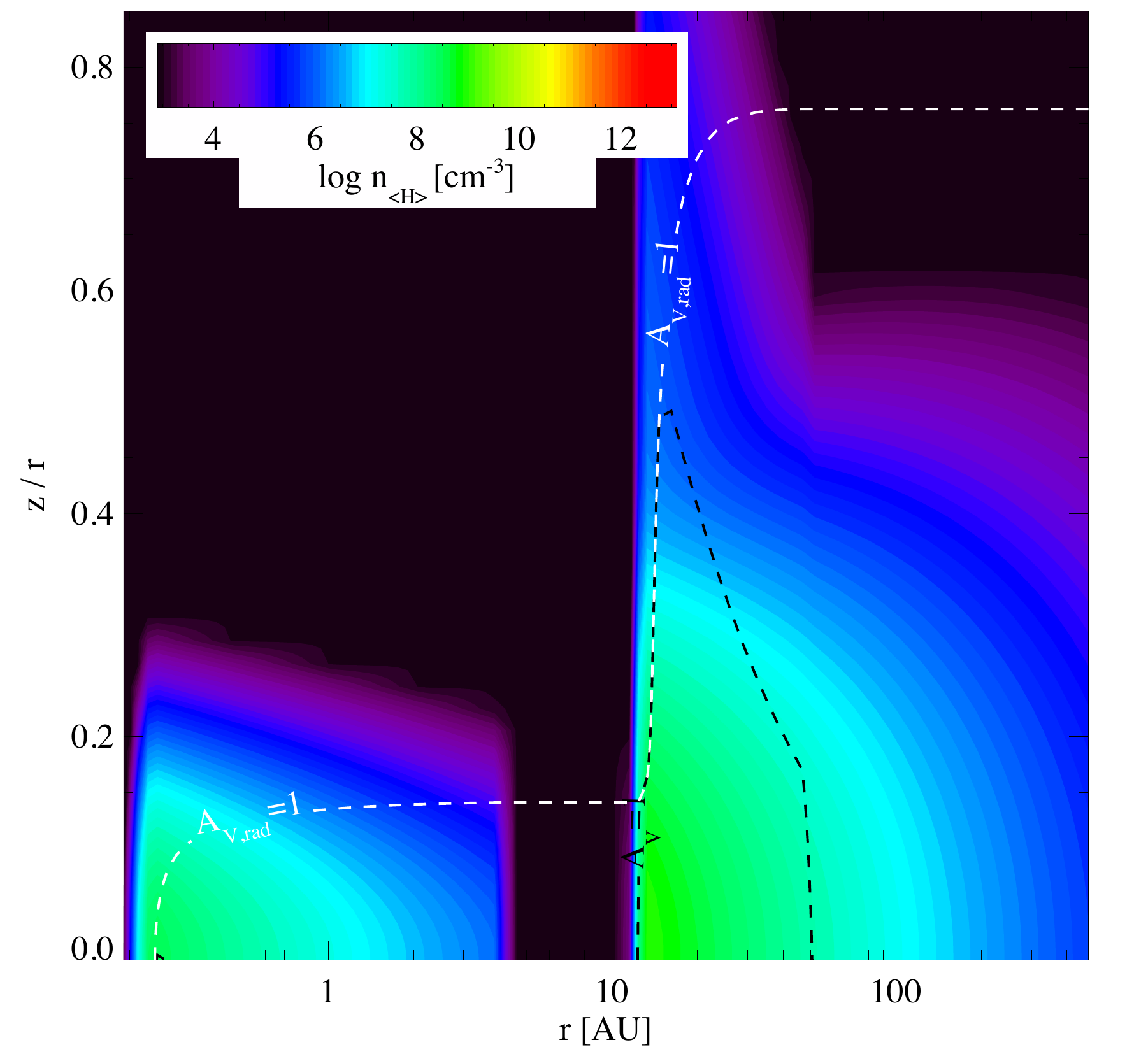}&
   \includegraphics[width=0.4\textwidth]{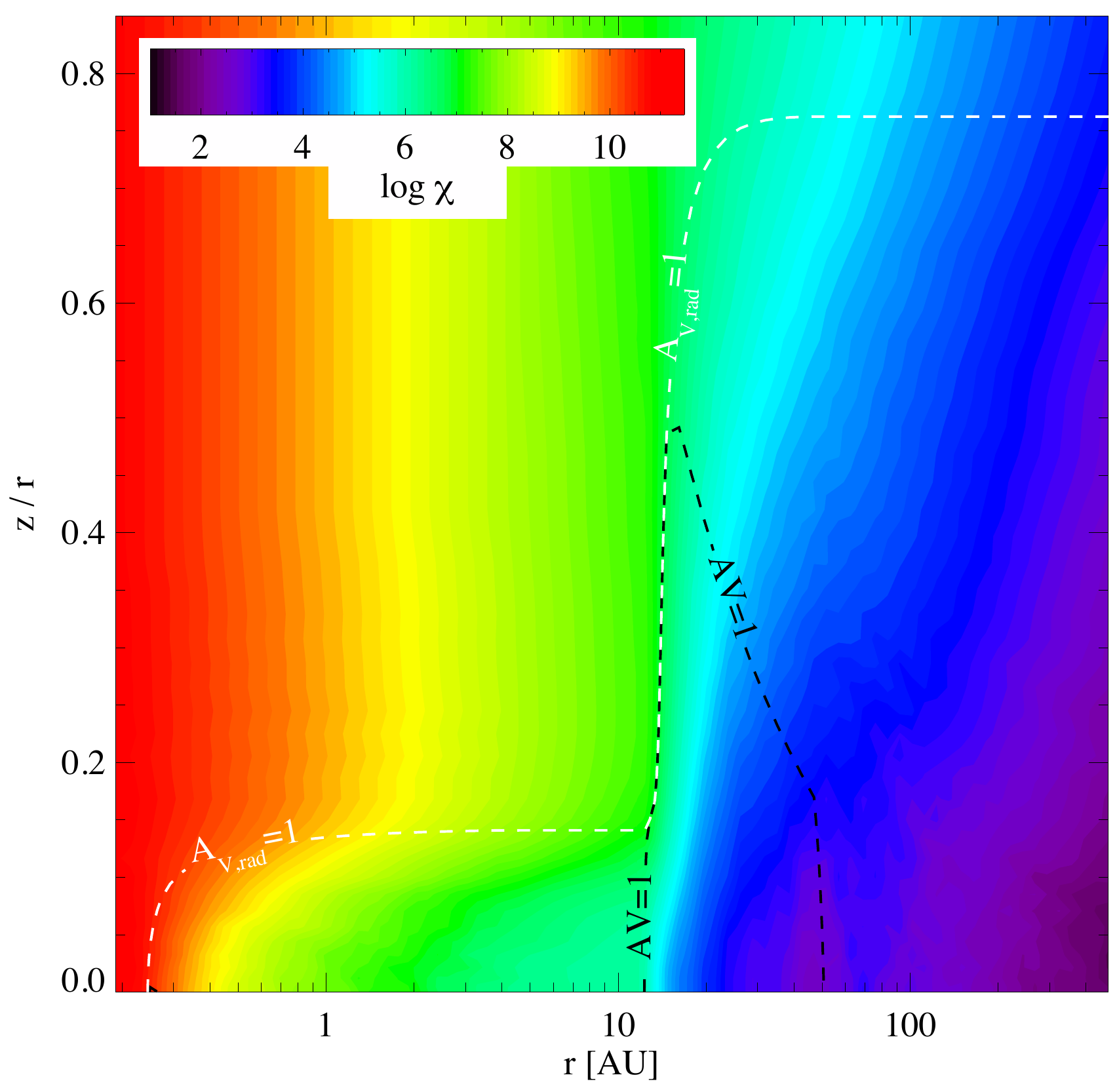}\\
  \includegraphics[width=0.4\textwidth]{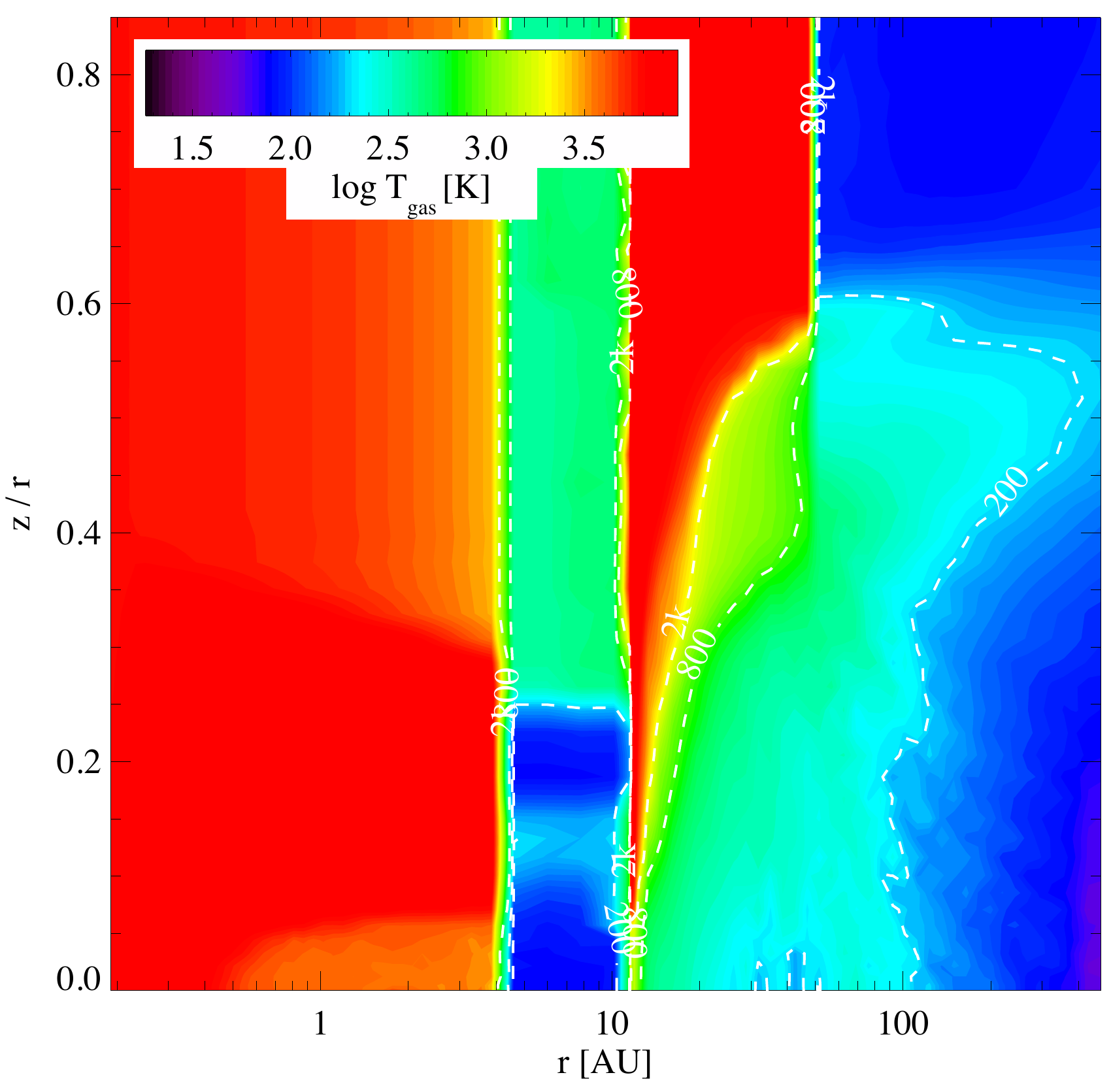}&
   \includegraphics[width=0.4\textwidth]{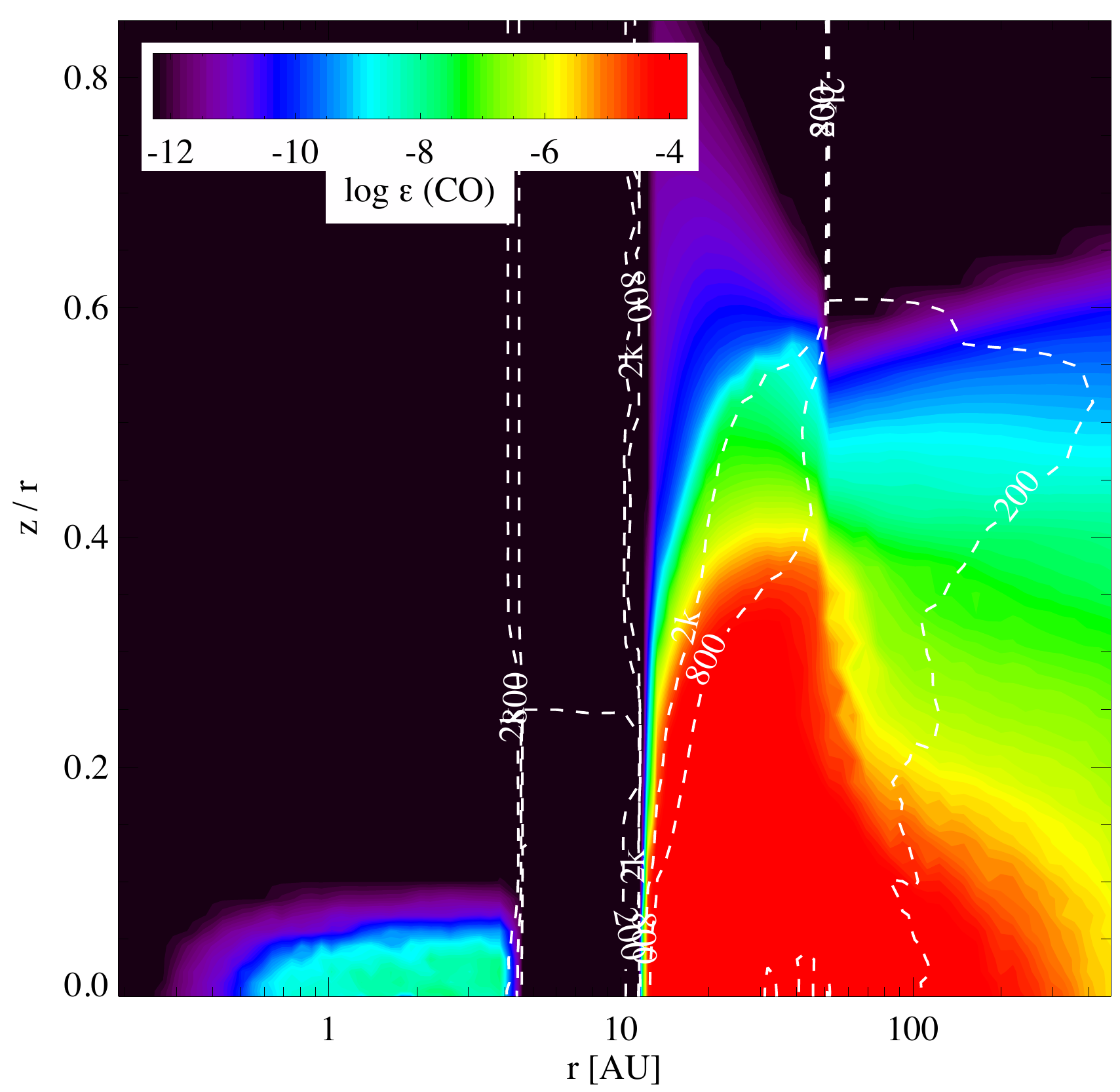}\\
   \includegraphics[width=0.4\textwidth]{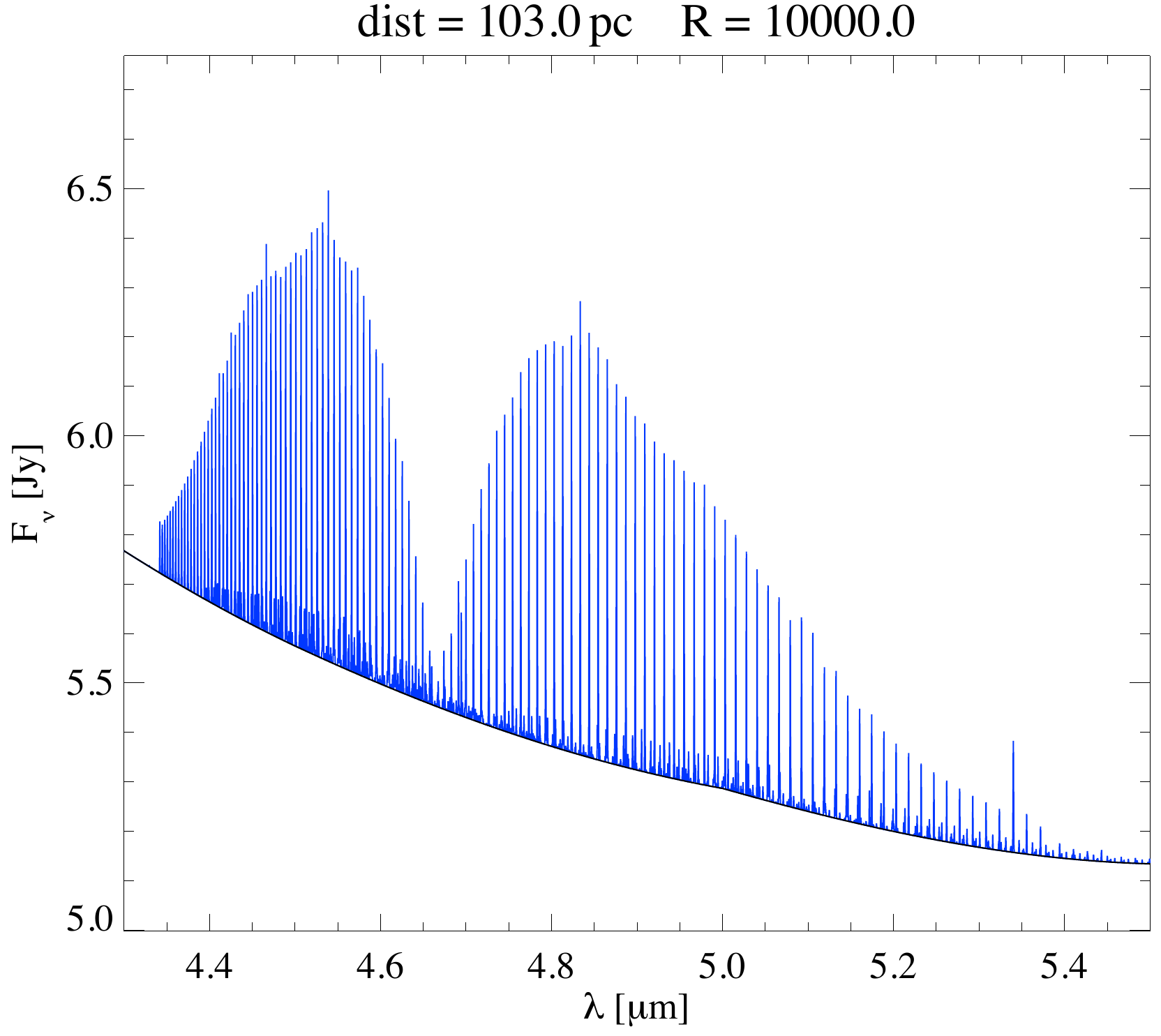}&
   \includegraphics[width=0.4\textwidth]{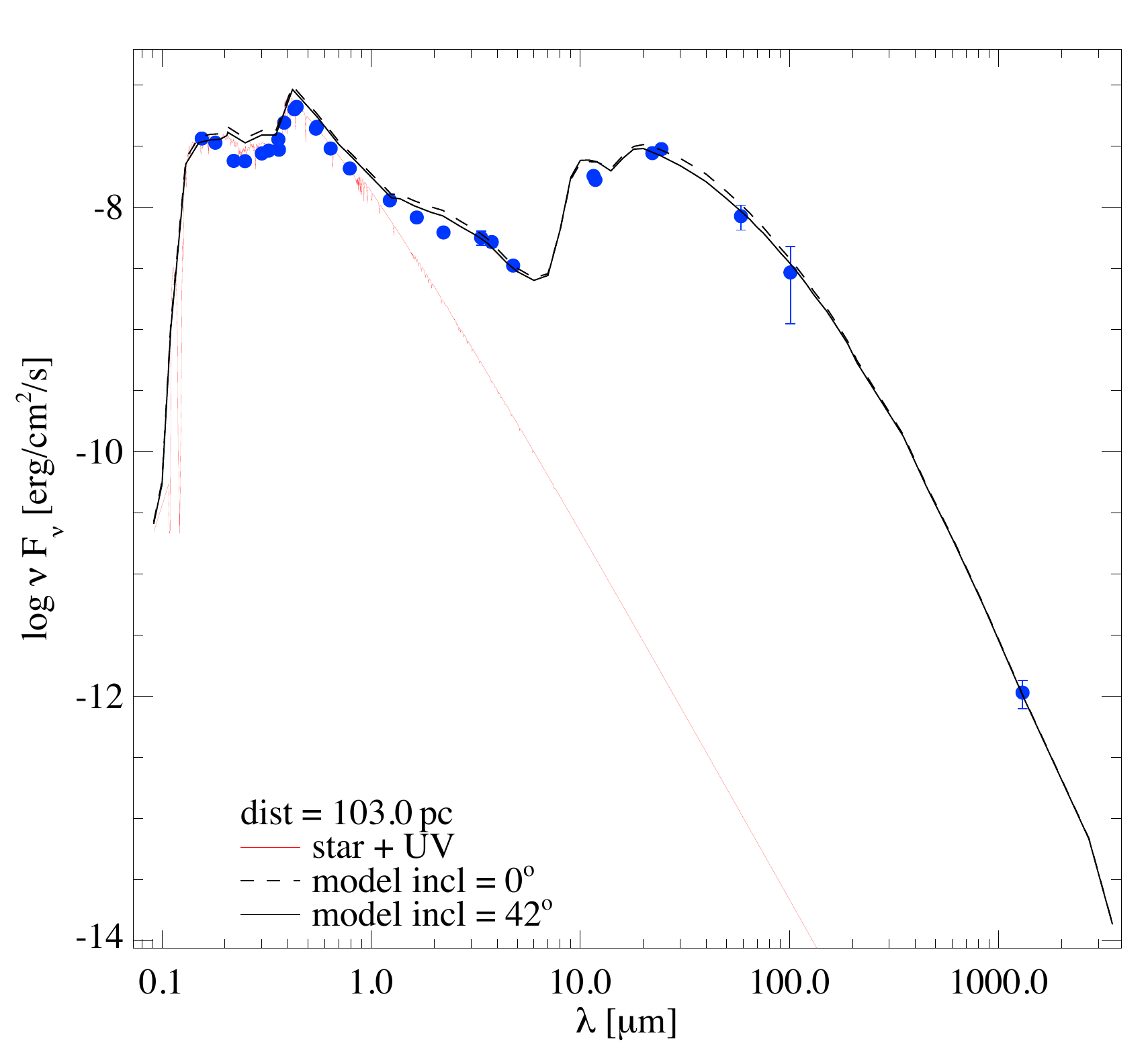}
 \end{array}$
\end{center}
\caption{Top: Gas density distribution on the left and strength of the UV radiation field log($\chi$), on the right. Contour lines showing $\rm{A_{v,rad}=1.0}$ (white) and $\rm{min(A_{v,ver},A_{v,rad})=1.0} $ (black) are overplotted on both. Middle: Gas temperature on the left and CO abundance on the right. Contour lines showing the gas temperatures of 200 K, 800 K, and 2k=2000 K (the temperatures in the gap should reflect those in a very low density ISM medium) are overplotted on both. Lower left: The ro-vibrational bands of the CO molecule. Lower right: The modelled SED with observational data overplotted as blue dots.}
         \label{fig:mod}
\end{figure*}

\begin{figure*}[!htbp]
\begin{center}$
\begin{array}{cc}
   \includegraphics[width=0.4\textwidth]{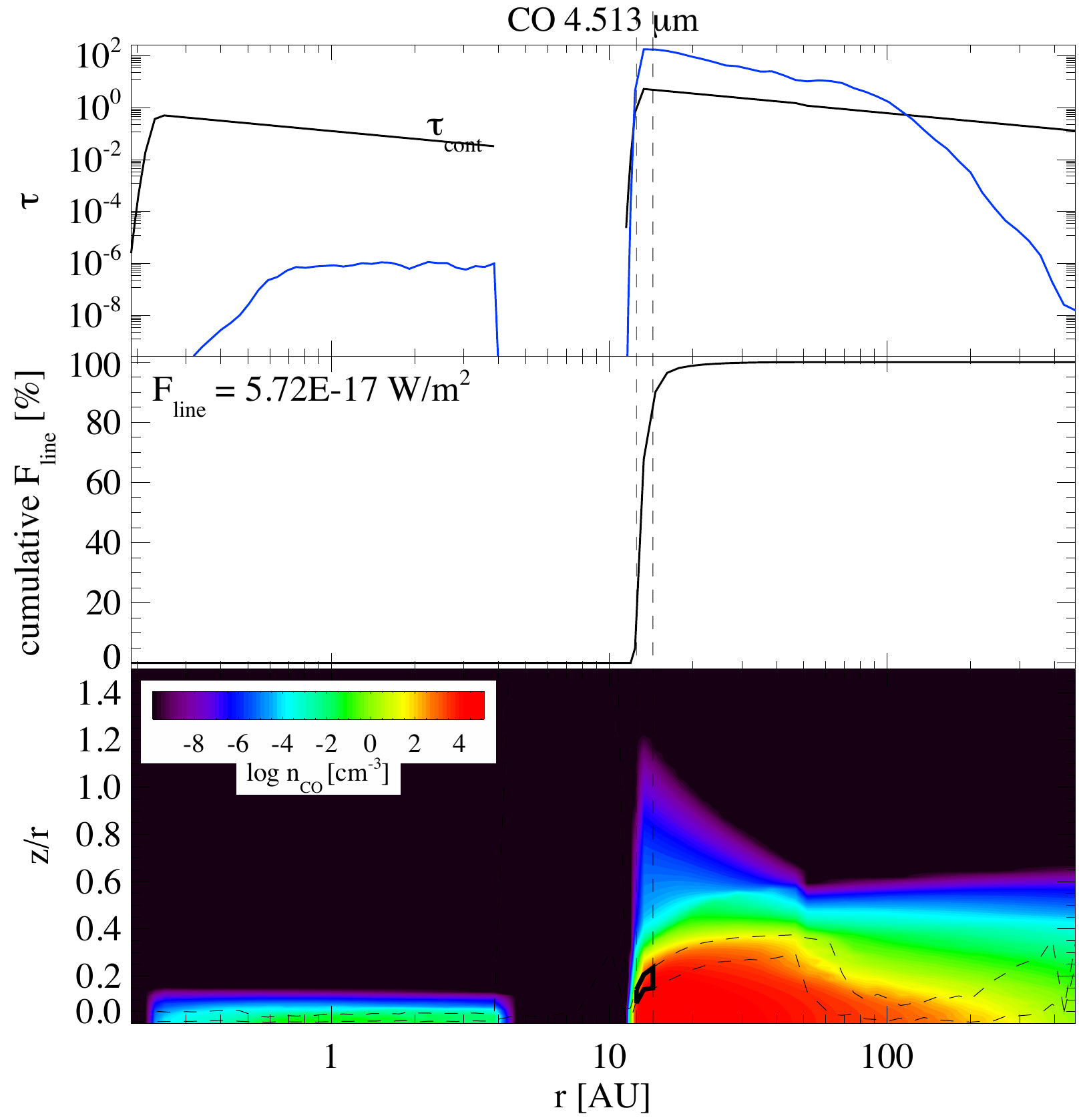}&
   \includegraphics[width=0.4\textwidth]{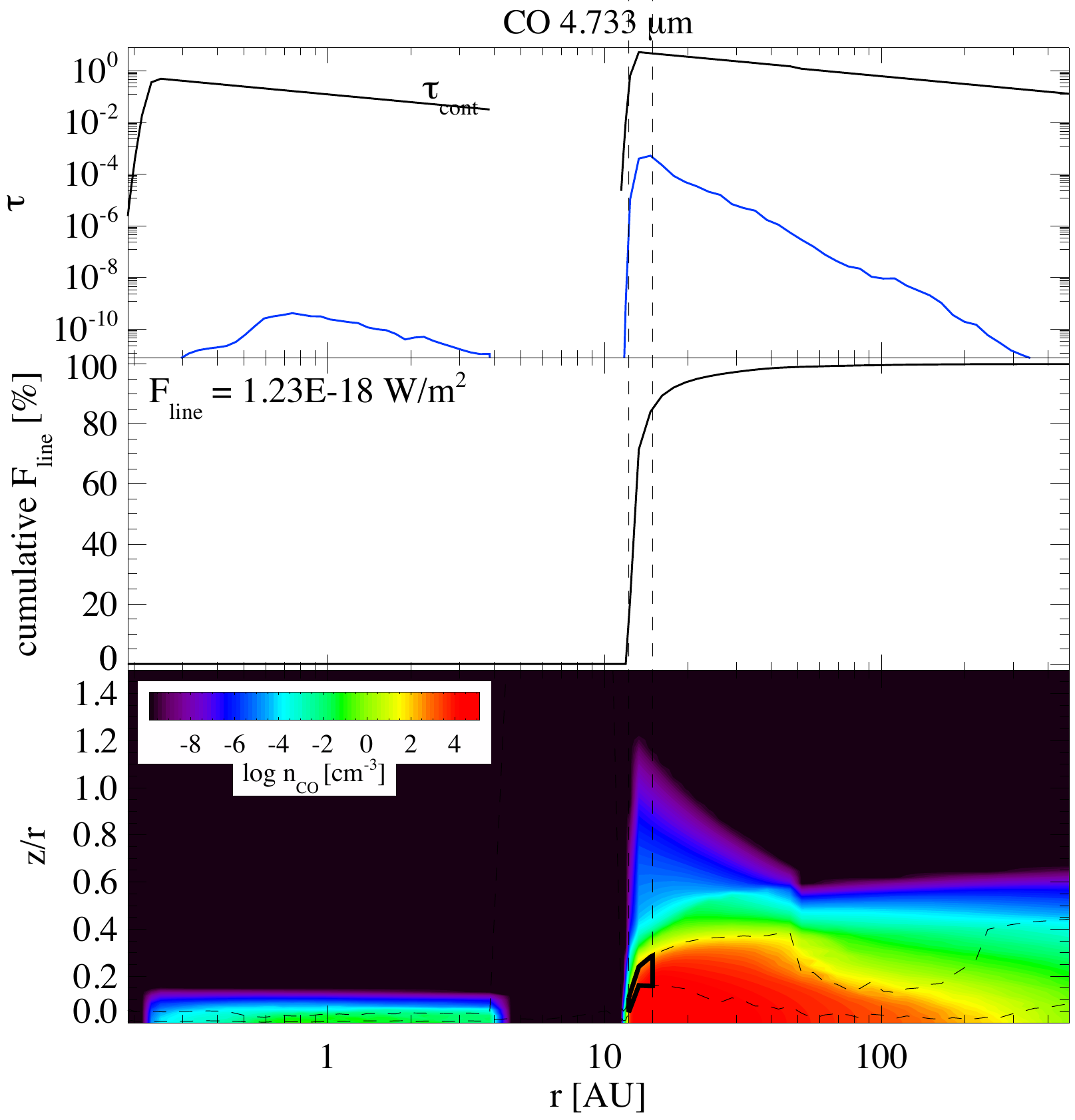}\\
   \includegraphics[width=0.4\textwidth]{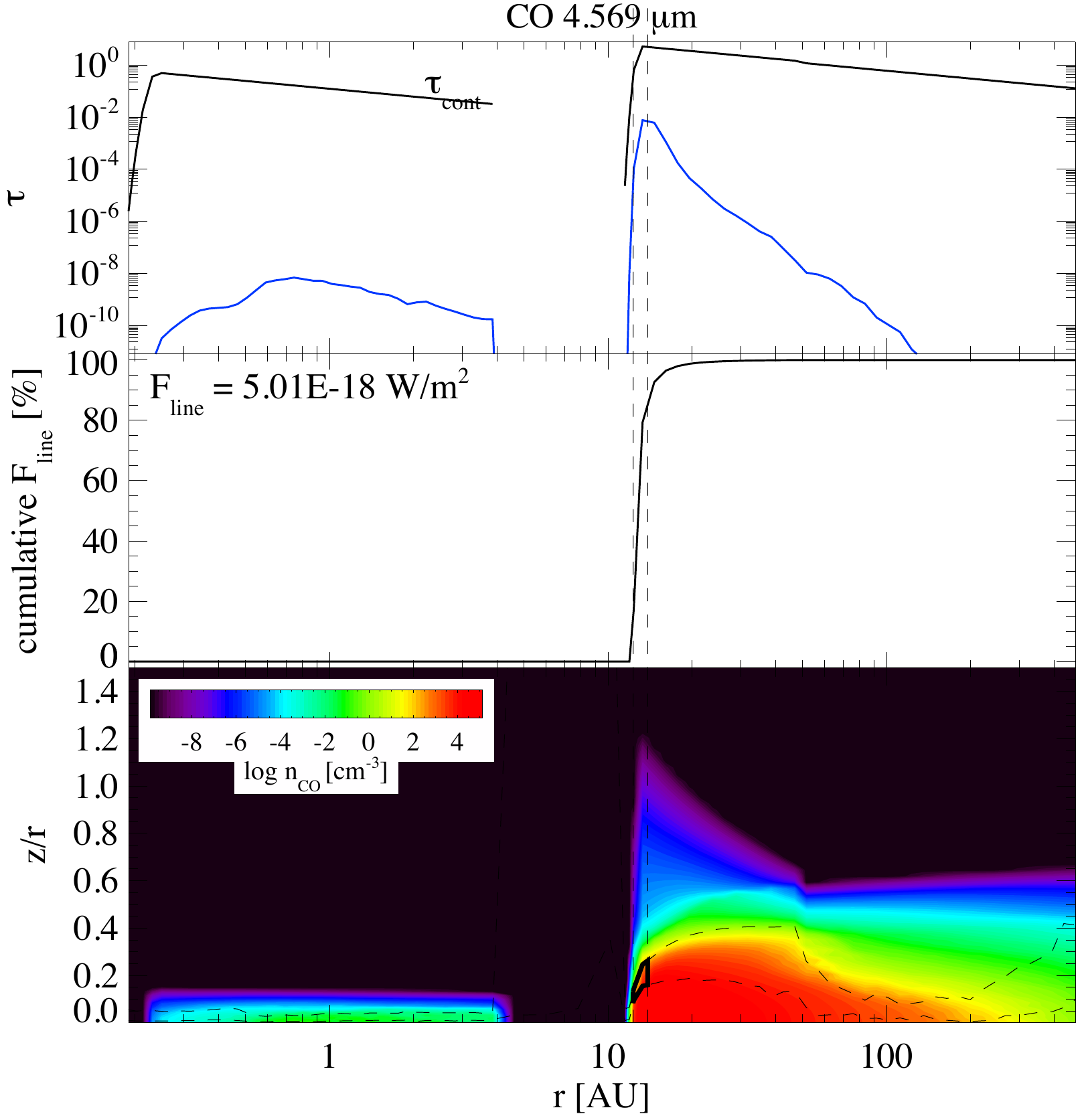}&
   \includegraphics[width=0.4\textwidth]{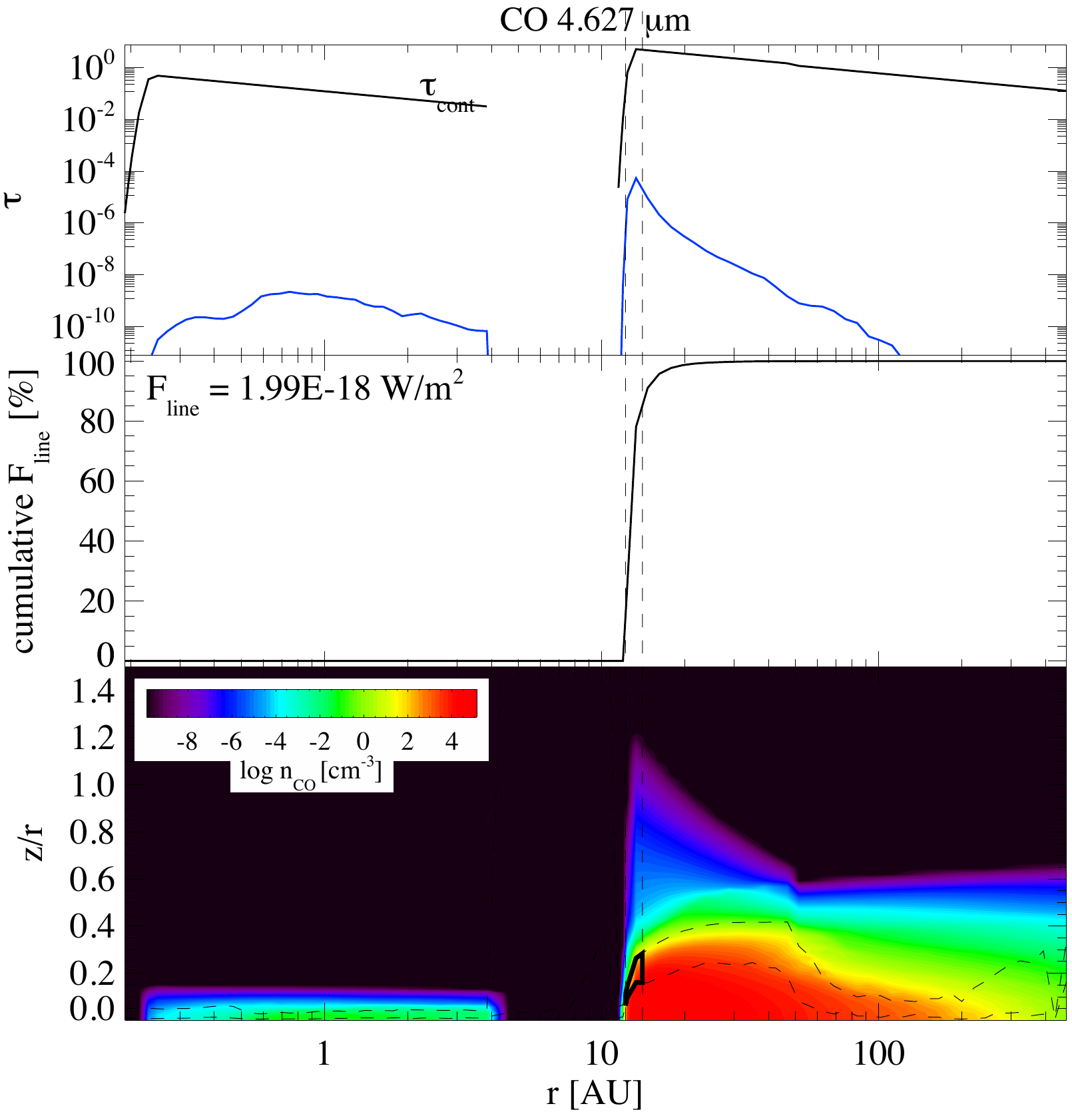}
 \end{array}$
\end{center}
\caption{{Four different  CO ro-vibrational transitions, from top left to bottom right: v(1-0)R20, v(2-1)P1, v(2-1)R20, and v(3-2)R20.} In each of the three panels, the optical depth,$\tau$, of the line (blue) and the continuum (black) (upper panel), the radial cumulative line flux (middle panel), and the CO density (bottom panel) are shown. The vertical dashed lines through all three panels show the radial region where 70\% of the line flux originates (calculated from vertical escape probability).}
         \label{fig:mod_cumul}
\end{figure*}

The gas density profile, the stellar UV field strength\footnote{The stellar UV field strength, $\chi$, is a dimensionless quantity defined as the integral over the radiation field (91-205 nm), normalized to that of the Draine field \citep{draine1996}.}, the CO abundance, the gas temperature and the CO ro-vibrational band are plotted in Fig. \ref{fig:mod}. The cumulative line fluxes of three representative lines, v(1-0)R20, v(2-1)R20, and (3-2)R20 are shown in Fig. \ref{fig:mod_cumul}. Our model indicates that most of the CO ro-vibrational line emission builds up in a narrow region at the disc wall (10-13 AU). {At radii <10 AU the CO abundance is too low and beyond 15 AU the fluorescence mechanism becomes less efficient. The gas temperature contour plot shows that the temperature reaches 1000-5000 K in the region around 10-13 AU at heights 1-10 AU.} If we look at the cumulative flux plot (Fig. \ref{fig:mod_cumul}) we see that about 50\% of the line flux is coming from this small region, while the total line flux builds up out to $\sim$20 AU {(in the case of low J lines up to 30-40 AU). {The spatial extent of the observed CO emission drops to 1/10th of the maximum value at about $\sim$40 AU (Goto et al. 2012).} The modelled CO emission is less extended (see Fig. \ref{fig:mod_cumul}), but convolution with a Gaussian beam of 0\farcs17 gives a spatial extent (1/10th of maximum) of 28 AU. We expect an ucertainty for the observed spatial extent of \textless12AU (the size of PSF). However, it is important to note that the observed spatial profile is calculated for a range of observed v=2-1 lines, while our modelled spatial profile is calculated from one line (v(2-1)R06)}.
The model predictions for the $\Delta$v=2, CO overtone emission at 2.3 $\mu$m show lines around $\sim2\cdot10^{-16}$erg/$\rm cm^2$/s, consistent with the non detection  of these lines (three sigma upper limit: $\sim8.08\cdot10^{-15}$erg/$\rm cm^2$/s, van der Plas et al. submitted).

\subsection{{Modelled line profiles}} \label{sec:model_slitsim}
ProDiMo outputs line data cubes for each chosen CO ro-vibrational transition. Each cube contains a 201x201 pixel coordinate grid in units of AU and the spectral intensity in $\rm{[erg/cm^2/s/Hz/sr]}$ at every spatial position in the described coordinate grid for each of the 91 velocity channels, covering -20 km/s to +20 km/s and  the continuum (see Appendix \ref{app:cube}). From this cube we create line images, profiles, and flux tables. They are our modelled 'observations'. {Each cube is piped through a slit filtering IDL procedure to add observational effects (Appendix \ref{app:slit}).}

\begin{table}[!htbp]
\caption{modelled and slit filtered line fluxes}             % title of Table
\label{tab:modflux}    
\centering                         
\begin{tabular}{c c c c}       

\hline 
	Transition&$\lambda_{\rm{line}}[\mu \rm{m}]$&	$F_{\rm line}$&	$F_{\rm cont}$\\ [1ex]
	&	&[10$^{-15}$$\rm{\frac{erg}{cm^2 s}}$]&	[10$^{-13}$$\rm{\frac{erg}{cm^2 s}}$]\\ [1ex]
\hline 
           v(1-0)P21&4.8652 &   35.22  &    5.17\\
           v(1-0)P22&4.8760    &   32.96  &    5.14\\
           v(1-0)P26&4.9204    &   28.04   &    5.03\\
           v(1-0)P30&4.9668    &   25.51   &    4.91\\
           v(1-0)P27&4.9318    &   27.23   &    5.00\\
\hline 
           v(2-1)P25&4.9716    &   2.87   &    4.90\\
           v(2-1)P21&4.9269    &   2.87   &    5.01\\
           v(2-1)P23&4.9490    &   2.86   &    4.96\\
           v(2-1)P27&4.9947    &   2.94   &    4.85\\
           v(2-1)R09&4.6448    &   2.97   &    5.81\\
           v(2-1)R10&4.6374    &   3.05     &  5.84\\
           v(2-1)R11&4.6301    &   3.11   &    5.86\\
           v(2-1)R12&4.6230    &   3.20   &    5.89\\
           v(2-1)R13&4.6160    &   3.25   &    5.92\\
           v(2-1)R14&4.6090    &   3.26   &    5.94\\
           v(2-1)R08&4.6523    &   2.77   &    5.78\\
           v(2-1)R04&4.6831    &   1.98   &    5.69\\
           v(2-1)R05&4.6752    &   2.19   &    5.70\\
           v(2-1)R06&4.6675    &   2.44   &    5.73\\
\hline 
           v(3-2)P21&4.9901    &   1.10   &    4.86\\
           v(3-2)P15&4.9257    &   1.14 &      5.01\\
           v(3-2)P14&4.9154    &   1.14 &      5.04\\
           v(3-2)P11&4.8853    &   1.13 &      5.12\\
           v(3-2)P07&4.8468    &   1.04 &      5.22\\
           v(3-2)R14&4.6670    &   1.30 &      5.74\\
           v(3-2)R17&4.6464    &   1.35 &      5.81\\
           v(3-2)R18&4.6398    &   1.35 &      5.83\\
           v(3-2)R20&4.6268    &   1.38 &      5.88\\
           v(3-2)R23&4.6080    &   1.41 &      5.95\\
\hline 
           v(4-3)P08&4.9185   &    0.66 &      5.03\\
           v(4-3)R29&4.6311  &     0.81 &      5.86\\
           v(4-3)R27&4.6425  &     0.81 &      5.82\\
           v(4-3)R23&4.6665  &     0.80 &      5.73\\
           
\hline            
   
\end{tabular}
\tablefoot{With a simulated slit at position angle P.A.=145$\degree$ corresponding to the observed lines collected on the 29th.}
\end{table}

{The final modelled line profiles, with the observations overplotted, are shown in Fig. \ref{fig:mod_prof}. Here, a slit pointing offset of 0$\farcs$06 is imposed (the choice of offset is discussed in Appendix \ref{app:slit}).}
{The upper two frames show that the line shape variations from the 29th cannot be explained by a single slit pointing offset of 0$\farcs$06 (the line profile at P.A.=235$\degree$ is leaning toward the wrong side).
The middle and lower frame of Fig. \ref{fig:mod_prof} show that the line shape variations observed on the 30th could be explained by one single consistent pointing offset of 0$\farcs$06.}
{Table \ref{tab:modflux} lists the line fluxes and continuum fluxes for each modelled slit filtered line. }

\begin{figure}[!htbp]
\begin{center}$
\begin{array}{cc}
%\cline{1-0}
  \includegraphics[width=0.5\textwidth]{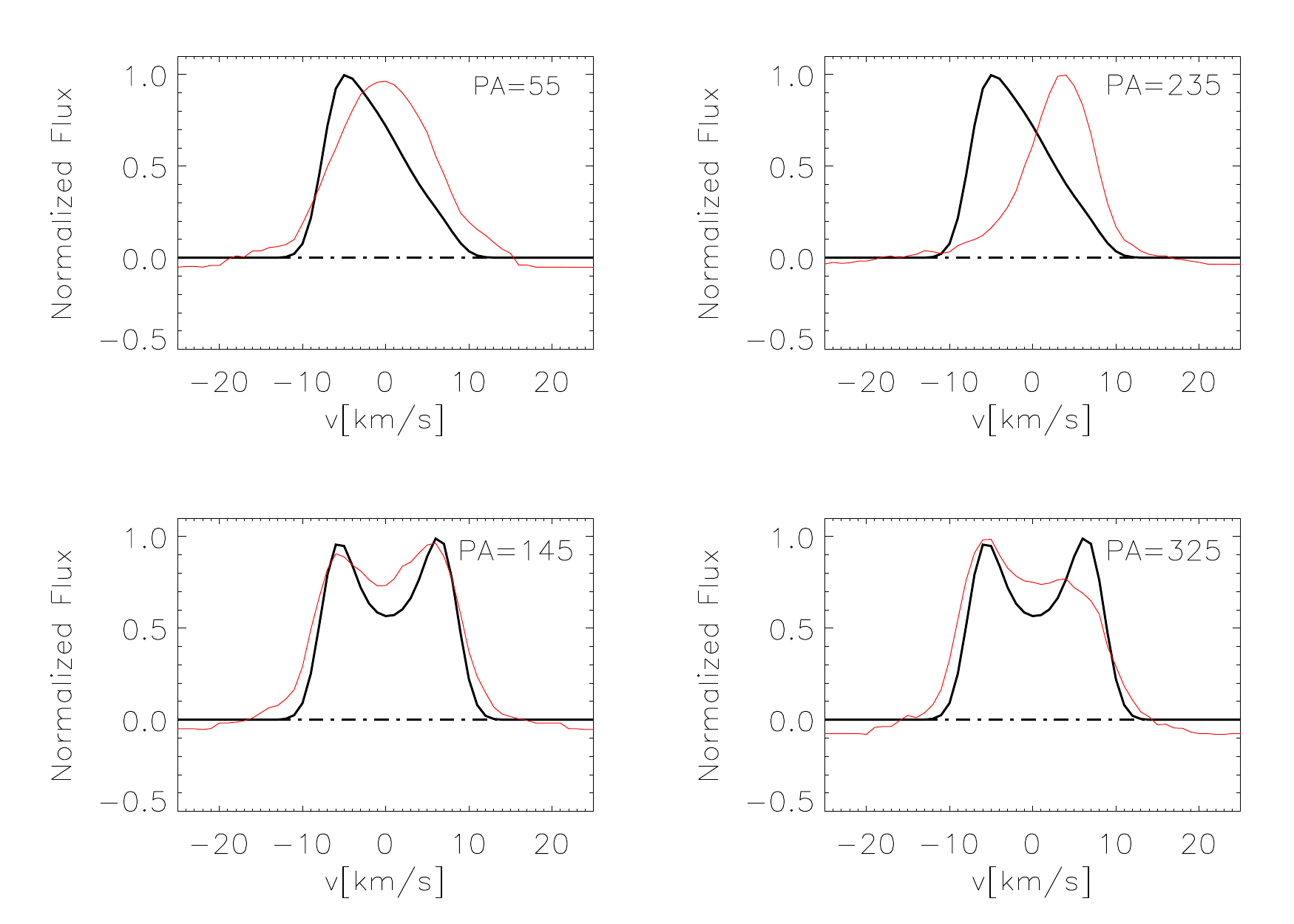}&\\
%\cline{1-1}
   \includegraphics[width=0.5\textwidth]{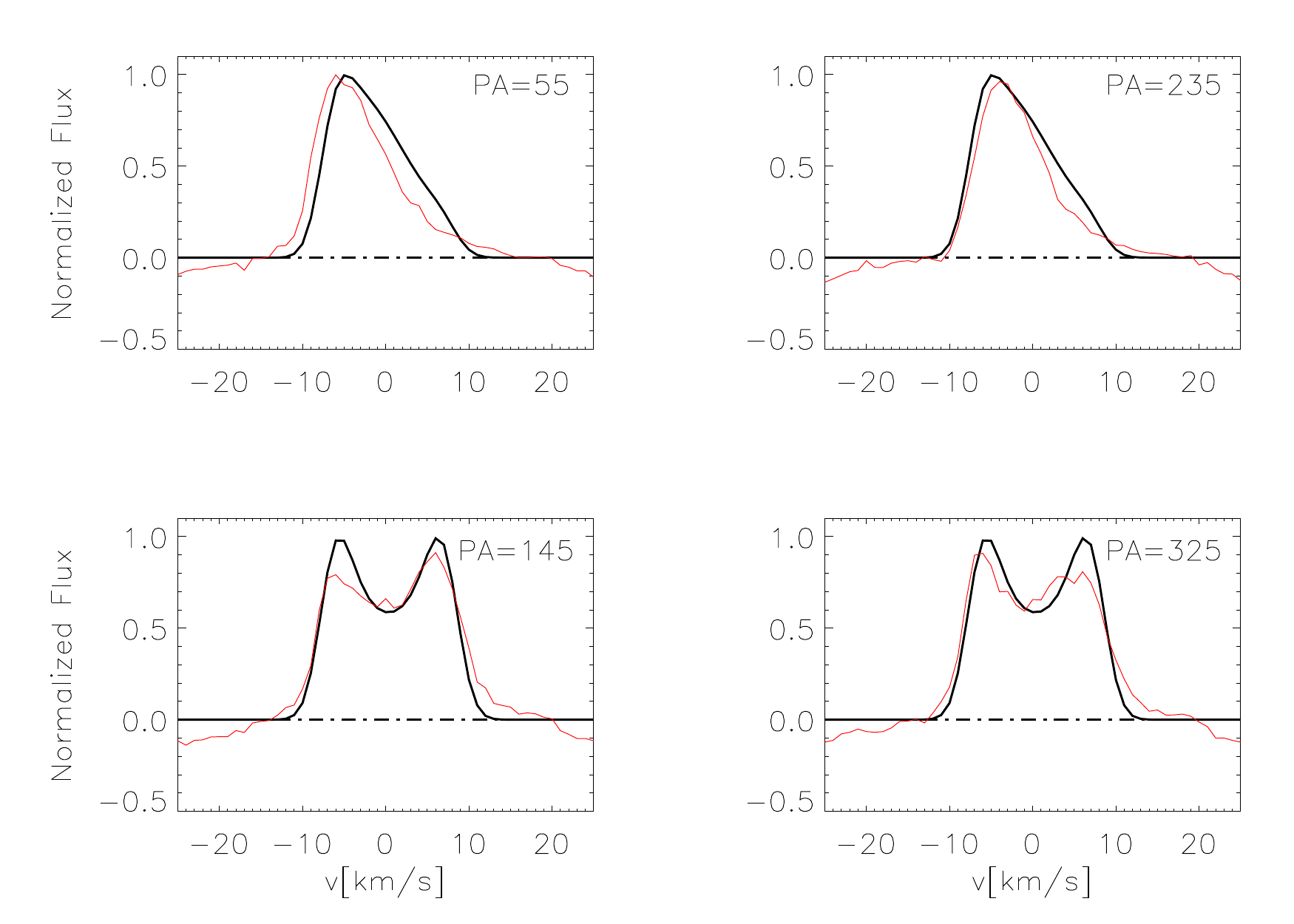}&\\
%\cline{1-1}
   \includegraphics[width=0.5\textwidth]{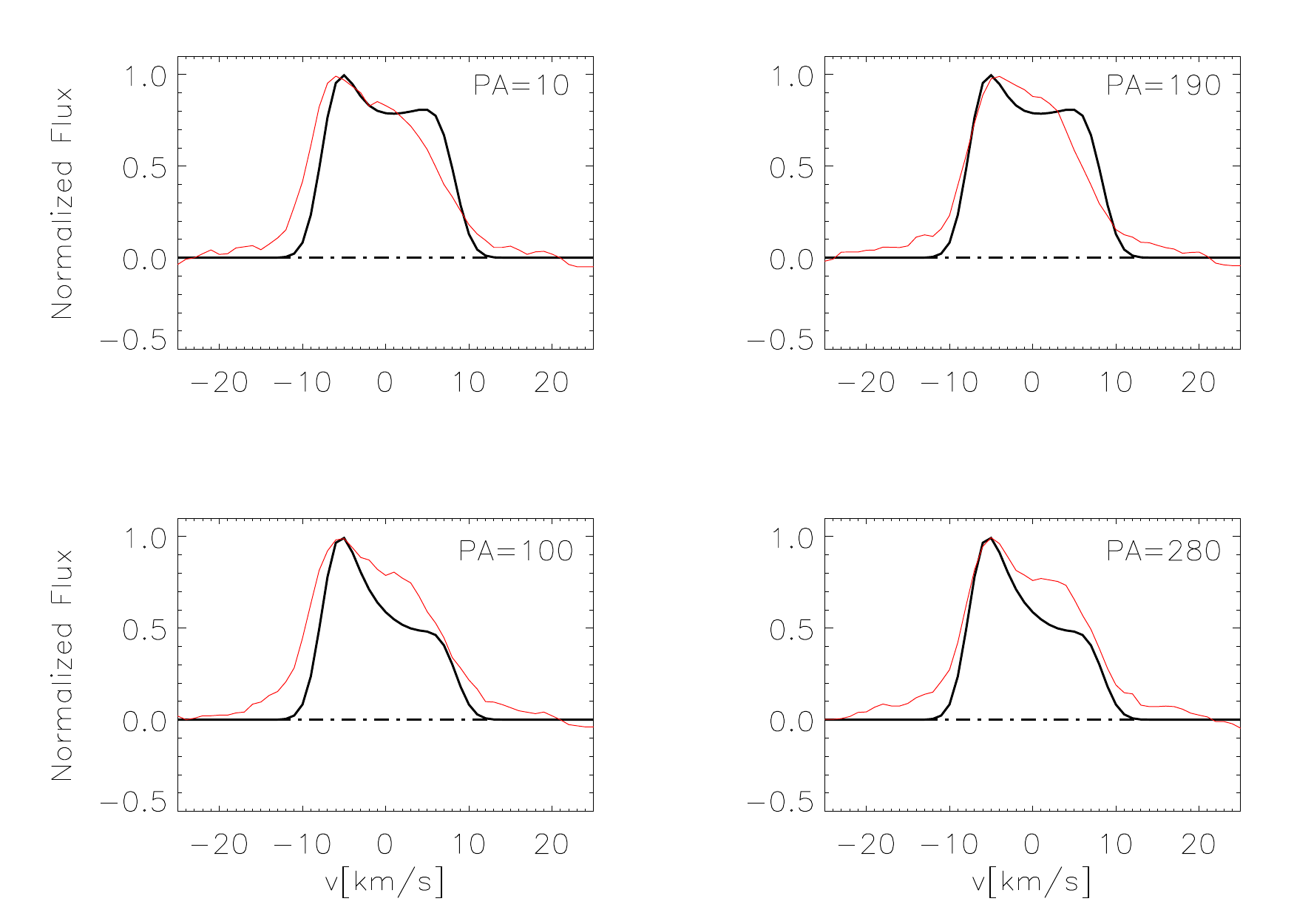}\\
%\cline{1-1}
\end{array}$
\end{center}
\caption{The top four panels are the average modelled and slit filtered comparison of the line sample collected on the 29th at P.A.=55/145/235/325, the middle four panels are the average modelled and slit filtered comparison of the line sample collected on the 30th at P.A.=55/145/235/325 and the bottom four panels are the average modelled and slit filtered comparison of the line sample collected on the 30th at P.A.=10/100/190/280. For all frames the pointing offset was 0$\farcs$6. The black lines are the median of all the individual modelled transistion lines (The individual lines are all quite similar and the median therefore looks almost identical to the individual lines). The red lines are the average observed line profile (described in Section \ref{sec:res}). The sample of lines plotted and used for the average are those listed in Table \ref{tab:obsline}.}
\label{fig:mod_prof}
\end{figure}

\subsection{Boltzmann plots/ population diagrams} \label{sec:mod_rot}
As described in the observational section, rotational diagrams are produced for the modelled line transitions for the two orthogonal position angles separately. The rotational diagrams are shown in the Appendix, Fig.\ref{fig:modbol1}.

\begin{figure}[!htbp]
\begin{center}$
\begin{array}{cc}
 \includegraphics[width=0.5\textwidth]{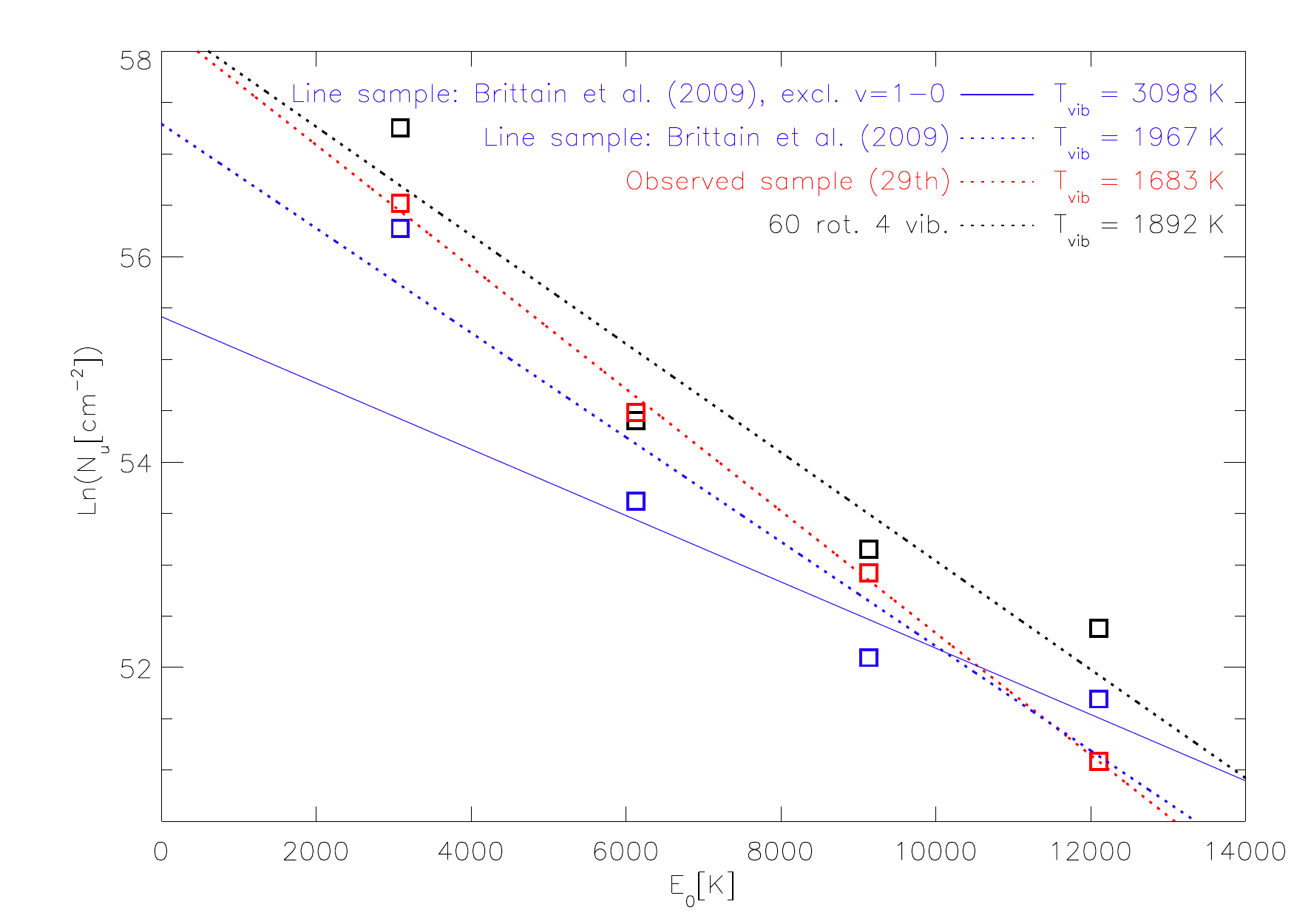}
   
\end{array}$
\end{center}
\caption{Vibrational diagram from the modelled lines of the lowest four v levels with all 60 rotational levels included. {The vibrational diagram from the modelled lines including only those in the sample observed on the 29th is overplotted. We also overplot two vibrational diagrams made using fluxes from our model but line samples equivalent to that used in \citet{brittain2009}. In one we use transitions from the first four v-bands while in the other we exclude the v=1-0 transitions from the fit. The limited samples have been shifted by a factor of 29.4 for the presentation.} The line fluxes used in this plot have not been slit filtered. Observing through a slit at various P.A. just shifts all the fluxes by the same constant factor and does not change the fitted $T_{\rm{vib}}$ substantially.}
\label{fig:modvib_full}
\end{figure}

Adding up all J levels for each v-band separately, modelled vibrational diagrams were made, following the scheme of the observed vibrational diagrams and plotted together with these in  Fig. \ref{fig:vibobs29} . We find a $T_{\rm{vib}}$=1691K at P.A.=145 and a $T_{\rm{vib}}$=1698K at P.A.=55 (with the lines and observational conditions from the 29th, see Table \ref{table:PA} and Table \ref{tab:obsline}). This underestimates the corresponding observed vibrational temperature by almost a factor of two (Section \ref{sec:boltz}), i.e. from our model we expect a much larger difference between line fluxes from differing v-bands than we actually observe.
For the 30th we find a $T_{\rm{vib}}$=1908K at P.A.=10 and $T_{\rm{vib}}$=1884K at P.A.=100. Here, we are close to the observed values (Section \ref{sec:boltz}) but the v=1 level is not included (no high quality v=1 lines where detected at these P.A. in the observations). For completeness, a full vibrational diagram including all 60 rotational levels present in the model, was produced for each of the four vibrational levels and is shown in Fig. \ref{fig:modvib_full}. { The vibrational diagram from our modelled sample, including only lines observed on the 29th, are overplotted for comparison. We find vibrational temperatures that are similar, but the sample from the 29th has been shifted up (by a factor of 29.4) for the presentation. In Fig. \ref{fig:modvib_full}, we also overplot two vibrational diagrams made using fluxes from our model but line samples equivalent to that used in \citet{brittain2009}. In one we use transitions from the first four v-bands while in the other we exclude the v=1-0 transitions from the fit (this approach is used in van der Plas et al. submitted). These two were also shifted by a factor of 29.4. It is clear from this comparison that varying the sample of the CO ro-vibrational lines included in the vibrational diagram can alter the derived vibrational temperature significantly.}

\subsection{Line fluxes and band ratios}
\begin{figure}[!htbp]
\begin{center}$
\begin{array}{cc}
  \includegraphics[width=0.5\textwidth]{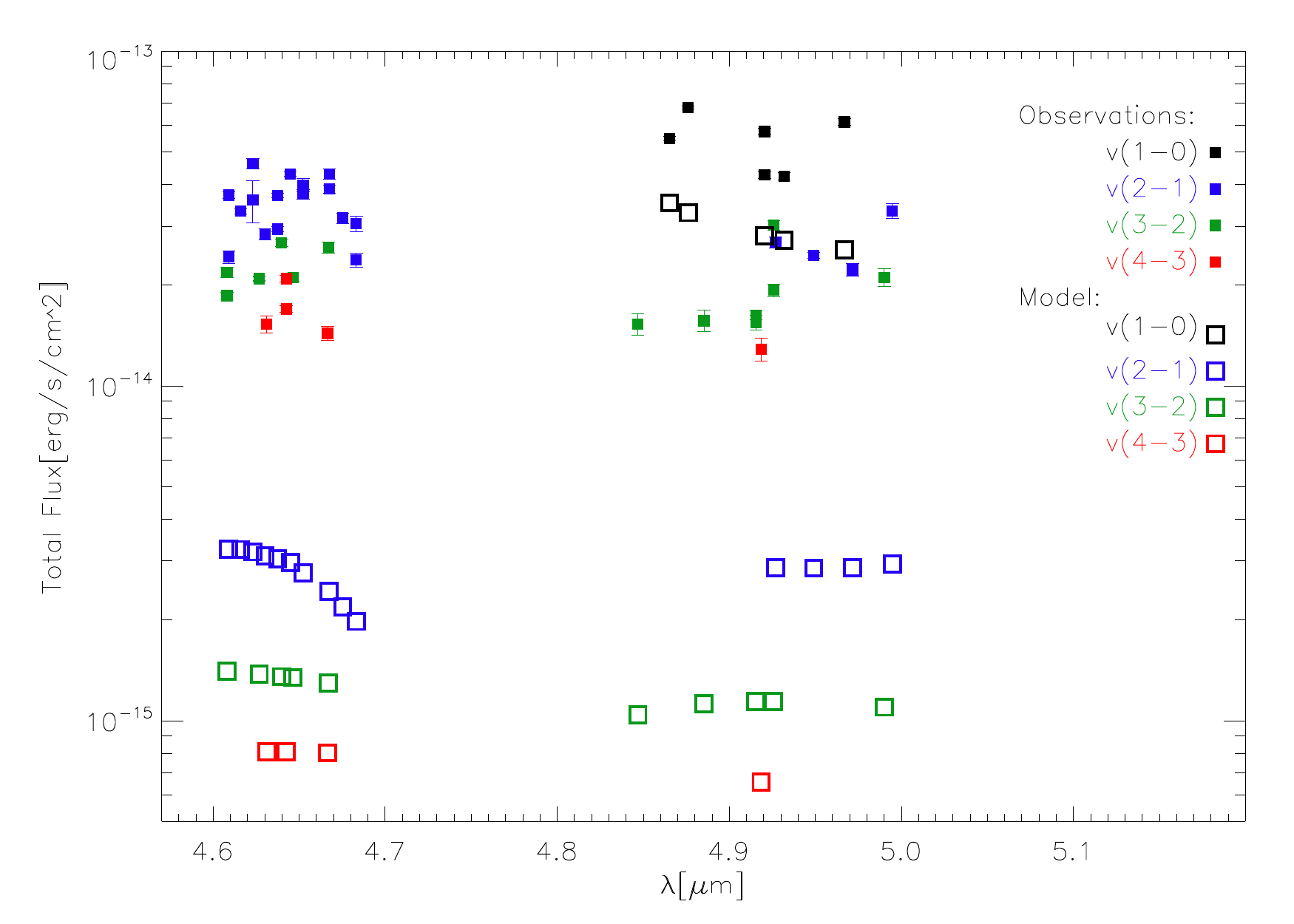}
 \end{array}$
\end{center}
\caption{Flux versus wavelength for all lines observed on the 29th at P.A.=145$\degree$ listed in Table \ref{tab:obsline}. The observed data is shown as triangles and for comparison the modelled fluxes are shown as stars. The different vibrational bands are colour coded: v=1-0 is black, v=2-1 is blue, v=3-2 is green, and v=4-3 is red. The error bars are in some cases within the size of the symbol.}
         \label{fig:wlFobs29}
\end{figure}

To further visualize our model to observation comparison, we overplot modelled and observed fluxes as a function of wavelength. We select the full line dataset collected at P.A.=145$\degree$ observed on the 29th (since this is the night with the best S/N and the position angle with least slit loss)(Fig. \ref{fig:wlFobs29}). The model line fluxes for the v=1-0 band are within a factor of two of the observed values. The model line fluxes of the higher v-bands fall a factor of 10 below the observed values. Hence the line ratio between higher and lower v-band lines is much higher in models than in observed data.

\subsection{UV fluorescence} \label{sec:modvsobs}
\begin{figure}[!htbp]
\begin{center}
 \includegraphics[width=0.5\textwidth]{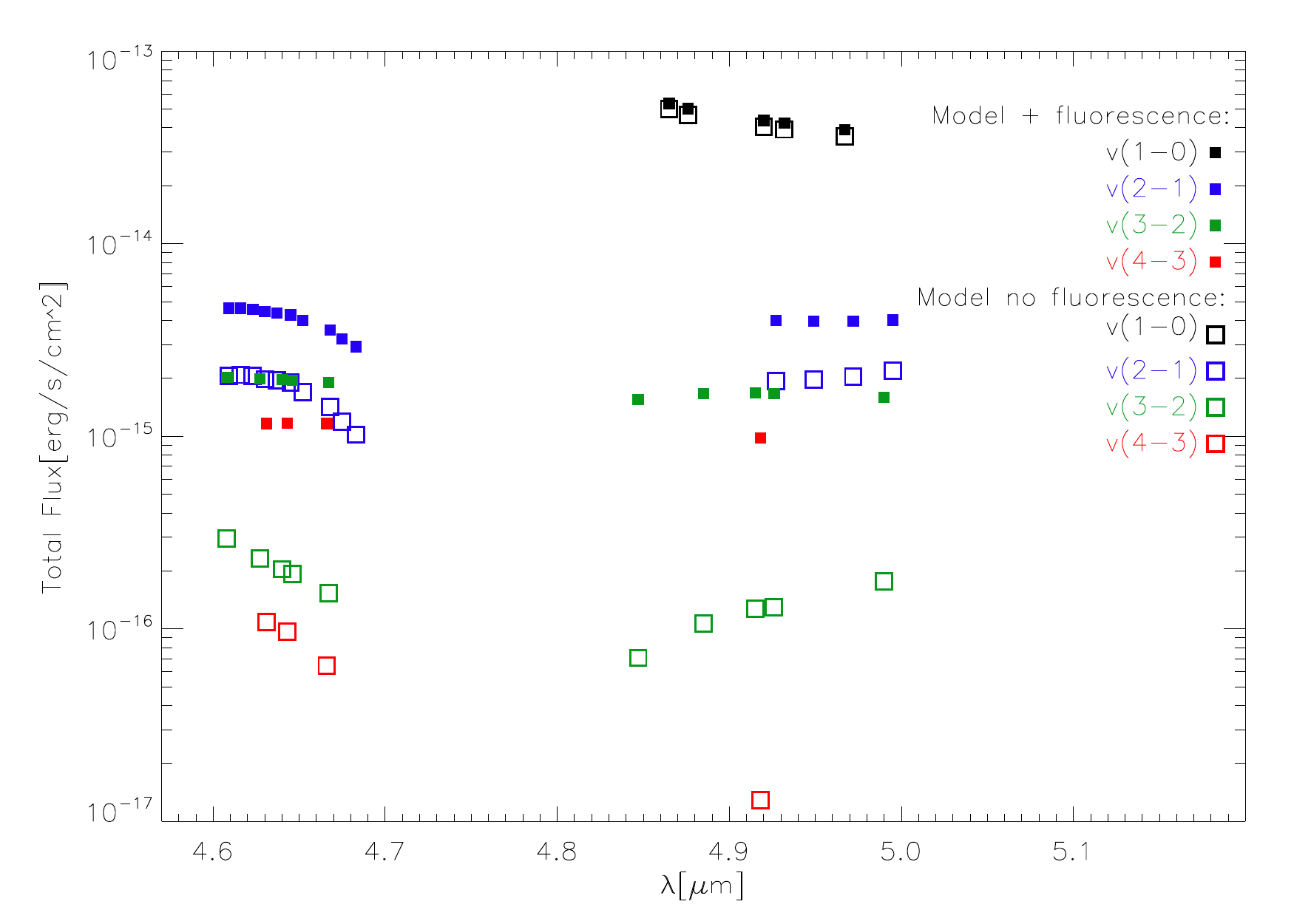}
 \end{center}
\caption{Line flux versus wavelength plot of the model with (filled squares) and without (hollow squares) UV fluorescence. The vibrational bands are colour coded (see legend on the figure).}
\label{fig:flu_off}
\end{figure}

From our model, we find that $T_{\rm{rot}}$ is approximately equal to $T_{\rm{vib}}$, while, from the observations we find a $T_{\rm{vib}}$ which is about 2-3 times $T_{\rm{rot}}$, indicating that fluorescence contributes strongly to the excitation.
To asses the performance of the UV fluorescence in the model, Fig. \ref{fig:flu_off} shows a line flux versus wavelength plot of  the model with and without UV fluorescence. For the v=1-0 lines the fluxes are similar, at v=2-1 the model with UV fluorescence is about a factor of two higher and finally the higher v-bands are about a factor of ten higher with UV fluorescence. The vibrational temperature calculated from the fluxes without UV fluorescence is $T_{vib}$=1099 K, while the previously derived value, for the model with UV fluorescence on was $T_{vib}$=1691 K. The UV fluorescence has significantly improved the modelled line fluxes and line ratios but just not enough.

The model with UV fluorescence piped through the slit simulator reproduces the observed line shapes and the line flux of the v=1-0 lines. However, the model underestimates the line fluxes of the remaining higher v-bands and the derived vibrational temperatures, indicating that the ratios of the v-bands with respect to one another are not correct (see Fig. \ref{fig:wlFobs29}). From our observational data, we find band ratios around 0.2-1 for the first four vibrational bands while the model predicts band ratios of about 10/1 between the first two vibrational bands.

\begin{figure}[!htbp]
\begin{center}
 \includegraphics[width=0.5\textwidth]{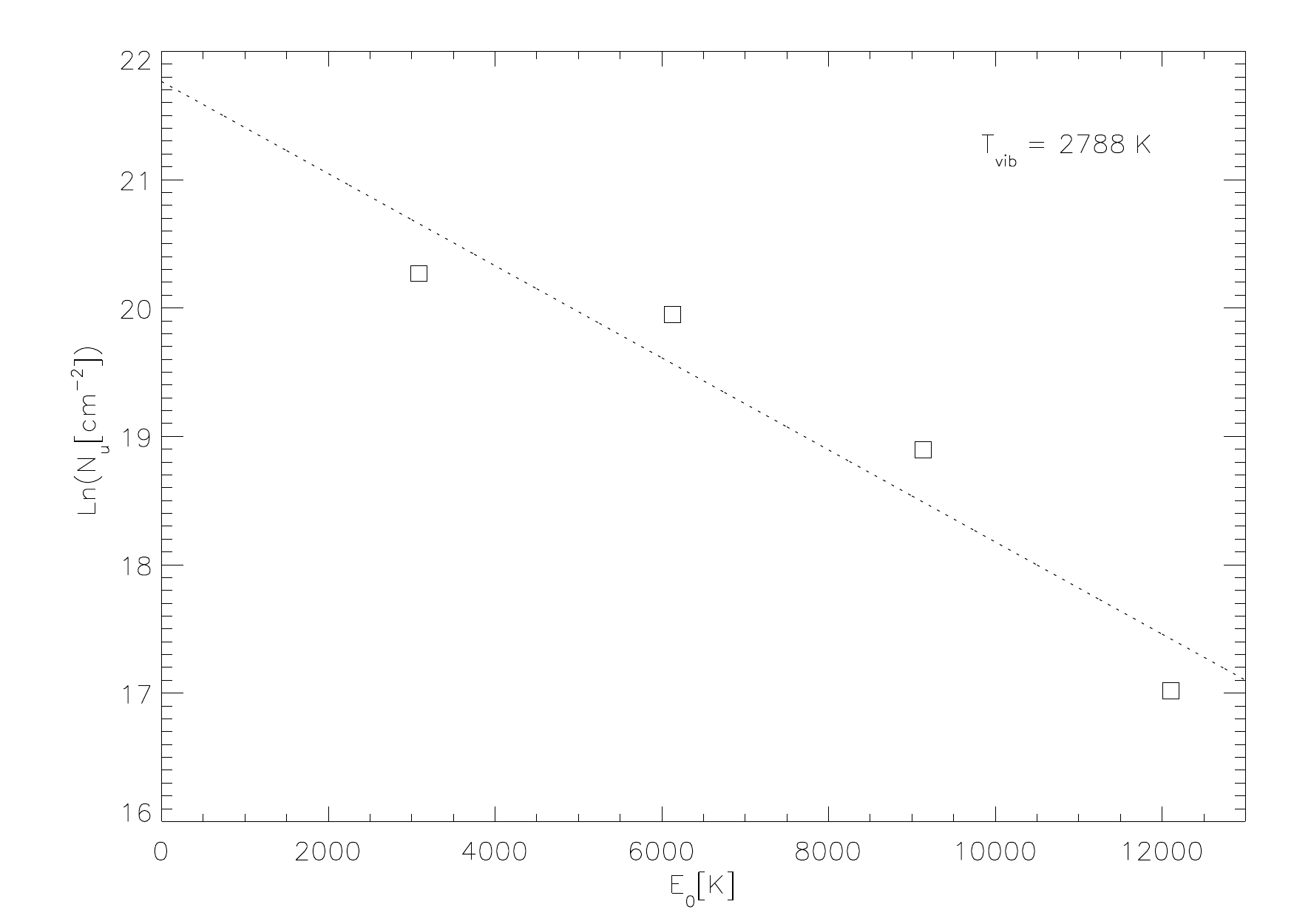}
 \end{center}
\caption{Vibrational diagram of the lowest four v levels. Same plot as in Fig. \ref{fig:vibobs29}, but with the alternative model with lower dust/gas ratio  (see Table \ref{tab:alt_models} model \#9).}
\label{fig:vib_lg22}
\end{figure}

To explore this discrepancy in line fluxes and ratios between model and observations, we ran several models with various parameters altered. These alternative models are listed in Table \ref{tab:alt_models}. 
The main idea in all of the tests is to get the CO warmer or exposed to more UV fluorescence. Since the gas heating/cooling and chemistry are calculated self-consistently we cannot change the position or the temperature of the CO directly but other parameters are not as constricted: 1) We increase the PAH abundance, since PAH is important in the heating of the gas. 2) We switch on CO self-shielding, since CO can then protect itself from photo dissociation by UV radiation and thereby sit higher up in the disc. 3) \& 4) We increase the grid size. Most of the CO emission comes from a narrow region at the disc wall and maybe we do not resolve this region well enough. 5) \& 6) We use a smaller gas to dust mass ratio. This allows us to test the effects of lower gas volume densities leading to less efficient depopulation of higher levels by collision.
The model that comes closest to reproduce the vibrational temperature and thereby the line flux ratio that we observe, is one with a lower gas mass. The vibrational temperature for this model is $T_{\rm{vib}}$=2788 K and the vibrational diagram is shown in Fig. \ref{fig:vib_lg22}. The absolute strength of the lines are underestimated by three orders of magnitude compared to observations (Table \ref{tab:alt_models}). 

Lowering the total gas mass leads to lower gas volume densities making collisional quenching of the population less efficient. This enables the higher v-levels to stay flourescently pumped. To obtain both the correct line fluxes and band ratios, we should keep the gas mass fixed, but ensure that the volume densities are much lower. One possibility could be that the gas has to be more vertically extended than assumed, i.e. larger scale heights at the secondary wall.
A low density CO layer on top of the disc wall.
In this test model, the profiles where only slightly wider (below CRIRES detection level), and the emitting region slightly farther out, still consistent with observations.

\begin{table}[!htbp]
\caption{Overview of the small grid of models}             % title of Table
\label{tab:alt_models}    
\centering                         
\begin{tabular}{c c c c c c c c}       
\hline            
	$$&$\frac{\rm{dust}}{\rm{gas}}$&CO&f$_{\rm{PAH}}$&grid&$T_{vib}$&\multicolumn{2}{c}{$F$[$\rm{\frac{erg}{cm^2 s}}$]}\\
	&&&&&[K]&\multicolumn{2}{c}{[10$^{-14}$]}	\\
\hline            

Obs	&---&---&---&---&2978&4.3&6.8	\\
Orig	&0.71&off&0.7	&100x100&1683&.36&5.0\\
\#1 	&0.71&off&7.0	&100x100&1438&.92&19.\\
\#2 	&0.71&on	&0.7	&100x100&1708&.39&5.1\\
\#3 	&0.71&off&0.7	&200x200&1786&.36&4.8\\
\#4 	&0.71&on	&0.7	&200x200&1806&.37&4.8\\
\#5 	&7.1&off&0.7&200x200&2385&.040&.19\\
\#6 	&71.&off&0.7&200x200&2788&.002&.006\\
 
\hline            
   
\end{tabular}
\tablefoot{We list line fluxes and vibrational band ratios. Varied parameters are: the dust/gas mass ratio, the PAH abundance, the grid size and switching CO self shielding on/off. The outcome of the various models (without slit filtering applied) are presented here by comparison of the $T_{vib}$ and the line flux for two representative lines: v(2-1)R06, v(1-0)P22. All model here are unfiltered. The line sample used for the computation of $T_{vib}$ is that collected on the 29th and the observational result is shown at the top.} 
\end{table}

%\begin{figure}[!htbp]
%\begin{center}
%  \includegraphics[width=0.5\textwidth]{wlflux_obs29_gdless2_200.pdf}
%\end{center}
%\caption{Same plot as in Fig. \ref{fig:wlFobs29}, but with the alternative model with lowered %dust/gas ratio (see Table \ref{tab:alt_models} model \#6). The observed data is shown as %triangles and the modelled fluxes are shown as stars. The different vibrational bands are colour %coded: v=1-0 is red, v=2-1 is green, v=3-2 is blue and v=4-3 is orange.}
%         \label{fig:altmodel}
%\end{figure}

\section{Discussion} \label{discus}
In our sample of observed lines, we see variations in line profile shape. Both symmetric double peaks, as expected for a disc in rotation, but also asymmetric line profiles are observed. The line shapes change with varying P.A., but stay the same through all v-levels and rotational quantum number J. Shape variations between P.A. are expected because of the similar angular size of the slit width and the disc wall. Meanwhile, in the two different nights, we see different profile shapes for the same P.A. (P.A.=235$\degree$), indicating a changing offset from one night to the next. {Figures \ref{off29} and \ref{off30} in Appendix \ref{app:slit},} visualize the changes in offsets seen from the line profile over time.

Through the first full night, where the telescope is re-centred several times, three of the four position angles are consistent with one roughly constant offset, i.e. the offset cannot be caused by a random error. The spectra from the remaining position angle would suggest a very different offset than the rest but was collected in between the rest.
On the second night the various position angles could be consistent with one offset. 
Pointing offsets can explain the observed line shape asymmetries well, but the source of the offset is not fully understood.
A test observation was recently performed with the CRIRES instrument (July 2013) observing HD100546 at antiparallel position angles. {The presence of an offset was confirmed and is most likely caused by a misalignment between slit rotation axis and the centring of the instrument. This type of misalignment can, for cases where the disc is spatially resolved (nearby discs) and the emission is coming from a limited region (e.g.  transitional discs), have a very large impact on the CO ro-vibrational line profiles (e.g. mimic disc asymmetries). However, for discs with no inner gap, the CO ro-vibrational lines are emitted much closer to the star and would therefore not be affected much by slit offsets. The same is true for more distant discs that are not spatially resolved.}

In general, line profile asymmetries could also be explained by disc asymmetries. {The line profile variations that originates from an elliptical emitting region with the star offset, would be qualitatively similar to those arising from a slit pointing offset along the semi-major axis of the disc.
In that sense our grid of offset versus P.A. variations (Fig. \ref{fig:paramtest}) can also give an indication about how disc asymmetry would affect the line profile shapes.} 
However, in this case the 'offsets' derived from the line profiles should not change between various position angles or from one night to the next, but rather indicate the same offset. One could imagine that the disc could have an uneven distribution of CO emitting regions. To explain asymmetric variations between all position angles, we could have several randomly distributed strong CO emitting 'spots' that by chance fall into some of the position angles. Even in this case, we should not see changes from one night to the next at the same P.A., or anti-parallel P.A. This would imply that the disc has undergone some physical change between such varying spectra. This kind of change is highly unlikely since the time between collection of spectra is too short for variations in the disc to happen(down to 20 minutes).

{\citet{brittain2013} found single peaked profiles for the CO ro-vibrational emission from HD100546. They use the Gemini South telescope with a slit width of 0$\farcs$34. The double-peaked shape could be lost due to their lower spectral resolution (R=50,000). Comparing several observations at different epochs, they find a small shifting asymmetry only present in the v=1-0 lines. They suggest this could be due to a circumplanetary CO component orbiting in the outer disc.} According to their analysis, this additional CO component would be at zero velocity at the epoch of our observations (Brittain, private communication). With our very narrow slit, the small CO component would easily be filtered out, meaning that our symmetric double peaked line profiles and the above conclusions are not inconsistent with their findings.

The sensitivity of the line profile shape to slit P.A. changes indicates that the main contribution to the CO ro-vibrational emission is coming from the disc wall. The shapes of the observed lines are similar for all v-levels, also supporting that these lines should all originate in the same region (the disc wall). The model prediction that most of the emission originates between 10 and 13 AU, is consistent with this. {Our modelled spatial extent underestimates the observed as found by \citet{goto2012} by a factor of 0.7. However, the modelled extent was calculated from one line while the observed was calculated from a range of v=2-1 lines. Furthermore, the comparison needs to be improved by convolving the modelled spatial profile with the observed PSF instead of just approximating by a Gaussian and by co-adding the same sets of lines (work in progress). }

Measuring line fluxes and flux ratios (vibrational temperatures), there are inconsistencies between model and observations. Our model line fluxes are consistent with observations within a factor of two for the first vibrational band. We underestimate the line fluxes of the higher v-bands by about a factor of ten. We therefore also underestimate the line flux ratios by up to a factor of ten. This is probably because the CO is not sufficiently pumped by UV fluorescence. At heights in the disc where the stellar UV field reaches directly, CO self-shielding and H$_2$ shielding play a minor role. 

We can reproduce the line flux ratios from the observations with lower gas volume densities, potentially connected to a larger vertical extent of the gas. A low density layer atop the inner rim. We checked that the scale height calculated from the gas temperature is not significantly different from what the input in the parameterized model. A low velocity disc wind as a cause for a low density vertically extended CO component can be ruled out since \citet{pontoppidan2008} and \citet{bast2011} showed that this would cause single-peaked instead of double-peaked line profiles.

\section{Conclusion}
In this paper, we present for the first time an extensive analysis of a large comprehensive set of observational data of the CO ro-vibrational emission lines from the disc around HD100546: Collected at eight different position angles covering six different grating settings at the CRIRES/VLT. The observations show line asymmetries and the line profile shapes vary with position angle. We also present modelled emission lines, using a ProDiMo model of HD100546 piped through a slit filtering tool. This model is in no way designed for the CO ro-vibrational lines, so the comparison provides us with a very strong test.

We find from our observational analysis that large parts of the CO ro-vibrational emission is emitted at the disc wall (10-13 AU). The total line flux builds up out to 40 AU (Goto et al. 2012).
Our observational data contains many CO ro-vibrational transitions and we find that the shape of the lines is similar for all v levels and all J levels. We find unexpected low band ratios between the v=1-0 and the other v-bands and calculate a relatively high vibrational 'temperature', $T_{\rm{vib}}\sim$3000K, for our observed sample of lines.

With our HD100546 model and the slit filtering tool, we reproduce the observed line shapes. Hence the model does an excellent job in predicting the correct CO ro-vibrational emission region for all v-bands. {The main part of the CO ro-vibrational emission in the model is emitted at the disc wall 10-13 AU, PSF convolved out to 28 AU (for one sample line).}
Our model reproduces the observed line fluxes for the v=1-0 lines within a factor of two. It underestimates the line fluxes for the higher v-bands by a factor of 10. The band ratios between v=1-0 and v=2-1 are not reproduced and neither is the vibrational temperature. From the model we find $T_{\rm{vib}}\sim$1700K which differs from the observationally derived value with about a factor of two. 

An explanation for the above discrepancies could be that the fluorescence pumping is not efficient enough in our standard model (factor $\sim$10).
Modelling tests suggest that lower gas volume densities can yield higher $T_{\rm{vib}}$, e.g. the same gas mass, but more vertically extended than the original model assumed.

In this particular case of a nearby (spatially resolved) transitional disc (inner cavity) the choice of slit position angle, the slit width and the centring of the slit can have a very large impact on the CO ro-vibrational line profiles. Therefore, these details need to be carefully considered when planning observations of CO ro-vibrational emission from such a disc. However, in the absence of an inner gap, the CO ro-vibrational lines will be originating much closer to the star and would not be affected by the slit in the same way. The same is true when the disc is more distant and no longer spatially resolved.

%__________________________________________________________________
%______________________________________________________________

\begin{acknowledgements}
The authors would like to thank Jonathan Smoker for performing tests on the CRIRES spectro-astrometry and Sean Brittain for discussions of the CO ro-vibrational data and a careful and constructive referee report. IK, WFT, and PW acknowledge funding from the EU FP7-2011 under Grant Agreement no. 284405 (PERG06-GA-2009-256513). GvdP acknowledges support from the Millennium Science Initiative (Chilean Ministry of Economy), through grant "Nucleus P10-022-F".
\end{acknowledgements}

\begin{appendix}

\section{Line profiles and slit loss} \label{sec:obs_com_prof}
In our dataset, we found clear variations in the shape of line profiles observed at different position angles. 
These variations are due to the relative size of the slit and the disc on the sky (Fig. \ref{fig:velocity}). At the distance of HD100546 (103 pc), a slit width of 0$\farcs$2 corresponds to a width of about 20 AU. This is coincidentally close to the diameter of the disc wall. Fig. \ref{fig:velocity} shows how varying the position angle of the slit results in {losing} parts of the emission originating at the disc wall (10-13 AU).
Thus different slit position angles will lead to different profile shapes. For HD100546 the resulting line profiles should show double peaks at P.A.$=145\degree$ and P.A.$=325\degree$, while showing flat topped profiles at P.A.$=55\degree$ and P.A.$=235\degree$. 

With a slit width that barely includes most of the disc wall, a poorly centred slit can affect the shape of the lines. 
If the slit is offset to one side, we {lose} emission from the velocity channels falling outside the slit.

These important effects need to be considered in both our observational analysis and in our modelling efforts.

\subsection{Observed line profiles} \label{sec:obs_line}
Table \ref{tab:obsline} presents a full listing of our selected unblended and clean CO ro-vibrational lines. For all these CO ro-vibrational transitions, line profiles (flux as a function of velocity) have been extracted. To investigate whether we can build high resolution average line profiles,
we checked the data for shape variations as a function of v, J, P.A., and different nights.

The individual line profiles are normalized by their fitted continuum, shifted to zero continuum, and the line profile is then normalized to the maximum flux value. Profiles are then combined to median average profiles.

\begin{figure}[!htbp]
\begin{center}$
\begin{array}{cc}
   \includegraphics[width=0.5\textwidth]{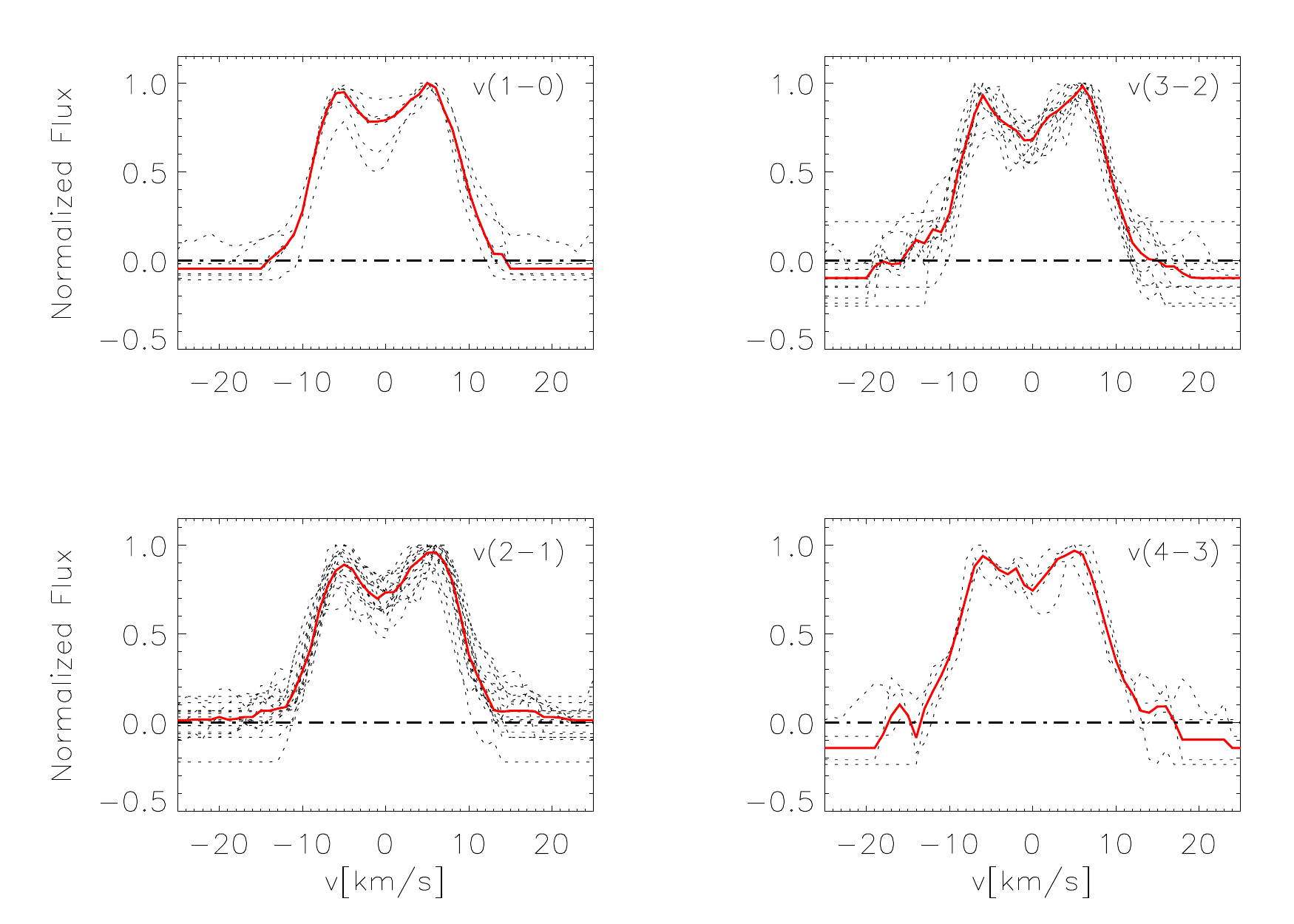}

\end{array}$
\end{center}
\caption{Averaged line profile at P.A.=145$\degree$ for lines observed on the 29th, for each v level separately. The lines included are those listed in Table \ref{tab:obsline}. The dotted lines are the normalized individual transitions and the red line is the median average of all these lines. }
         \label{fig:v_comp145}
\end{figure}

The profile shapes vary between the two nights and also between the four/eight different position angles. We find no significant variations when comparing v-levels and wavelength settings (see Fig. \ref{fig:v_comp145} for an example at P.A.=145$\degree$ for the night of the 29th). For a more quantitative view, the FWHM and the peak separations, $\rm{v_{sep}}$, for P.A.=145$\degree$ are shown for all v-bands in Table \ref{table:FWHM}. The average for each v-band is build over the individual J transitions at that v-band. The 1$\sigma$ standard deviation listed in the table is the deviation of the values at each J-level from the mean. The average from all v-bands agree with each other within 1$\sigma$.
%6.67*10^(-11)*2.4*2*10^(30)/3*10^7
%=6.7*2.4*2*10^12/3=10.72*10^12=28.7

The difference between line shapes observed at the same P.A. on the 29th and on the 30th could be explained by different pointing offsets present for each night. Within the individual position angles there are no shape variations during one night, i.e. data from different wavelength settings taken at different times with same P.A. are very similar.

\begin{figure}[!htbp]
\begin{center}$
\begin{array}{cc}
  \includegraphics[width=0.5\textwidth]{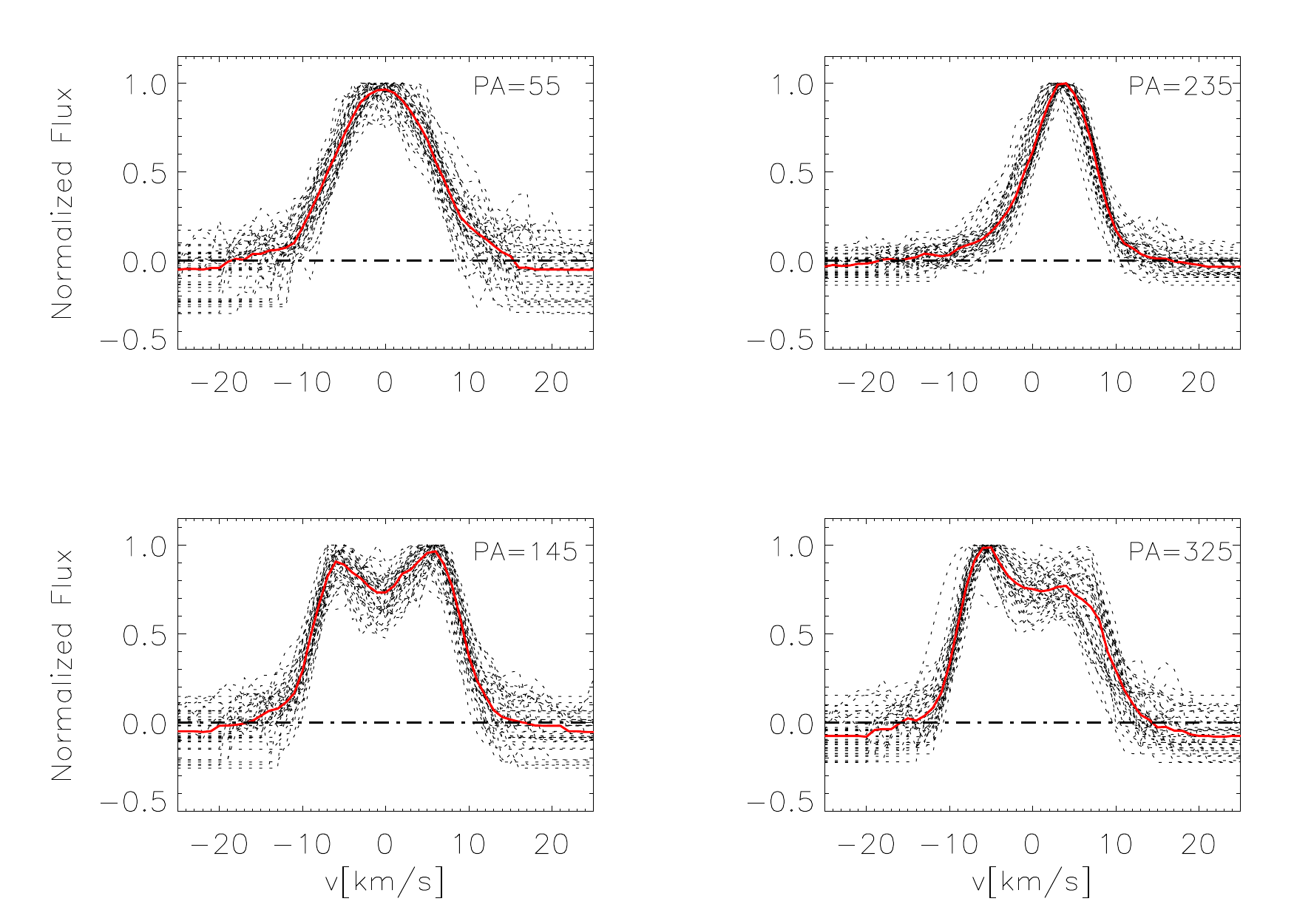}\\

  \includegraphics[width=0.5\textwidth]{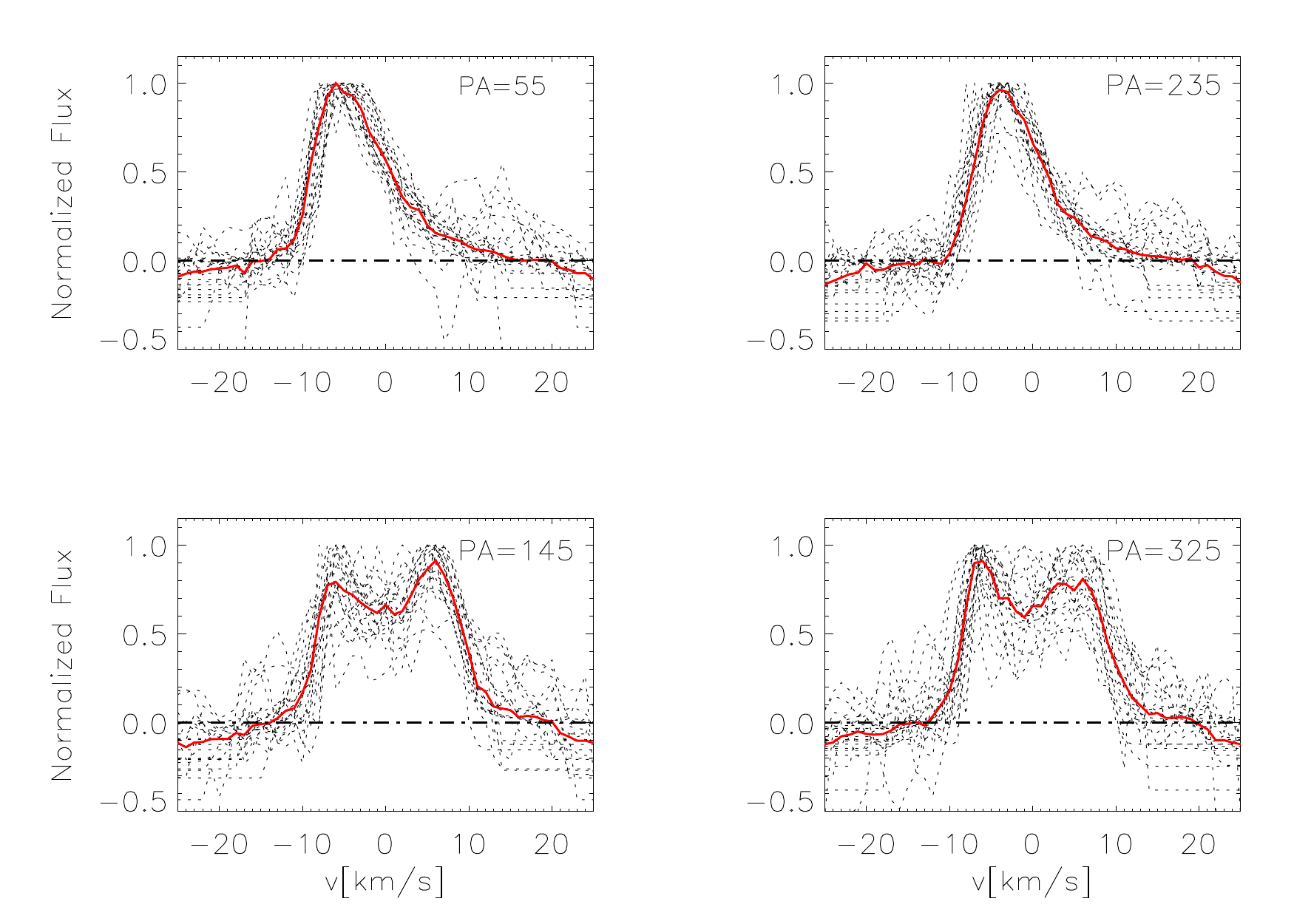}\\
  \includegraphics[width=0.5\textwidth]{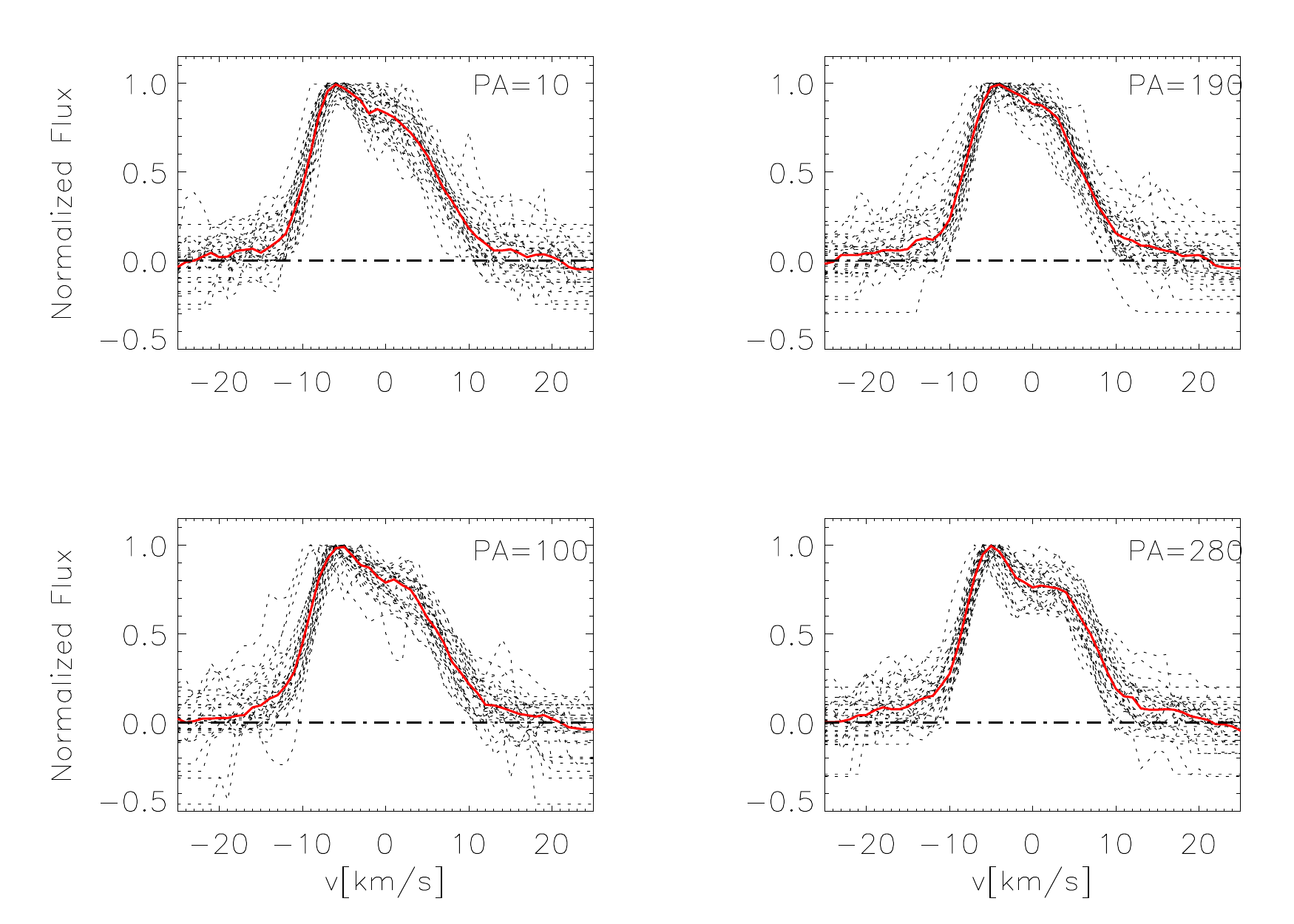}
\end{array}$
\end{center}      
\caption{Averaged line profiles at eight different position angles containing transitions from all v-levels observed on the 29th and 30th, listed in Table \ref{tab:obsline}. The dotted lines are the normalized individual transitions and the red line is the median of all these lines. The upper four panels show the line profiles collected at P.A.=$55\degree/145\degree/235\degree/325\degree$ on the 29th. The middle four panels show the line profiles collected at P.A.=$55\degree/145\degree/235\degree/325\degree$ on the 30th. The lower four panels show the line profiles collected at P.A.=$10\degree/100\degree/190\degree/280\degree$ on the 30th. }
         \label{fig:avprofileb_obs}
  \end{figure}

The above found stability of the line profile shapes justifies building averaged line profiles over velocity for each P.A. and each night (independent of v, J). 
These co-added line profiles are shown in Fig. \ref{fig:avprofileb_obs}. For the 29th, we find double peaked profiles at both P.A.=145$\degree$ and P.A.=325$\degree$, as we would expect given the disc itself has a position angle of 145$\degree$. At P.A.=55$\degree$ and P.A.=235$\degree$, we find single peaked asymmetric profiles, indicating an offset (with no offset we should find flat topped profiles). The profiles for the anti-parallel positions P.A.=55$\degree$ and P.A.=235$\degree$, though very similar, do not match exactly but lean in opposite directions, which should not be the case for a true antiparallel case. For the 30th, the four repeated position angles (P.A.=55$\degree$, P.A.=235$\degree$, P.A.=145$\degree$, and P.A.=325$\degree$) show slightly altered shapes indicating a change in the offset with respect to the first night. The four new position angles (P.A.=10$\degree$, P.A.=190$\degree$, P.A.=100$\degree$, and P.A.=280$\degree$) all show asymmetric single peaks with a hint of a shoulder.

\begin{table}[!htbp]
\caption{FWHM and peak separation of the average line profiles}             % title of Table
\label{table:FWHM}      % is used to refer this table in the text
\centering                          % used for centering table
\begin{tabular}{c c c c c }        % centered columns (4 columns)
\hline            
 
 v-band	&FWHM [km/s]&$\sigma$&v$_{\rm{sep}}$ [km/s]&$\sigma$\\   
\hline            
 v=1-0 & 	18.30&1.05 &10.70&0.57 \\
 v=2-1 & 	18.16&0.76	 &10.50&2.47 \\   
 v=3-2 & 	18.25&0.68	 &11.97&1.34 \\   
 v=4-3 & 	18.35&0.64	 &11.14&2.10 \\   
\hline            
   
\end{tabular}
\tablefoot{Done for each v-band separately for lines observed at the P.A.=145$\degree$ on the 29th. The error listed is the 1$\sigma$ standard deviation of the individual J-levels included in each average.}   
\end{table}

\subsection{Modelled line profiles}
ProDiMo outputs line data cubes for each chosen CO ro-vibrational transition. Each cube contains a 201x201 pixel coordinate grid in units of AU and the spectral intensity in $\rm{[erg/cm^2/s/Hz/sr]}$ at every spatial position in the described coordinate grid for each of the 91 velocity channels, covering -20 km/s to +20 km/s and  the continuum (see Appendix \ref{app:cube}). From this cube we create line images, profiles, and flux tables. They are our modelled 'observations'.

\subsubsection{Applying a slit filter to line data cubes} \label{app:slit}
The line cubes do not contain any observational or instrumental effects. Hence, they need to be convolved with the instrumental PSF (we approximate the PSF by a Gaussian), rotated 
to the position angle of the disc on the sky and corrected for slit loss and pointing offsets before comparison with the actual observations.
Each cube is therefore piped through a slit filtering IDL procedure \citep{carmona2011} adapted for the purposes in this paper. The slit filter applies the described corrections and produces line profiles that can be directly compared to our observed data.

Each spectral slice of the ProDiMo data cube is convolved in the spatial ($x,y$) direction with a Gaussian two dimensional PSF using a FWHM of 0$\farcs$169 and the idl routines 'convolve.pro' and 'psf\_gaussian.pro'). In the spectral direction, the cube is convolved with the CRIRES instrumental resolution (R = 90000) using the routine 'gaussfold.pro'. We rotate the cube in the $x,y$, direction to match the position angle of HD 100546 by 145 degrees using the routine 'rot.pro' and cubic interpolation with an interpolation parameter of -0.6.

\begin{figure}[!htbp]
\begin{center}$
\begin{array}{cc}
   \includegraphics[width=0.3\textwidth]{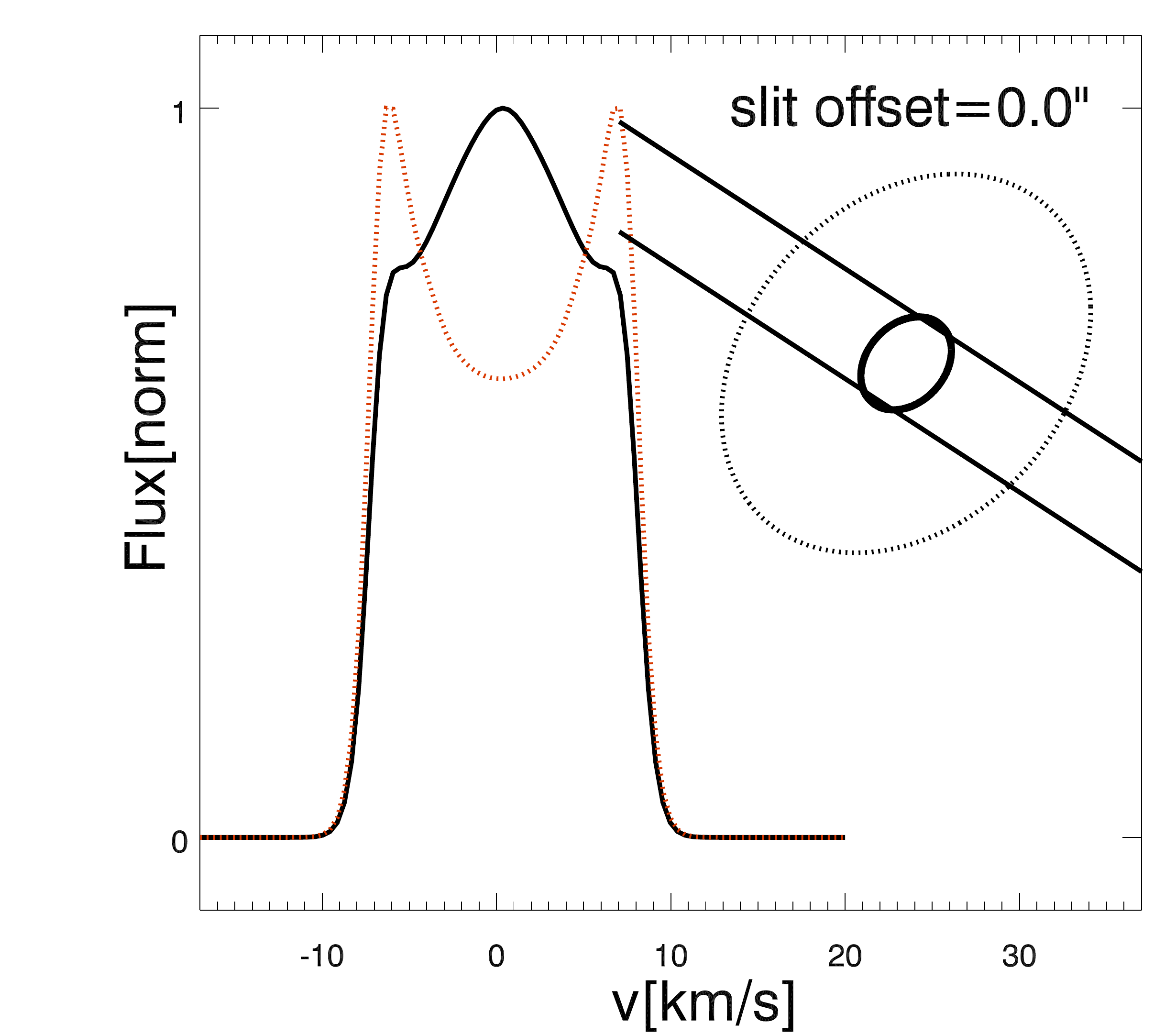}\\
   \includegraphics[width=0.3\textwidth]{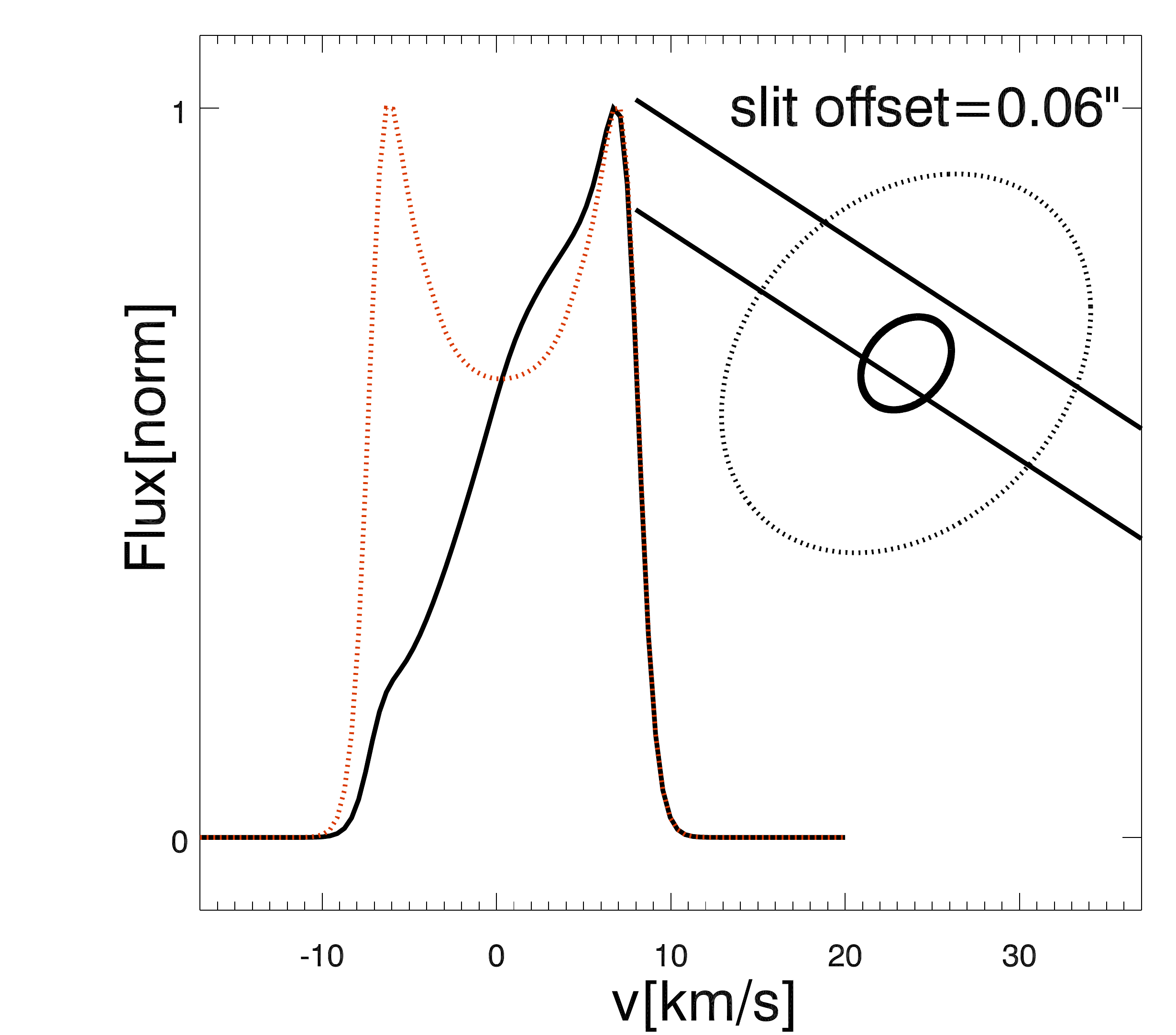}\\
   \includegraphics[width=0.3\textwidth]{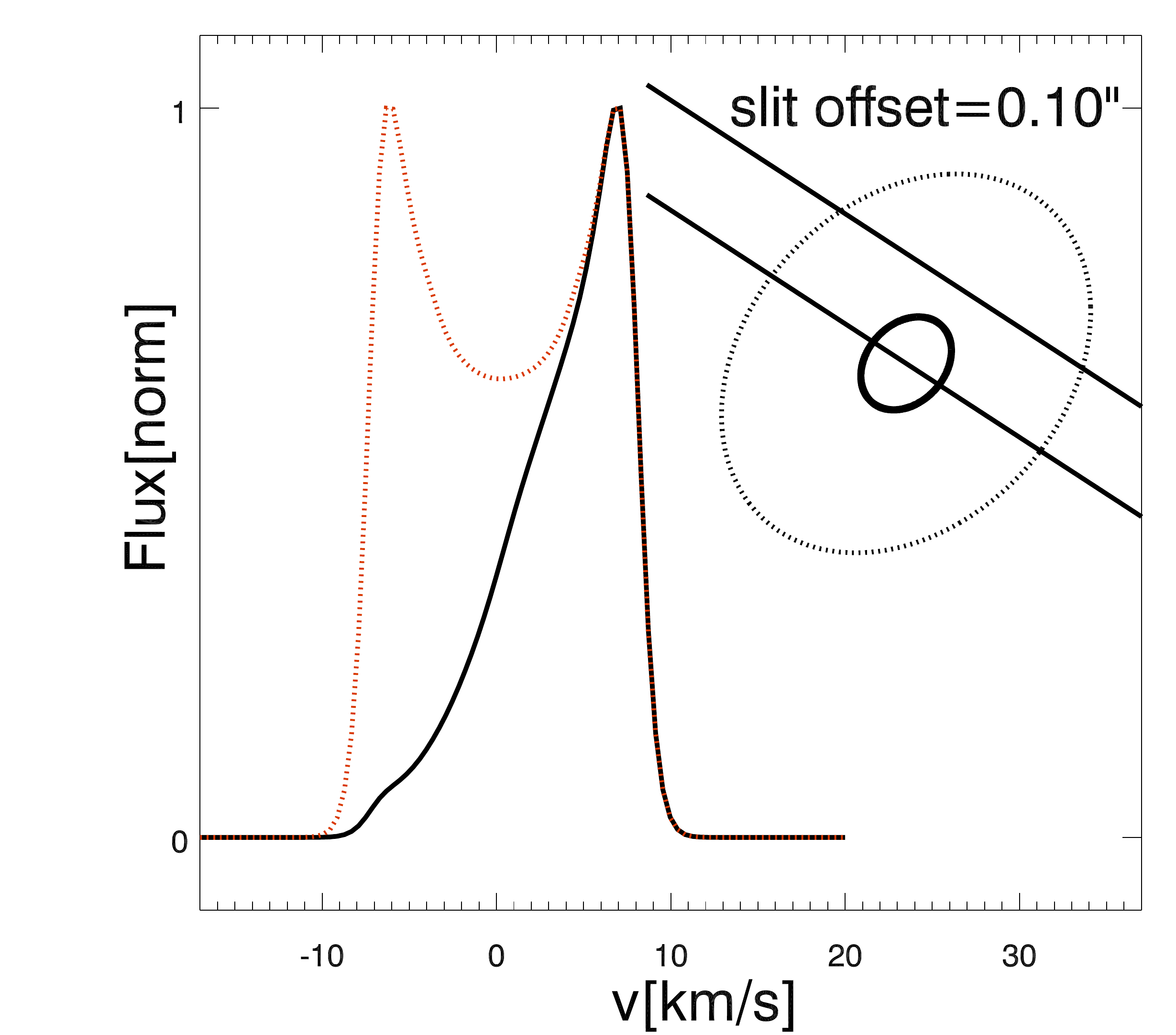}\\
   \includegraphics[width=0.3\textwidth]{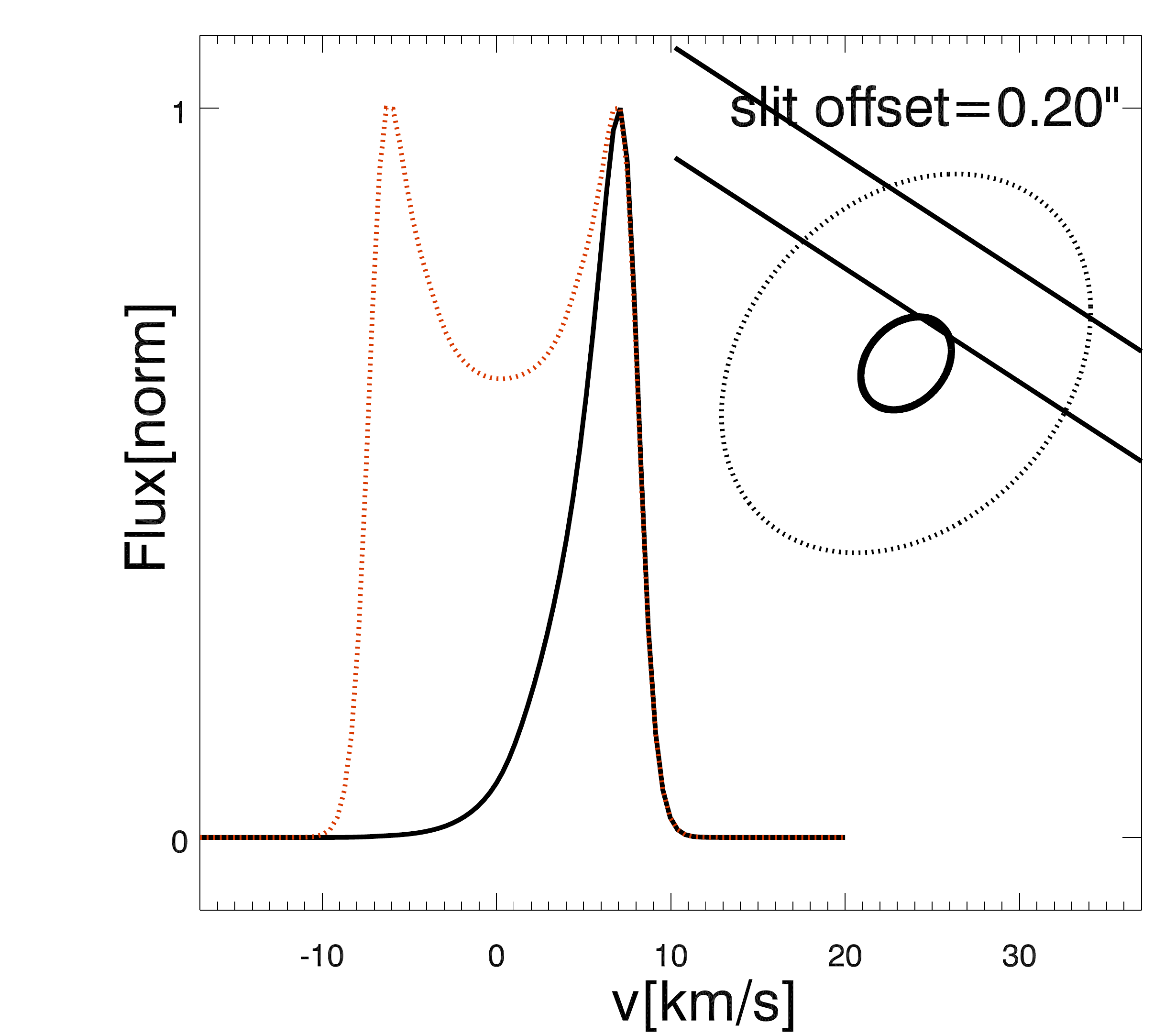}
 \end{array}$
\end{center}
      \caption{Output from the IDL slit filtering tool, showing the unfiltered normalized line profile (red line) and the slit filtered normalized line profile (black line) for three different pointing offsets (from the top first frame: 0", second frame: 0$\farcs$06, third frame: 0$\farcs$10, third frame: 0$\farcs$20). Next to each profile, a sketch of the disc with the slit on top is drawn. The dashed ellipse is a 50 AU annulus, while the solid ellipse represents the disc wall. The line profile shown is the v(2-1)R06 transition with the slit at a position angle of 55$\degree$.}
         \label{fig:astrometry_astrocube}
   \end{figure}

Each spectral slice of the data cube is now covered with a two dimensional mask (x,y) representing the spectrograph slit (with arbitrary orientation and/or offset from the disc centre and centred on $x_0$,slit, $y_0$,slit). From the resulting signal we calculate the line profile and astrometric signal (SPP and FWHM). The SPP is calculated as the shortest distance between the centre of mass (centroid) of the CO + continuum emission and a line perpendicular to the slit running through the slit origin. The FWHM is calculated from a Gaussian fit to the CO + continuum emission.  Fig. \ref{fig:astrometry_astrocube}, shows for three different offsets the effect that slit filtering has on a line profile, together with a sketch of the slit on disc. {The line profiles are normalized in the same way as the observed profiles (see Sect. \ref{sec:obs_line}).}

\begin{figure}[!htbp]
\begin{center}$
\begin{array}{cc}
   \includegraphics[width=0.5\textwidth]{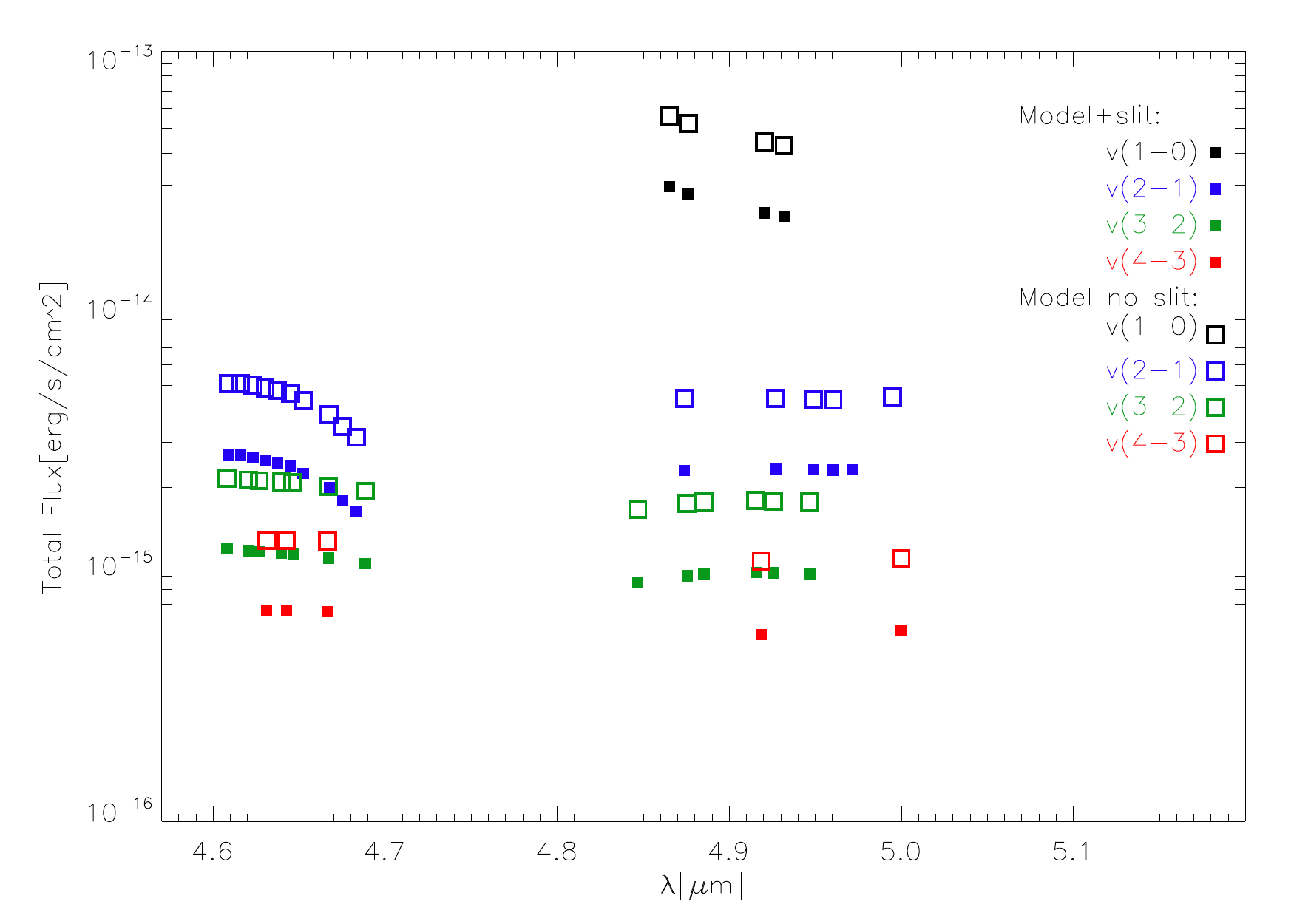}
 \end{array}$
\end{center}
\caption{Flux versus wavelength comparing the slit filtered model to the non slit filtered model for all lines observed on the 29th at P.A.=145$\degree$ listed in Table \ref{tab:obsline}. The slit filtered modelled fluxes are shown as diamonds and the modelled fluxes without slit filter are shown as stars. The different vibrational bands are colour coded: v=1-0 is black, v=2-1 is blue, v=3-2 is green and v=4-3 is red.}
         \label{fig:wlfl_unfilt}
\end{figure}

To asses the importance of slit loss and telescope effects, we have also compared the unfiltered model line fluxes to the slit filtered model line fluxes (Fig. \ref{fig:wlfl_unfilt}). We find that the slit affects all transitions in the same way and flux ratios are conserved through the slit filtering. This was expected, because the spatial profiles of the line emission are very similar.

Because our observed line profiles change on very short time scales (see Appendix \ref{sec:obs_line}) it is virtually impossible to explain the shape asymmetries and variations with disc asymmetries alone and this calls for the shape variations to be caused by pointing offsets.

\begin{figure*}[!htbp]
\begin{center}$
\begin{array}{cc}
   \includegraphics[width=1.25\textwidth, angle=90]{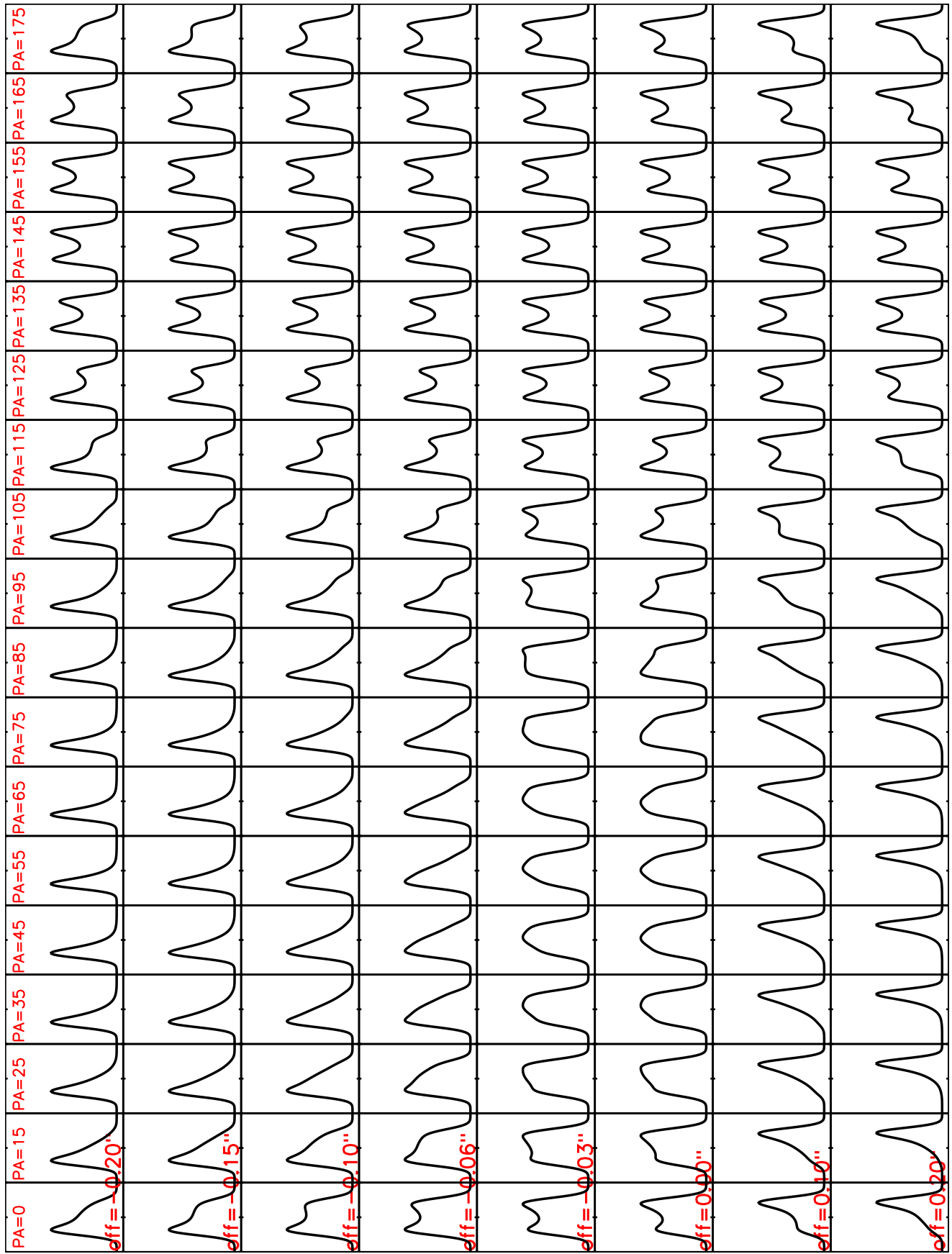}
\end{array}$
\end{center}
\caption{Parameter exploration made with our slit filtering tool. The position angle of the slit was varied from 0$\degree$ to 175$\degree$ in steps of 10$\degree$. The offset values are -0$\farcs$20, -0$\farcs$15, -0$\farcs$10, -0$\farcs$06, -0$\farcs$03, 0", 0$\farcs$10, and 0$\farcs$20. The slit position angle is varied when moving horizontally in the figure (the P.A. values are printed at the top) and the offset is varied when moving vertically in the figure (the offset values are printed on the left). The y values are normalized fluxes from -0.1 to 1.7, while the x values are velocities from -15 km/s to +15 km/s.}
\label{fig:paramtest}
\end{figure*}

\begin{figure}[!htbp]
\begin{center}
 \includegraphics[width=0.5\textwidth]{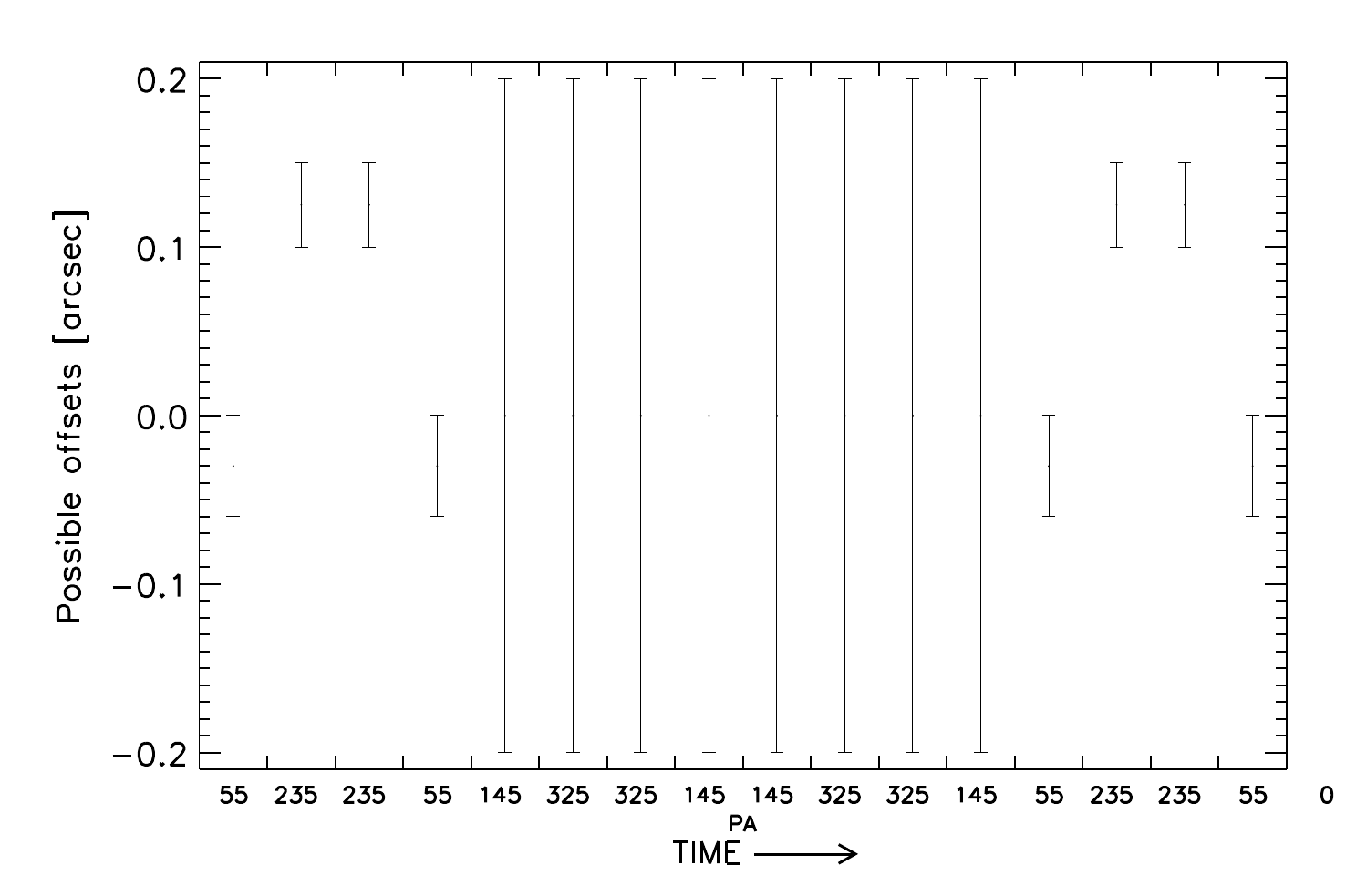}
 \end{center}
\caption{Derived offset versus time for our individual observed spectra collected on the 29th. Time moves forward along the positive direction of the x-axis. Along the x-axis the position angle of each individual spectrum is written.}
\label{off29}
\end{figure}

\begin{figure}[!htbp]
\begin{center}
 \includegraphics[width=0.5\textwidth]{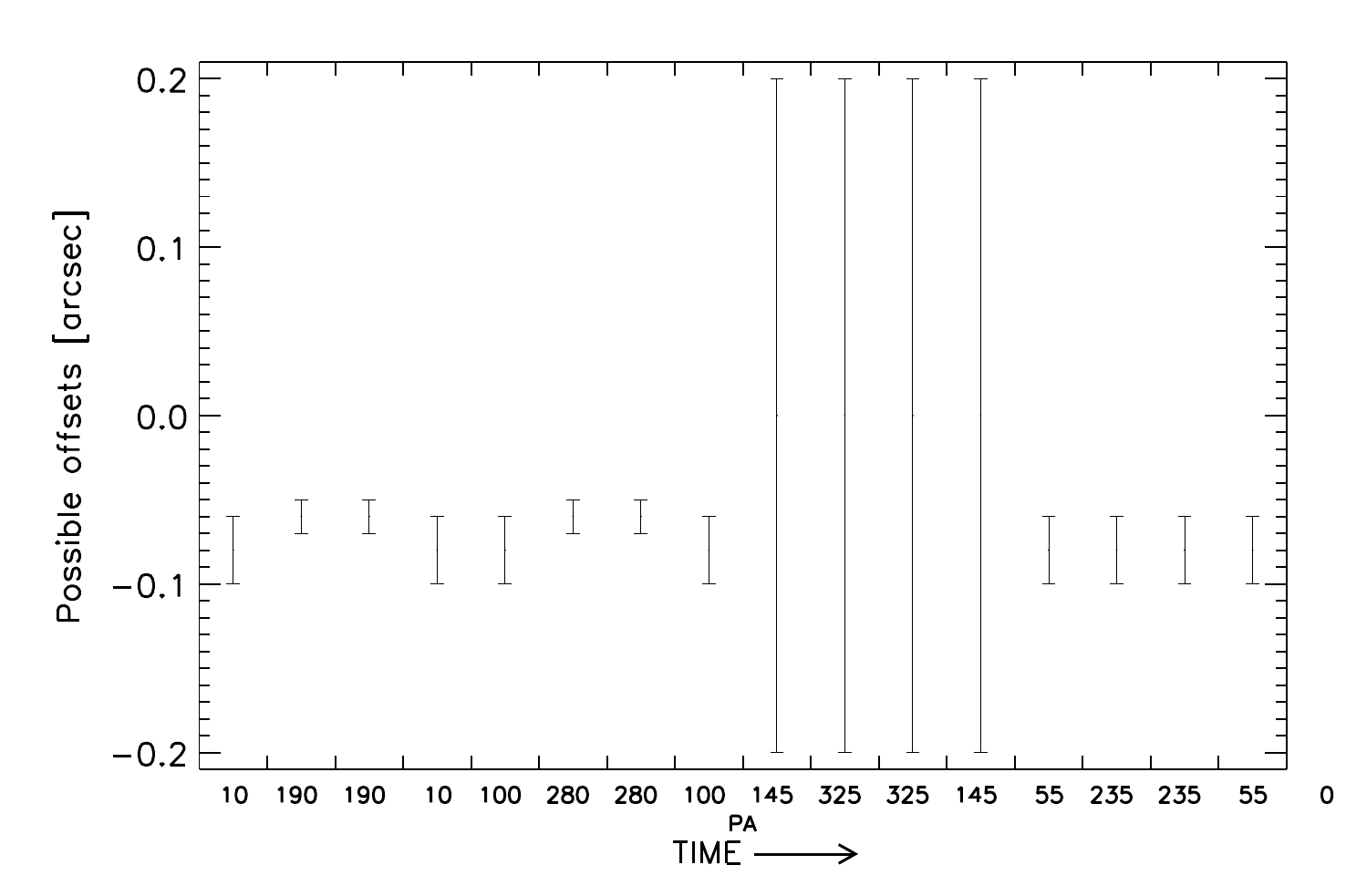}
 \end{center}
\caption{Derived offset versus time for our individual observed spectra collected on the 30th. Time moves forward along the positive direction of the x-axis. Along the x-axis the position angle of each individual spectrum is written.}
\label{off30}
\end{figure}

To explore the variety in line profiles due to varying P.A. and pointing offsets, we select a representative line, v(2-1)R06, and study seven different offsets (-0$\farcs$20; -0$\farcs$15; -0$\farcs$10; -0$\farcs$06; -0$\farcs$03; 0"; 0$\farcs$10; and 0$\farcs$20), measured along the semi-major axis of the disc (positive to the right) while varying the slit position angle from 15$\degree$ to 175$\degree$ with a step size of 10$\degree$.
The full parameter grid of line profile shapes is presented in Fig. \ref{fig:paramtest}.
These tests show that the shape variations seen in our observed data are fully consistent with the variations from a combination of P.A. and/or offset covered in our theoretical grid. 
From our theoretical grid we furthermore derive the plots shown in Figures \ref{off29} and \ref{off30}. Here the pointing offsets needed to explain our observed line profile variations are plotted as a function of Time and P.A. These figures reveal that a pointing offset of -0$\farcs$06 would be an appropriate assumption for most position angles.
%\begin{landscape}
%\end{landscape}
 This offset was imposed on the slit filtering procedure for the simulated lines of the night of the 29th and 30th. For our full selection of lines the slit filtering procedure is run for various position angles based on the settings and the PSF of the observations made on the 29th and 30th of March 2010, listed in Table \ref{table:PA}. 

For comparison purposes, the final modelled line profiles are normalized co-added and averaged for each P.A. setting separately, following the exact same scheme as the combining of the observed line profiles (see Appendix \ref{sec:obs_line} for more details). These normalized combined line profiles are shown in Fig. \ref{fig:mod_prof} with the corresponding average observed line profiles overplotted. Table \ref{tab:modflux} lists the line fluxes and continuum fluxes for each modelled slit filtered line at one particular position angle (P.A.=145$\degree$).

The upper frame of Fig. \ref{fig:mod_prof} shows that the line shape variations, that we see in our observations from the 29th, cannot be explained well by a single slit pointing offset of -0$\farcs$06 (the line profile collected at P.A.=235$\degree$ is leaning toward the wrong side).
The middle and lower frame of Fig. \ref{fig:mod_prof} show that the line shape variations observed on the 30th could in principle be explained by one single consistent pointing offset: The slit filtered model (with a pointing offset of -0$\farcs$6) reproduces the observed line shapes reasonably well. 

\section{A flux-conserving scheme to convert images in polar coordinates
  to regular cartesian grids} \label{app:cube}

\begin{figure}[!htbp]
\centering
\includegraphics[width=85mm]{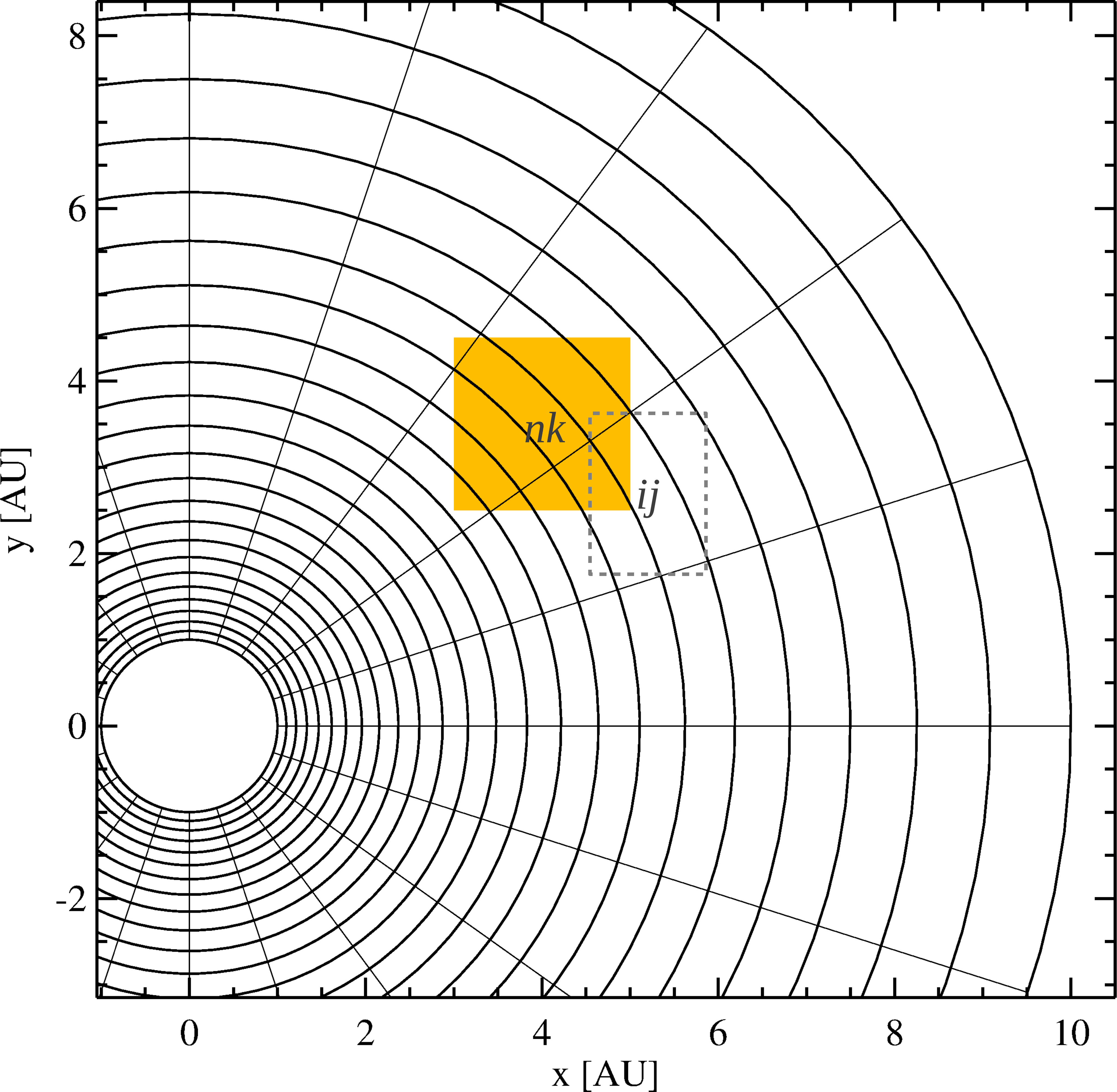}
\caption{Conversion from polar intensities $I_\nu(r,\theta)$ to a
  regular cartesian grid $I'_\nu(x,y)$. The $r_i$ and $\theta_j$
  grid points are drawn as black lines, the yellow box represents a
  regular image pixel between cartesian grid points $x_n$ and
  $y_k$. The dashed box shows the construction of a rectangle that
  exactly contains the polar cell $(i,j)$.}
\label{GridConvert}
\end{figure}

Continuum images and channel maps in ProDiMo are initially calculated
on a polar grid (see \citet{thi2011} for details) with
roughly\footnote{Some adjustments inside the inner rim and increased
  resolution towards the outer disc radius and beyond.}  logarithmic
equidistant radial grid points $\{r_i\,|\,i\!\in\![0...N_r]\}$ and linear
equidistant angular grid points $\{\theta_j\,|\,j\!\in\![0...N_\theta]\}$.
This is necessary to resolve the tiny inner disc rim (important for
the short wavelengths) as well as resolving the outer regions
(important for the long wavelengths) at the same time. The polar intensities
$I_\nu(i,j)$ have associated areas in the image plane as
\begin{equation}
  A_{ij} = \pi(r_i^2-r_{i-1}^2)/N_\theta \qquad\mbox{for\ }
                                       i\!\in\![1...N_r] ,
                                       j\!\in\![1...N_\theta]
\end{equation}
The intensities are assumed to be constant on the polar
"pixels'', i.e. the area $A_{ij}$ bracketed by $r_i$, $r_{i-1}$,
$\theta_j$ and $\theta_{j-1}$, see Fig.~\ref{GridConvert}.

We seek a fast numerical method to convert these polar images 
onto a regular grid with equidistant cartesian image coordinates 
$(x_n,\ n\!\in\![0...N_x])$ and $(y_k,\ k\!\in\![0...N_y])$
with associated pixel areas 
\begin{equation}
  A_{nk} = (x_n-x_{n-1})(y_k-y_{k-1}) \qquad\mbox{for\ } n\!\in\![1...N_x], 
                                               k\!\in\![1...N_y]
\end{equation}
The method is flux-conservative if
\begin{equation}
   \sum_{i=1}^{N_r}\sum_{j=1}^{N_\theta} A_{ij}\,I_\nu(i,j)
 = \sum_{n=1}^{N_x}\sum_{k=1}^{N_y} A_{nk}\,I'_\nu(n,k)
\end{equation}
where $I'_\nu(n,k)$ are the desired intensities on the regular
cartesian pixels.  The exact solution of this problem would be to
calculate the overlap areas, $O(A_{nk},A_{ij})$, between the regular
pixels $A_{nk}$ and any polar pixel $A_{ij}$, then sum up the fluxes 
and divide by the pixel area as
\begin{equation}
  I'_\nu(n,k) = \sum_{i=1}^{N_r}\sum_{j=1}^{N_\theta} 
                \frac{O(A_{nk},A_{ij})}{A_{nk}} \,I_\nu(i,j) \ ,
\end{equation}
but to calculate those overlap areas is painful, see
Fig.~\ref{GridConvert}. A more practical idea is to create a
cartesian rectangular area that exactly contains the polar pixel, see
Fig.~\ref{GridConvert}, by taking the minimum and maximum of
the four corner points of $A_{ij}$
\begin{eqnarray}
  \begin{array}{rp{1mm}lp{1mm}rp{1mm}l}
  x_l &=& \min\{x_1,x_2,x_3,x_4\} && y_l &=& \min\{y_1,y_2,y_3,y_4\} \\
  x_r &=& \max\{x_1,x_2,x_3,x_4\} && y_r &=& \max\{y_1,y_2,y_3,y_4\} \\
  x_1 &=& r_{i-1}\sin(\theta_{j-1}) && y_1 &=& r_{i-1}\cos(\theta_{j-1}) \\
  x_2 &=& r_{i-1}\sin(\theta_j)    && y_2 &=& r_{i-1}\cos(\theta_j) \\
  x_3 &=& r_i\sin(\theta_{j-1})    && y_3 &=& r_i\cos(\theta_{j-1}) \\
  x_4 &=& r_i\sin(\theta_j)        && y_4 &=& r_i\cos(\theta_j) 
  \end{array}
  \label{ConstructPixel}
\end{eqnarray}
The area of this rectangular pixel, 
\begin{equation}
  A'_{ij} = (x_r-x_l)(y_r-y_l) \ ,
\end{equation}
is always larger than the area of the original polar pixel $A_{ij}$,
resulting in a correction factor in Eq.~(\ref{PixelConvert})
below. The area overlaps between $A_{nk}$ and $A'_{ij}$ are now easy
to calculate, and the resulting conversion formula is
\begin{eqnarray}
  C_{nk} &=& \sum_{i=1}^{N_r}\sum_{j=1}^{N_\theta} 
             O(A_{nk},A'_{ij})\frac{A_{ij}}{A'_{ij}}\nonumber\\
  I'_\nu(n,k) &=& \frac{1}{C_{nk}}\sum_{i=1}^{N_r}\sum_{j=1}^{N_\theta} 
             O(A_{nk},A'_{ij})\frac{A_{ij}}{A'_{ij}}\,I_\nu(i,j) \ ,
  \label{PixelConvert}
\end{eqnarray}
The area coverage $C_{nk}$ is close, but not exactly equal to
$A_{nk}$, because some regular pixels will be slightly oversampled 
by the polar pixels, and others slightly under sampled, and this 
effect is automatically corrected for in Eq.~(\ref{PixelConvert}).

This conversion method is fast, reliable, and exactly flux
conservative.  However, the resulting spatial resolution in the
regular image will suffer somewhat in areas where the rectangular
pixels $A'_{ij}$, as constructed from the polar pixels $A_{ij}$, are
larger than the regular pixels $A_{nk}$.  Here, the simulation can be
improved by introducing polar sub-pixels in the first place, i.e.\ by
subdividing large polar pixels into a suitable number of smaller polar
sub-pixels, equally spaced in $r$ and $\theta$, assuming the
intensities to be constant over all sub-pixels, before applying
Eqs.~(\ref{ConstructPixel}) to (\ref{PixelConvert}).

\begin{figure}[!htbp]
\begin{center}$
\begin{array}{cc}
   \includegraphics[width=0.413\textwidth]{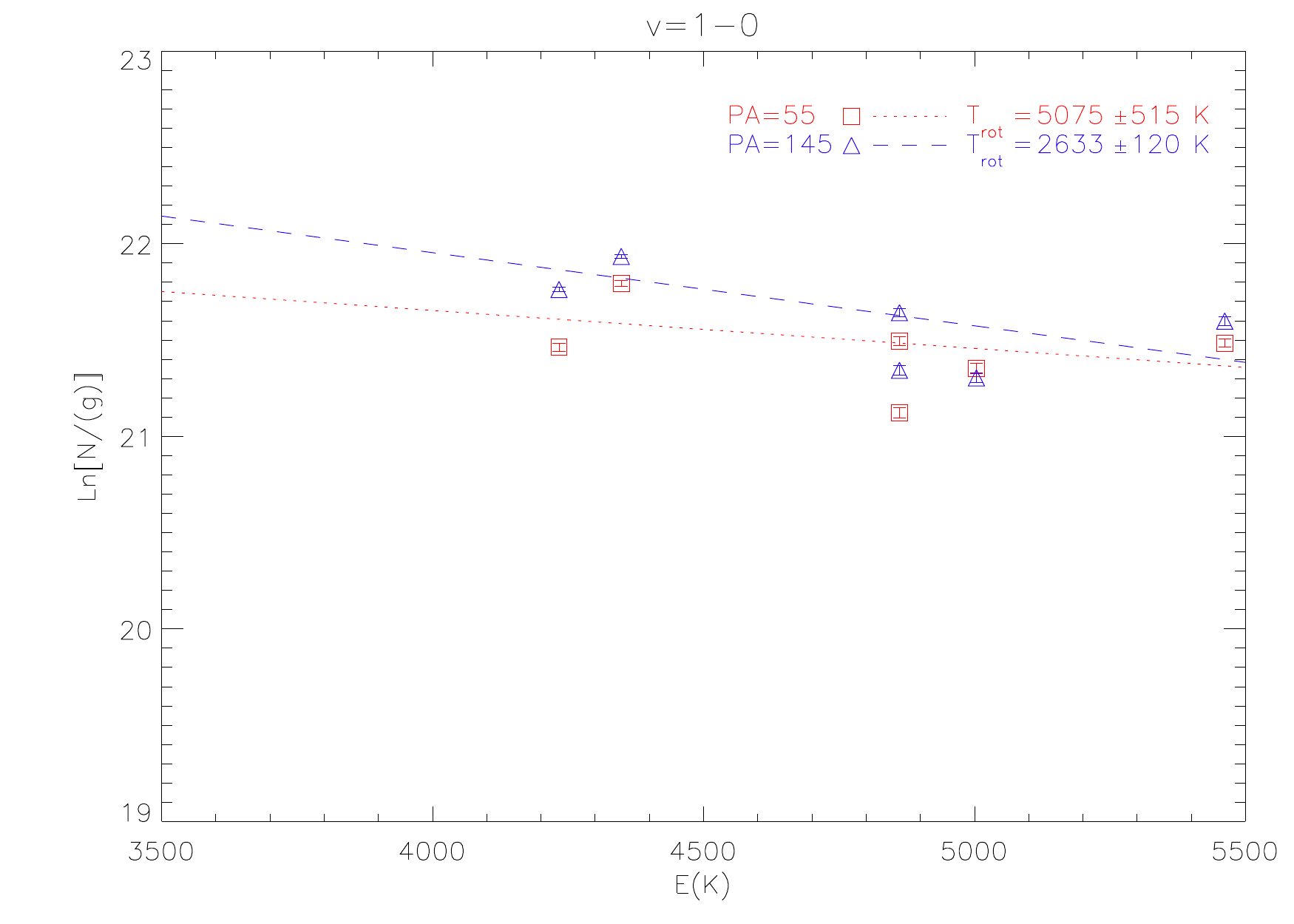}\\
   \includegraphics[width=0.413\textwidth]{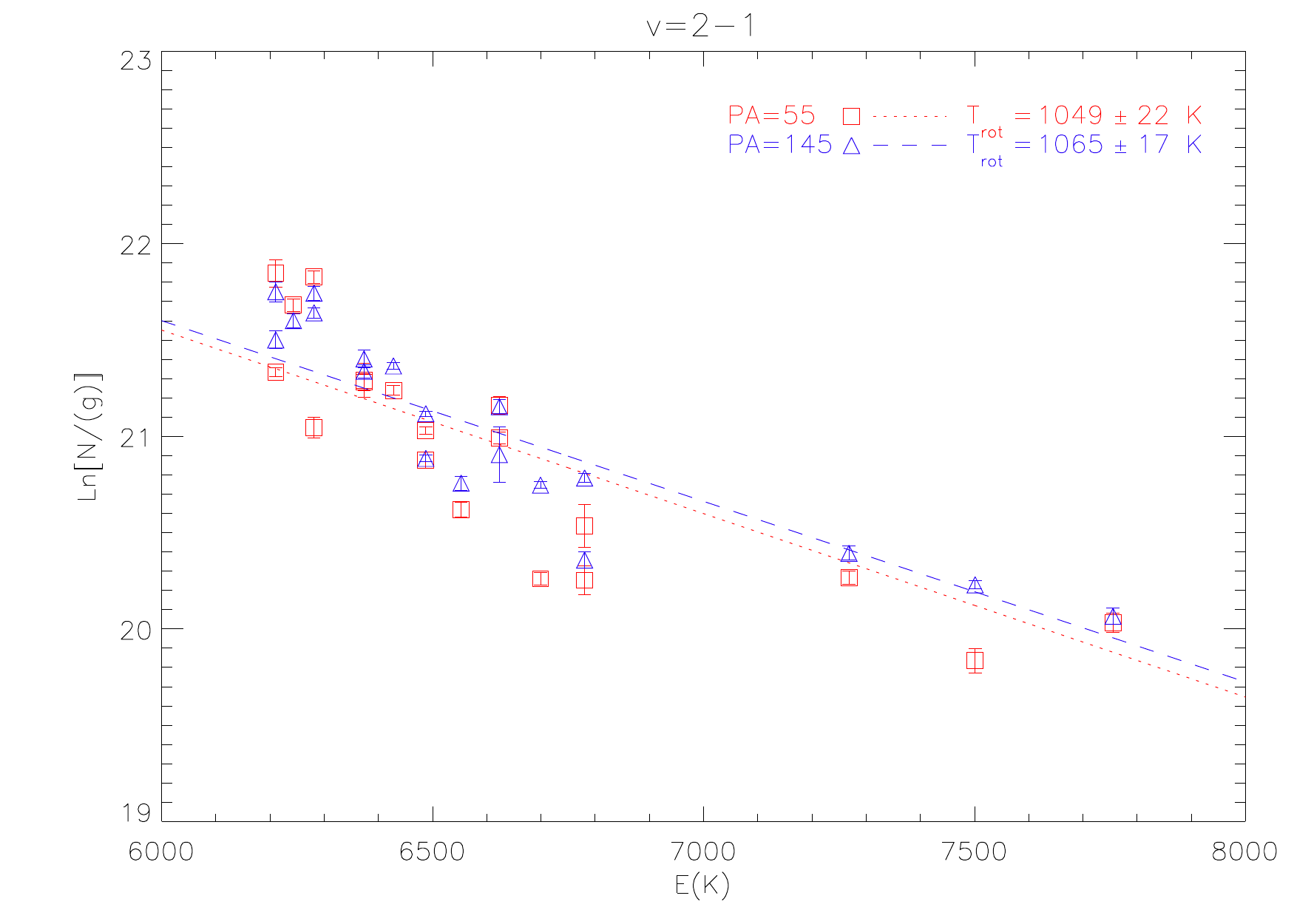}\\
   \includegraphics[width=0.413\textwidth]{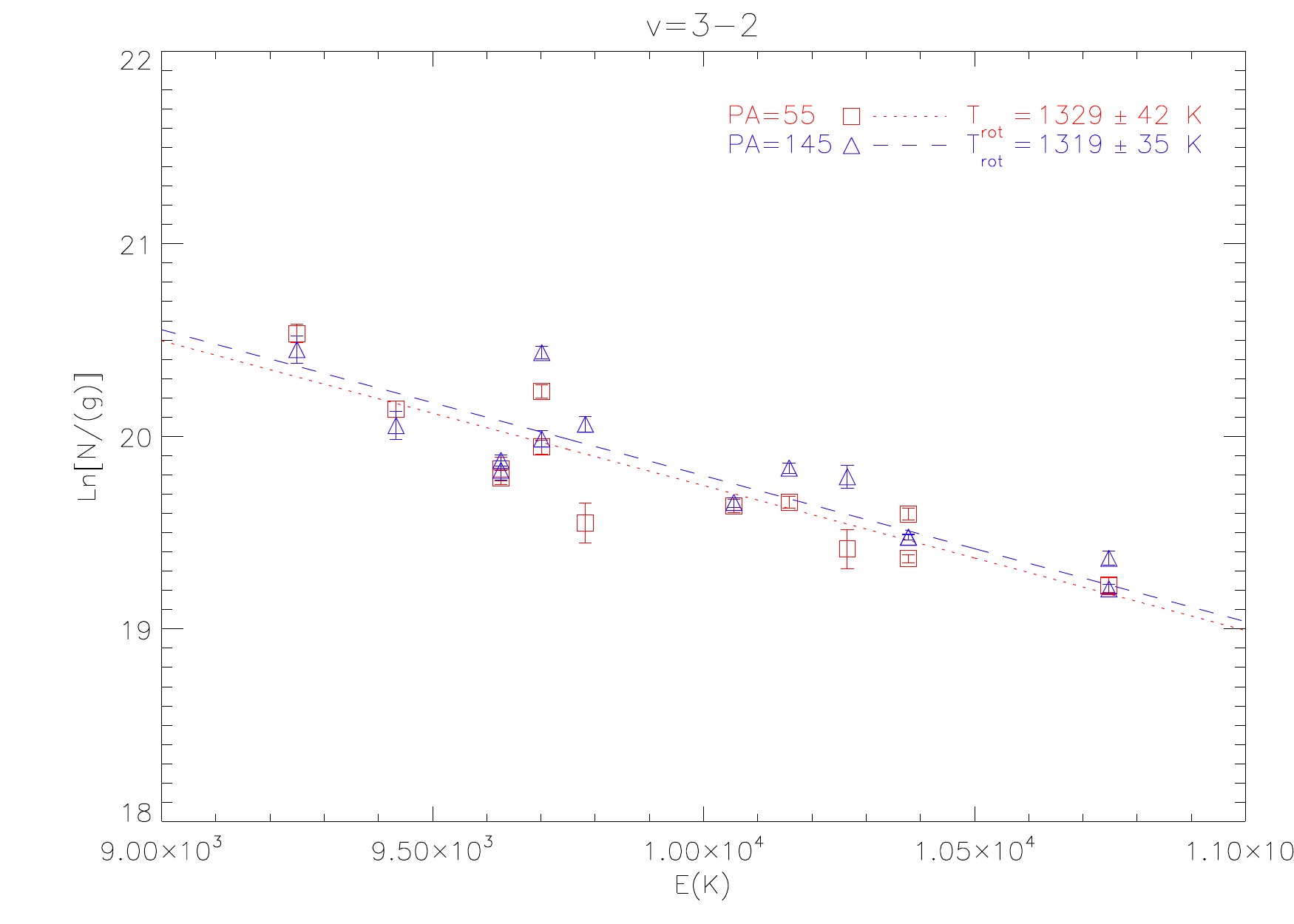}\\
    \includegraphics[width=0.413\textwidth]{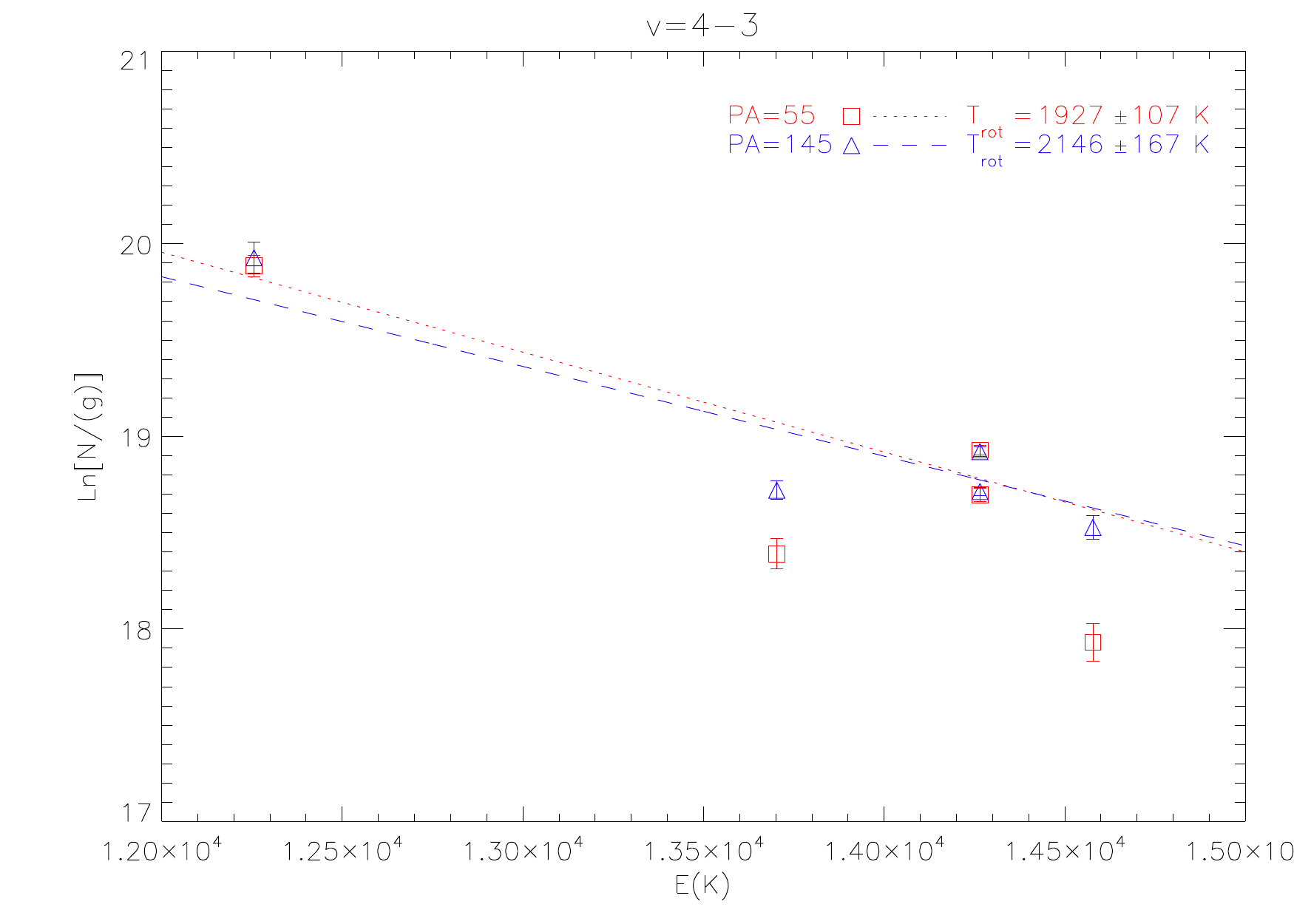}
 \end{array}$
\end{center}
\caption{Rotational diagram for each of the first four v levels made from observational data gathered on the 29th. In each plot, data from two position angles are plotted for comparison (P.A.=55$\degree$ shown as red squares and P.A.=145$\degree$ shown as blue triangles). A linear fit to find $T_{\rm{rot}}$ is made separately for each P.A. The fitted values are printed on the plots together with their formal error bars. However, we expect a somewhat larger error from the limited J range.}
         \label{fig:boltzmannobs29}
\end{figure}

\begin{figure}[!htbp]
\begin{center}$
\begin{array}{cc}
  \includegraphics[width=0.4435\textwidth]{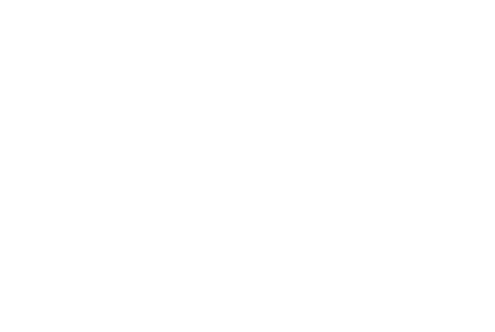}\\
   \includegraphics[width=0.413\textwidth]{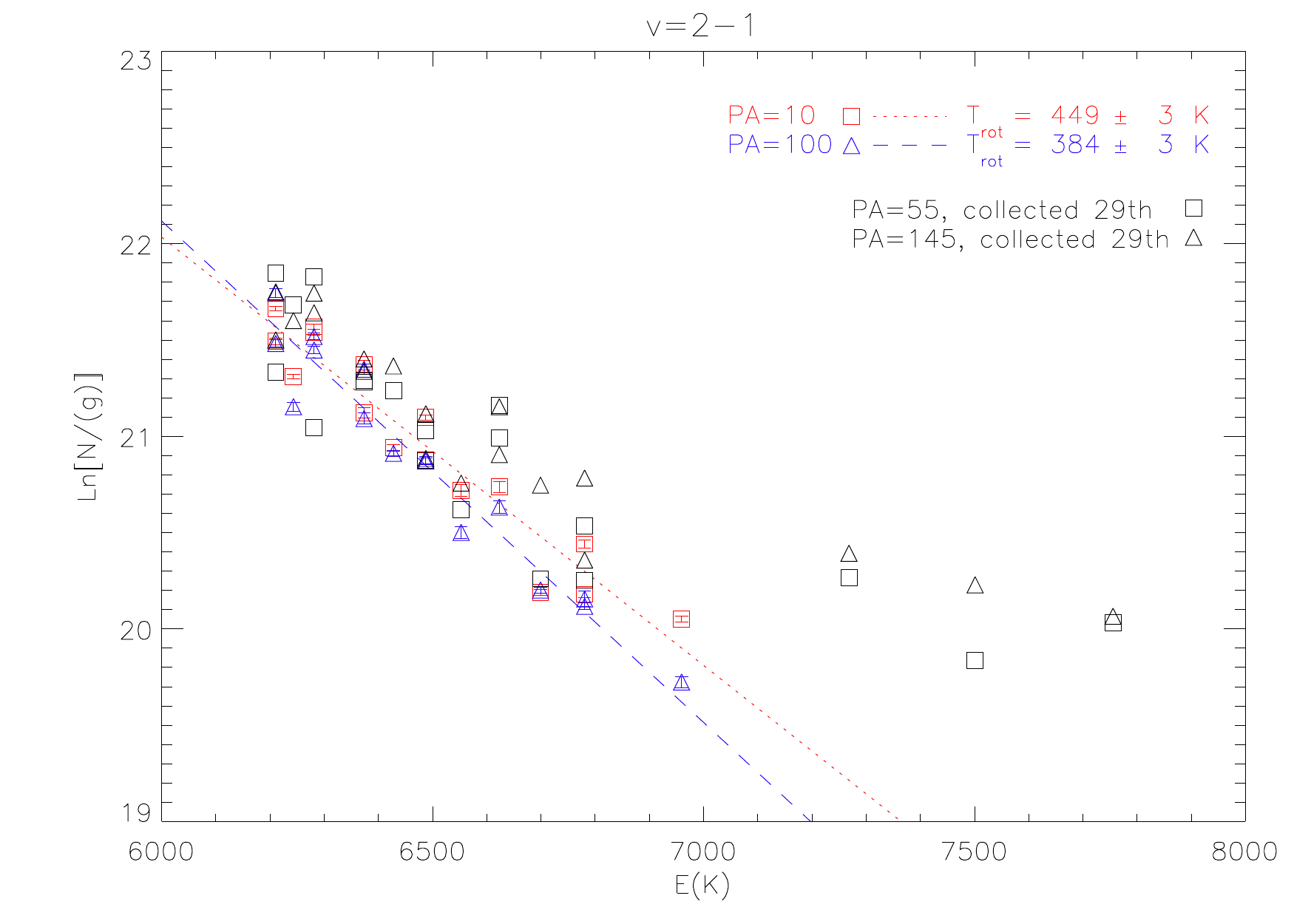}\\
   \includegraphics[width=0.413\textwidth]{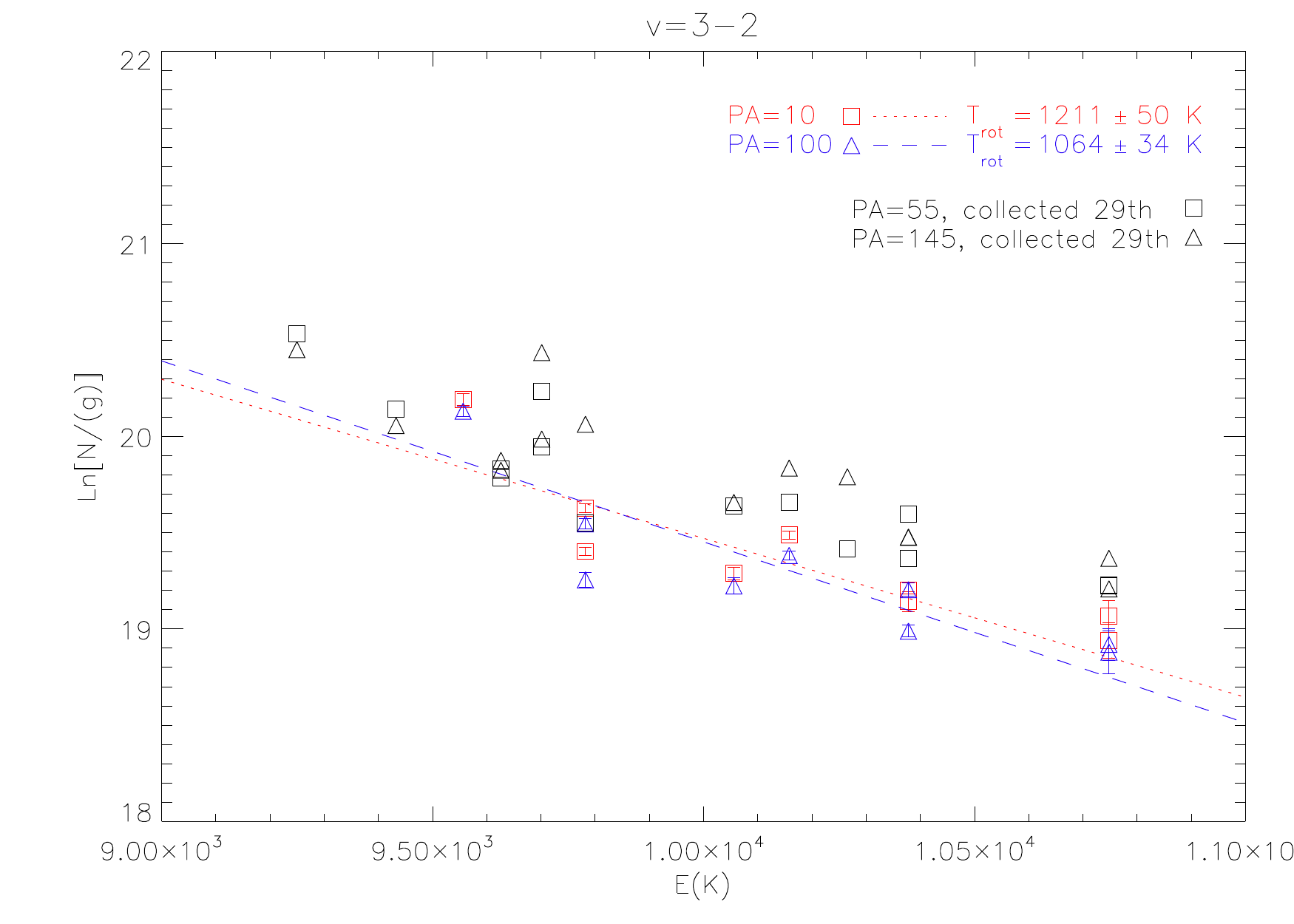}\\
   \includegraphics[width=0.413\textwidth]{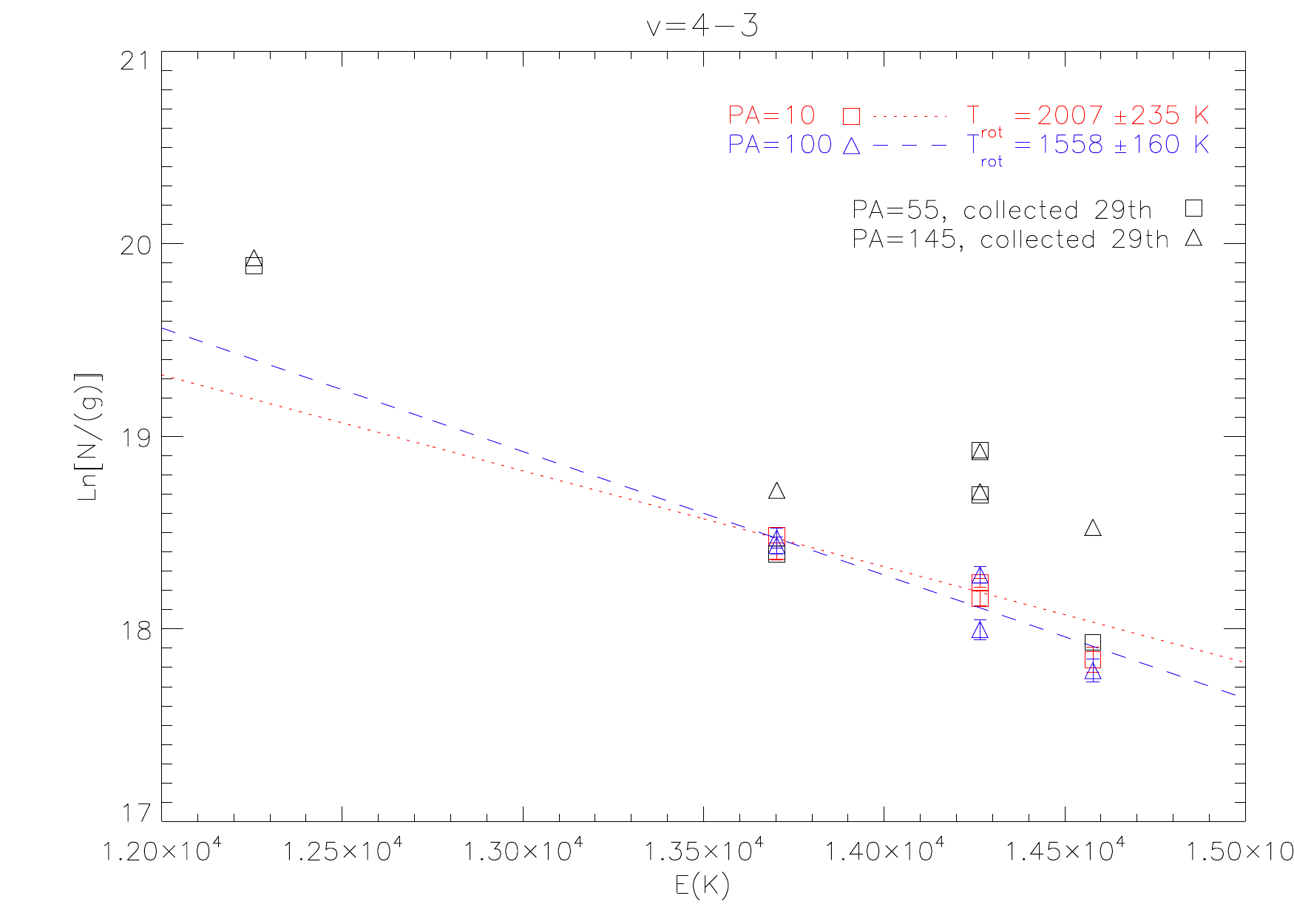}
 \end{array}$
\end{center}
\caption{Rotational diagram for each of the first four v levels made from observational data gathered on the 30th. In each plot, data from two position angles are plotted for comparison (P.A.=55$\degree$ shown as red squares and P.A.=145$\degree$ shown as blue triangles). A linear fit to find $T_{\rm{rot}}$ is made separately for each P.A. The fit is strongly affected by the limited line sample. To visualize this effect the rotational diagrams from the 29th are overplotted in black.}
         \label{fig:boltzmannobs30}
\end{figure}

\section{Population diagrams}

\subsection{Observed rotational diagrams}
Rotational diagrams were compiled, for each vibrational level, using the integrated line fluxes derived from the CRIRES data separating data from different position angles and observing nights. We note that in our sample the most optically thick lines, like the v=1-0 low J lines, are missing (these lines coincide with strong telluric absorption features and are therefore excluded). The rotational diagrams are shown in Figures \ref{fig:boltzmannobs29} and \ref{fig:boltzmannobs30}.
 The details of the equations and quantities used are described in Sect. \ref{sec:boltz}.
We see that there is no big variation between the fit for two position angles taken on the same night. The rotational temperatures found from each of the two observing nights are different, but this is mainly caused by a narrower fitting range in J levels for the second night. The v=2-1 and v=4-3 rotational diagrams are to sparsely populated and no v=1-0 lines were collected in the second night. For the first observing night the v=2-1 rotational diagram shows a T$_{\rm{rot}}\sim$1100 K and the v=3-2 rotational diagram shows a T$_{\rm{rot}}\sim$1300 K.
And for the second night the v=3-2 rotational diagram shows a T$_{\rm{rot}}\sim$1100/1200 K. In these diagrams, the individual lines all fall on a curve typically seen in the presence of optical depth and non-LTE effects, making a linear fit questionable. \citet{thi2012} showed that it is hard to fit a single temperature and that the real rotational temperature can be found in a 'second' turn over at higher J levels. We do not have a wide enough J range for this. The formal error bars for the rotational temperatures are printed on the rotational diagrams but we expect that there is a somewhat larger error from the limited J range. 

\subsection{Modelled rotational diagrams}
\begin{figure}[!htbp]
\begin{center}$
\begin{array}{cc}
   \includegraphics[width=0.415\textwidth]{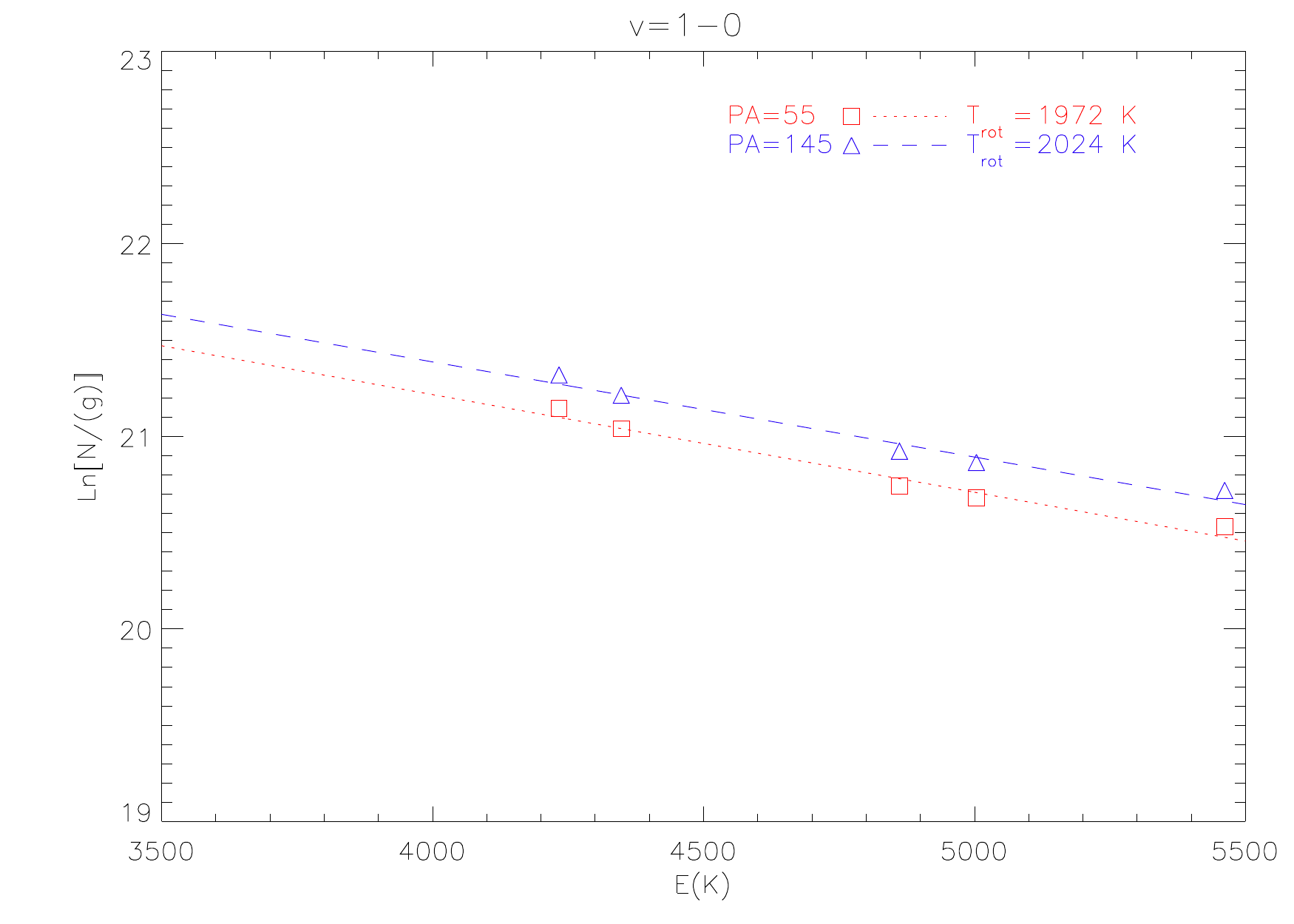}\\
   \includegraphics[width=0.415\textwidth]{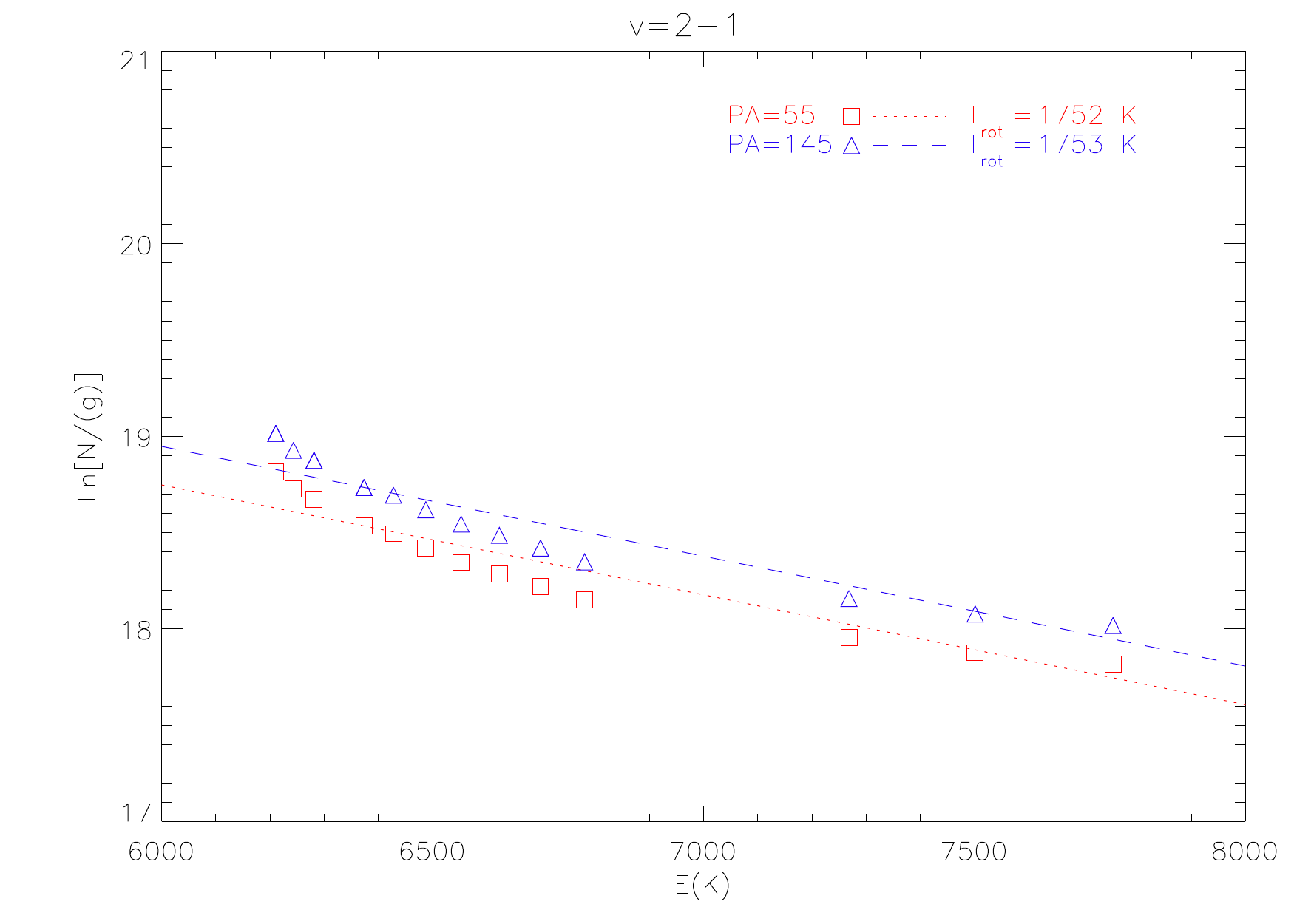}\\
 \includegraphics[width=0.415\textwidth]{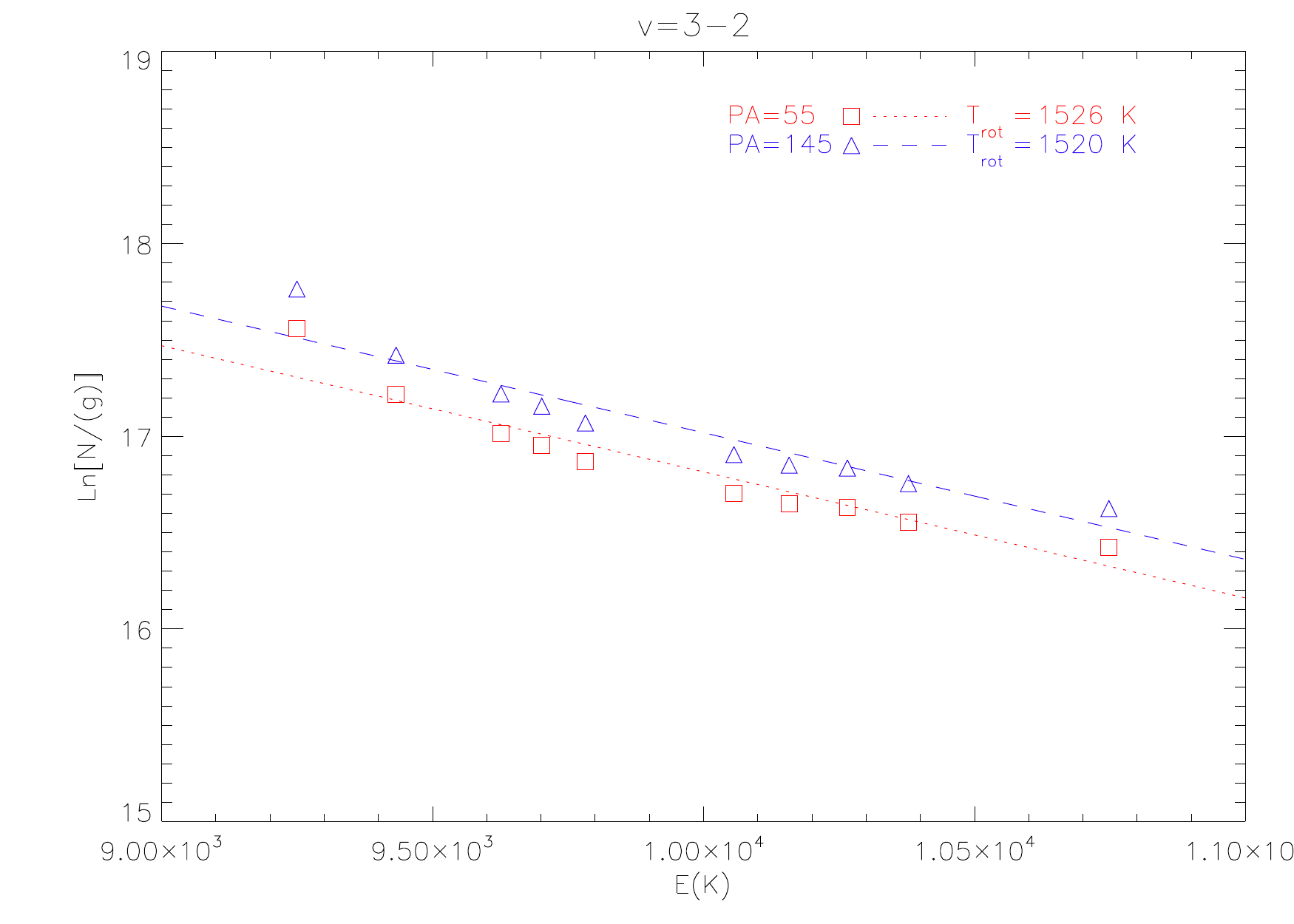}\\
  \includegraphics[width=0.415\textwidth]{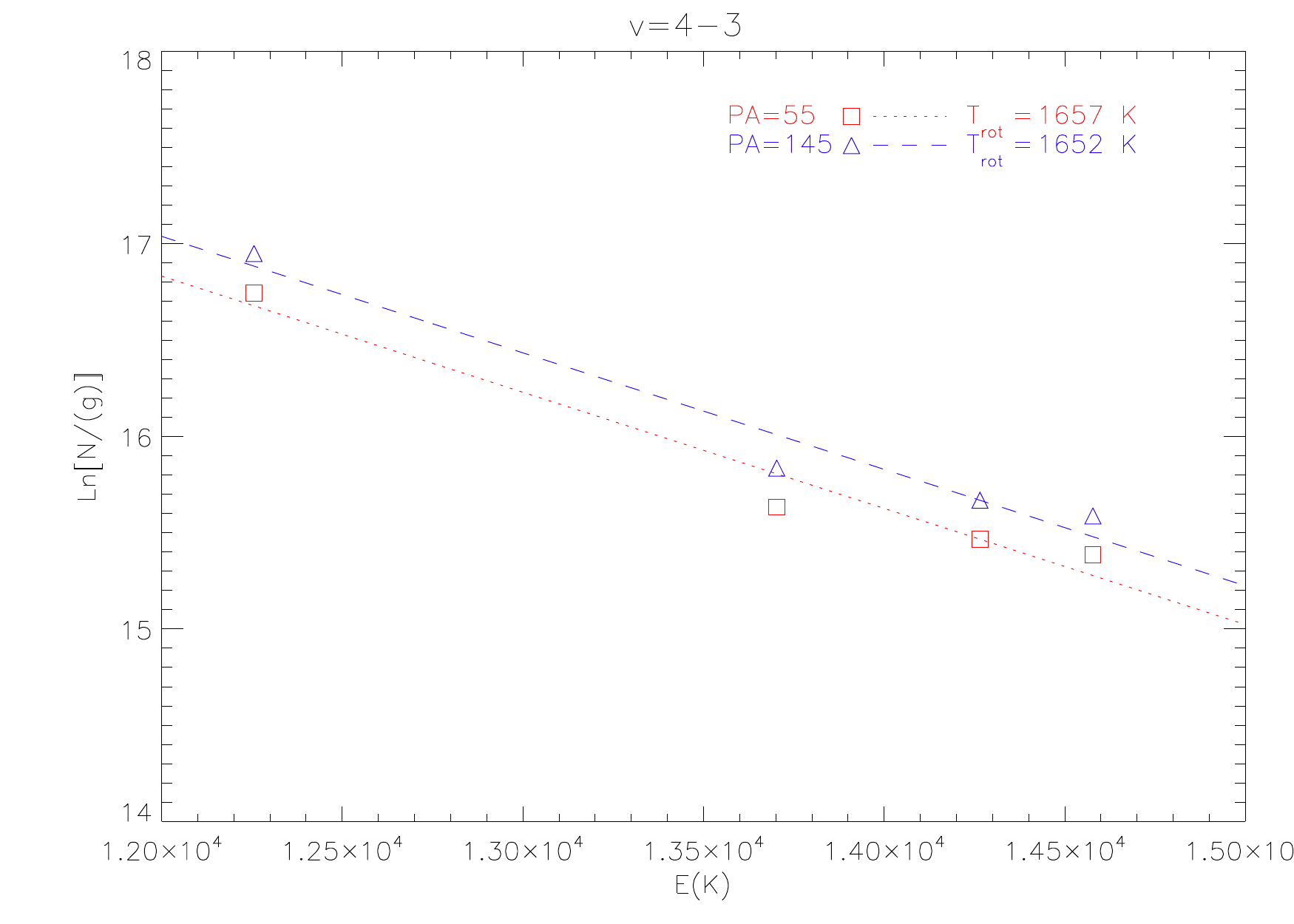}
\end{array}$
\caption{Rotational diagram of the lowest four v levels for the modelled line sample. The fitting is done with the exact same range and line selection as the observational data set collected on the 29th, see Table \ref{tab:obsline}.}
\label{fig:modbol1}
\end{center}
\end{figure}

As described for the observational data, rotational diagrams were produced for the modelled line transitions (comparable to the data collected on the 29th) for two orthogonal position angles separately. The rotational diagrams are shown in Fig.\ref{fig:modbol1} and the details of derivations are described in Sect. \ref{sec:boltz}.
The rotational temperatures found from the modelled lines are higher than those observed, but the lines fall on a curve and it is hard to get a good linear fit. The fit depends heavily on the range of lines included.

\end{appendix}

\end{document}